\documentclass[fleqn,usenatbib]{mnras}

\usepackage{newtxtext,newtxmath}

\usepackage[T1]{fontenc}

\DeclareRobustCommand{\VAN}[3]{#2}
\let\VANthebibliography\thebibliography
\def\thebibliography{\DeclareRobustCommand{\VAN}[3]{##3}\VANthebibliography}

\usepackage{graphicx}	
\usepackage{amsmath}	
\usepackage{bbold}      
\usepackage{commath}
\usepackage{bm}  
\usepackage[normalem]{ulem}

\usepackage{xtab,afterpage}

\title[Carbon Depletion in the Early Solar System]{Carbon Depletion in the Early Solar System}

\author[F. Binkert \& T. Birnstiel]{
Fabian Binkert,$^{1,2}$\thanks{E-mail: fbinkert@usm.lmu.de}
and Til Birnstiel$^{1,2}$
\\
$^{1}$University Observatory, Faculty of Physics, Ludwig-Maximilians-Universität München, Scheinerstr. 1, 81679 Munich, Germany\\
$^{2}$Exzellenzcluster ORIGINS, Boltzmannstr. 2, D-85748 Garching, Germany
}
\date{Accepted XXX. Received YYY; in original form ZZZ}

\pubyear{2022}

\begin{document}
\UseRawInputEncoding
\label{firstpage}
\pagerange{\pageref{firstpage}--\pageref{lastpage}}
\maketitle

\begin{abstract}
Earth and other rocky objects in the inner Solar System are depleted in carbon compared to objects in the outer Solar System, the Sun, or the ISM. It is believed that this is a result of the selective removal of refractory carbon from primordial circumstellar material. In this work, we study the irreversible release of carbon into the gaseous environment via photolysis and pyrolysis of refractory carbonaceous material during the disk phase of the early Solar System. We analytically solve the one-dimensional advection equation and derive an explicit expression that describes the depletion of carbonaceous material in solids under the influence of radial and vertical transport. We find both depletion mechanisms individually fail to reproduce Solar System abundances under typical conditions. While radial transport only marginally restricts photodecomposition, it is the inefficient vertical transport that limits carbon depletion under these conditions. We show explicitly that an increase in the vertical mixing efficiency, and/or an increase in the directly irradiated disk volume, favors carbon depletion. Thermal decomposition requires a hot inner disk (> 500 K) beyond 3 AU to deplete the formation region of Earth and chondrites. We find FU Ori-type outbursts to produce these conditions such that moderately refractory compounds are depleted. However, such outbursts likely do not deplete the most refractory carbonaceous compounds beyond the innermost disk region. Hence, the refractory carbon abundance at 1 AU typically does not reach terrestrial levels. Nevertheless, under specific conditions, we find photolysis and pyrolysis combined to reproduce Solar System abundances. 
\end{abstract}



\section{Introduction}
\label{sec:introduction}
The rocky bodies in the Solar System, i.e., rocky planets, cores of gas giant planets and small bodies, have formed from solid material inherited from the interstellar medium (ISM). This material had been delivered to the protosolar disk during the infall process in the early phase of star and disk formation before it was available to be incorporated into the rocky Solar System objects and their parent bodies. Therefore, the composition of rocky bodies in the Solar System today is linked to the composition of the refractory material available during their formation and the physical/chemical alteration processes which have acted on the bodies since their formation. \newline
Without any processing, the composition of the rocky bodies in the Solar System should be identical to the composition of the refractory material present in the ISM. Focusing specifically on carbon, we expect about 50 per cent of the carbon in the ISM to be bound in solid form, i.e. dust, and thus to be potentially refractory \citep[][]{Zubko04}. Relative to silicon, the solid component of the ISM has a carbon-to-silicon elemental abundance ratio of C/Si $\sim 6$ \citep[][]{Bergin15}. However, the primary form and morphology of the solid carbonaceous material remains uncertain. Suggested components include amorphous carbon or hydrocarbon grains and/or aromatic and aliphatic compounds. \cite{Gail2017} summarized the available information on the interstellar carbonaceous material and constructed a representative model which consist of 60 per cent organic material containing large amounts of H, O and N atoms, 10 per cent of pure carbon dust in amorphous form, 20 per cent of more moderately volatile materials which include aromatic and aliphatic compounds and 10 per cent other components. \newline
The solid interstellar carbonaceous material which has not been incorporated into Solar System bodies after the infall onto the early solar disk has likely been accreted by the Sun. This is also the case for the volatile components, which have not been accreted by gas-giant planets or have not been lost due to dissipation. The solar photosphere shows a comparable carbon-to-silicon abundance ratio to that of the ISM with a value of C/Si $\sim 8$ \citep[][]{Bergin15,Grevesse10}. The similarity to the value in the ISM suggests that the reported C/Si ratio is somewhat universal in the solar neighborhood, and thus also in the solid material that was available at the beginning of planet/planetesimal formation in the Solar System.
Therefore, one would expect to find all the components which are at least moderately refractory in planetesimals and subsequently also in the rocky components of the Solar System, except in regions where high temperatures or irradiation lead to the destruction of the refractory compounds. This is typically only the case in the inner disk region \citep[$\lesssim 0.5$ AU][]{Alessio15} or in the directly illuminated disk atmosphere. However, measurements of the carbon abundance in rocky Solar System objects reveal a significant depletion in carbon compared to the solid components in the ISM. The carbon content in the bulk silicate Earth (BSE), that is the Earth's mantle and crust without the core, is with a C/Si ratio of $1.1\cdot 10^{-3}$ more than three orders of magnitude depleted compared to ISM and solar values \citep[][]{Bergin15}. Even if one considers additional carbon to be incorporated in Earth's core, the entire planet is depleted in carbon by several orders of magnitude \citep[][]{Li21,Allegre01}. There is also evidence that ISM material has survived unprocessed to be incorporated into meteorites originating in the asteroid belt \citep[][]{Alexander17}. However, the amount of refractory carbon relative to silicon in these meteorites is also deceased by 1-2 orders of magnitude compared to what is available in the ISM. This is the case even in the least processed and most primitive meteorites, the carbonaceous chondrites \citep[][]{Bergin15,Geiss87}. Beyond the asteroid belt, where comets are expected to have formed, the C/Si ratio is comparable to that in the ISM. There is a clear radial gradient in the Solar System, with the inner Solar System being depleted in refractory carbon relative to silicon (see e.g. Figure 2 in \cite{Bergin15} or \autoref{fig:C_fraction_overview} in this work). \newline
This trend does not only exist in the Solar System, but is also observed in systems of polluted white dwarfs. Analysis of the spectra of white dwarfs gives insight into elemental abundances of their atmosphere, and due to the strong gravity, heavy elements are not expected to be present there. Therefore, the observed traces of heavy elements in the atmosphere can be explained by tidally disrupted, carbon depleted rocky objects that are accreted by the star \citep[][]{Jura06}. Based on a spectral analysis, \cite{Xu2014} report the elemental composition of extrasolar rocky planetesimals, based on traces in the atmospheres of polluted white dwarfs, to resemble, to zeroth order, the composition of bulk Earth (also see section \ref{sec:CoE}). In a more extended study, \cite{Xu2019} confirm their previous results and report the detailed elemental composition of 19 polluted white dwarf atmospheres and a carbon mass fraction in the range between $\sim3\cdot 10^{-4}$ and $\sim10^{-1}$ in their samples. Overall, this points to the conclusion that the mechanism, responsible for refractory carbon destruction, might be universal and not unique to the Solar System. Furthermore, the depletion mechanism must be active before planetesimal formation or the formation of the parent bodies of current Solar System objects, to explain the observed carbon depletion. It is likely that this mechanism is either a thermally-induced (e.g. pyrolysis, oxidation, evaporation/sublimation) or photo-induced (e.g. photochemical) process (or a combination of both). Both options have been studied extensively. Depletion via pyrolysis (thermal decomposition without oxidation) and oxidation has recently been studied by \cite{Gail2017} who studied refractory carbon destruction by short period flash heating events. \cite{Li21} suggest sublimation to be responsible for the observed carbon depletion. As for the photo-induced mechanisms, \cite{Lee2010} studied the erosion of carbon grains via hot oxygen atoms in the UV-illuminated region of the early solar disk. Refractory carbon aggregates are also turned into more volatile compounds if they react with oxygen-bearing species, such as OH or free atomic O \citep[][]{Gail02_b,Gail01,Finocchi97}. \cite{Siebenmorgen12} and \cite{Siebenmorgen10} studied the destruction of polycyclic aromatic hydrocarbons (PAHs) by X-ray and extreme ultraviolet (EUV) photons. Additionally, refractory carbon grains are directly photochemically destroyed by FUV photons via photolysis \citep[][]{Anderson2017,Alata15,Alata14}. While early studies had pointed to the conclusion that photo-induced processes can explain the observed carbon deficiencies in the Solar System, \cite{Klarmann2018} found that radial and vertical transport of carbon grains in the disk constitutes an obstacle to photo-induced refractory carbon depletion when explaining Solar System abundances. In this study, we build upon the work of \cite{Klarmann2018} by developing an analytical model of refractory carbon depletion via photolysis to show that the barriers imposed by grain transport are not insurmountable if the total dust mass (or the global dust-to-gas ratio) is low enough. Further, we extend our analytical model to include the refractory carbon depletion via sublimation. 

\section{Carbon on Earth}
\label{sec:CoE}
The carbon content in the bulk silicate Earth (BSE), that is the Earth's mantle and crust without the core, is more than three orders of magnitude depleted compared to ISM and solar values \citep[see compilations in e.g.][]{Bergin15, Lee2010}. The mass fraction of carbon in the BSE is estimated to be $(1.4\pm0.4) \cdot 10^{-4}$ \citep{Hirschmann18}, but the carbon content in the core is not known exactly. We summarize estimates for Earth in \autoref{table:table1}. The core is less dense than pure iron, thus, it must contain lighter elements. If the density difference is fully compensated with carbon alone, \cite{Li21} estimate the carbon mass fraction in the core to be less than $5.0$ per cent. However, this estimate is very generous as the core contains significant amounts of other lighter elements (e.g. sulfur, silicon, oxygen). More realistic estimates yield a carbon mass fraction of $\sim0.5-1$ per cent \citep[e.g.][]{Wood13, Allegre01}. For bulk Earth, i.e. including the core, \cite{Allegre01} arrive at a carbon mass fraction of 0.17 per cent up to 0.39 per cent. \cite{Li21} argue for a more realistic upper bound of $(0.4\pm0.2)$ per cent by mass for bulk Earth. The upper bound of \cite{Li21} is higher than, and thus consistent with, geochemical estimates of $0.053\pm0.021$ per cent by mass \citep[][]{Marty12} or the results of \cite{Fischer20} who estimate a range of 0.037-0.074 per cent by mass for the carbon mass fraction in the bulk Earth. On the other hand, the mass fraction of silicon in the BSE is about 21 per cent \citep[][]{Li21}. Considering the BSE carbon mass fraction of $0.014$ per cent quoted above, this value is in agreement with \cite{Bergin15} who estimate a C/Si atomic ratio in the BSE of $(0.11\pm0.04)$ atomic per cent. Silicon is less abundant in Earth's core than in the mantle and the crust. \cite{Wade2005} estimate a mass fraction of 5-7 per cent. Combining the estimates of \cite{Li21} and \cite{Wade2005}, we estimate the silicon mass fraction of the bulk Earth at $15.9-16.5$ per cent. This result is also in rough agreement with \cite{Allegre01} who estimate the bulk Earth silicon mass fraction as $17.1\pm0.1$ per cent. We use the former range and the carbon mass fraction from \cite{Li21} to calculate an upper bound for the C/Si atomic ratio in the bulk Earth of $5.7-5.9$ per cent. This value is more than 50 times larger compared to the BSE because of the large amounts of carbon which are possibly present in Earth's core. We also calculate the C/Si atomic ratio in the bulk Earth using the lower estimate of the carbon mass fraction of \cite{Fischer20} which yields a range of $0.53-1.1$ atomic per cent. This value is only five to ten times larger than in the BSE. These results are summarized in \autoref{table:table1}. \newline

\begin{table*}
\begin{center}
\caption{\label{table:table1} Overview of the carbon and silicon abundances on Earth as described in section \ref{sec:CoE}. Listed are literature values for the bulk silicate Earth (BSE), Earth's core and bulk Earth. The last column lists the carbon fraction $f_c$, the fundamental quantity which we evolved in this work.}
\begin{tabular}{ l l l l l } 
 \hline
 \hline
   &  C mass fraction (wt\%) & Si mass fraction (wt\%) &  C/Si atomic ratio (at.\%) & $f_c$   \\ 
  \hline
 BSE &  $0.014\pm0.004\mathrm{^a}$  & $21\mathrm{^b}$  & $0.11\pm0.04\mathrm{^c}$ & $(8\pm1)\cdot10^{-5}$\\
  \hline
 Earth's core & $\sim0.5-1\mathrm{^d}$ & $5-7\mathrm{^e}$  & $16.7-23.4$ & $(1.1-1.6)\cdot10^{-2}$\\
  \hline
 Bulk Earth & $<0.4\pm0.2\mathrm{^b}$ &  $15.9-16.5$ & $<(5.7-5.9)$ & $<(4.0-4.1)\cdot10^{-3}$ \\
  & $0.037-0.074\mathrm{^f}$ & & $0.53-1.1$  & $(3.7-7.7)\cdot10^{-4}$ \\
 \hline
\end{tabular}
\end{center}
Sources: $\mathrm{^a}$\cite{Hirschmann18}, $\mathrm{^b}$\cite{Li21}, $\mathrm{^c}$\cite{Bergin15}, $\mathrm{^d}$\cite{Wade2005},  $\mathrm{^e}$\cite{Wood13}, $\mathrm{^f}$\cite{Fischer20}
\end{table*}

\section{Model and dust transport}
In this section, we present the details of our dust model, describe the stationary disk model and introduce the equations that describe the transport of the dust components within the disk. Further, we introduce the local carbon depletion timescale $\tau_c$ to describe the timescale a process depletes the disk of refractory carbon.

\subsection{Dust Model}
\label{sec:dust_model}

\begin{figure}
\includegraphics[width=0.95\columnwidth]{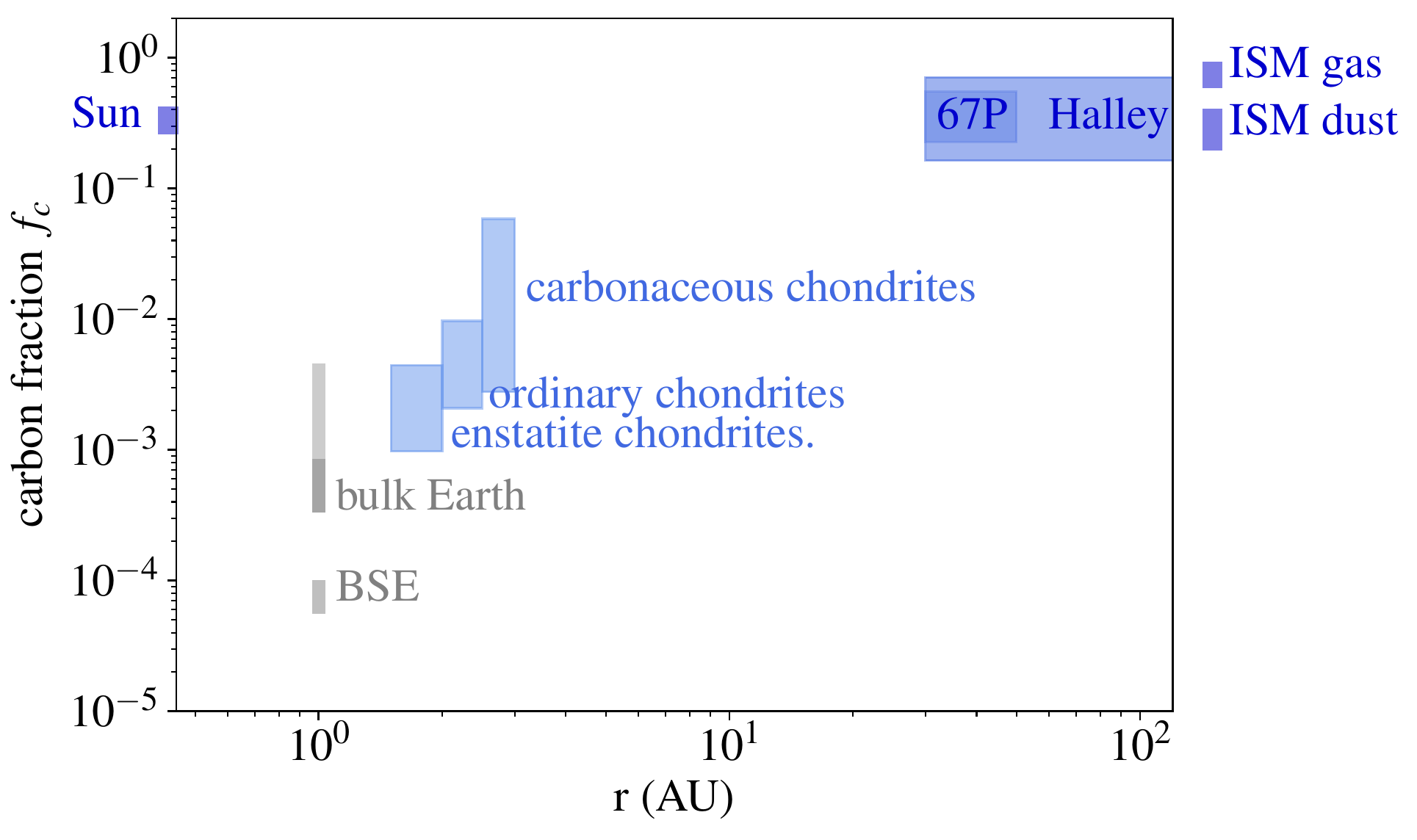}
\caption{Carbon abundance $f_c$, i.e. the mass fraction of refractory carbon, in some Solar System bodies and their heliocentric distance. In horizontal axes, the width of the boxes represent the expected formation regions of individual objects. For chondrites, the vertical height of the boxes represent the spread in measured values within individual classes. For the other objects, the vertical height represents uncertainties in the expected carbon fraction. For bulk Earth, we show estimates based on detailed geochemical modelling in dark gray and possible upper bounds in light gray. Further, we show estimates for bulk silicate Earth (BSE), i.e. Earth's mantle and crust without the core, and two comets, 67P/Churyumov-Gerasimenko and 1P/Halley. For comparison, values of the Sun and the ISM are added. See section \ref{sec:CoE} and \ref{sec:dust_model} for further explanation and references.}
\label{fig:C_fraction_overview}
\end{figure}
\begin{figure*}
\includegraphics[width=1.8\columnwidth]{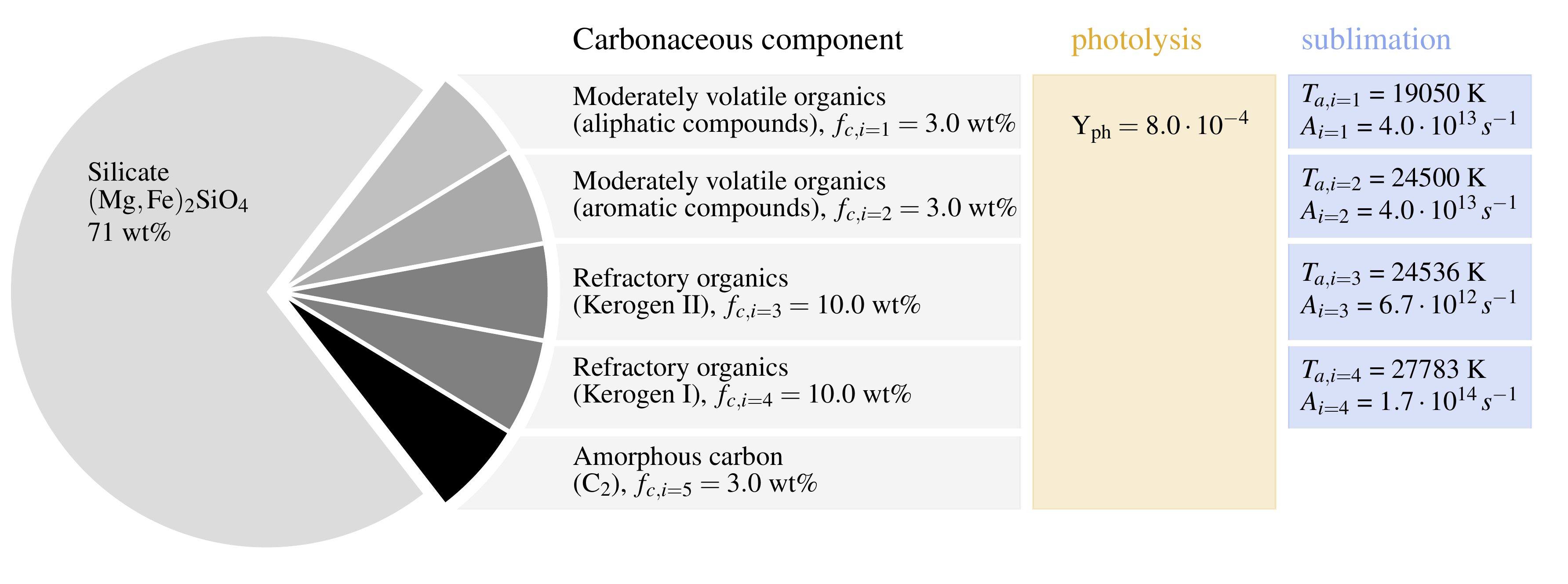}
\caption{Refractory dust composition (Note, for better visualization, the size of the individual wedges of the carbonaceous components are all equal in size and not proportional to their respective fractional abundance. Only their combined size accurately represents the total initial fractional abundance of carbonaceous material (29 \%). The yellow and blue boxes contain the model parameters for photolysis and irreversible sublimation. Note that we do not consider sublimation of the amorphous carbon compound in this study.}   
\label{fig:refarctory_dust_composition}
\end{figure*}

In sections \ref{sec:introduction} and \ref{sec:CoE}, we have summarized the detailed accounting of the C/Si atomic ratio and/or the individual elemental mass fractions of carbon and silicon obtained in remote observations and/or direct measurements of objects in the Solar System. In the Solar System, it is expected that all the solid bodies that are currently observed (i.e. planets, asteroids, comets, etc.), have formed from solid refractory components that already existed in the early Solar nebula. In the remainder of this study, we model the evolution of the early solid refractory compounds (carbonaceous and non-carbonaceous) in a young prestellar disk, in conditions similar to the early Solar System. Particularly, we model the refractory compounds as two distinct dust grain populations, a non-carbonaceous component, and a carbonaceous component, of which we further divide the latter into five individual carbonaceous compounds which all are subject to photo- and thermal decomposition processes. We model the non-carbonaceous dust component as silicate and assume, for simplicity, that every silicon atom in this component is locked up in bare silicate $\big(\mathrm{(Mg,Fe)_2SiO_4}\big)$, similar to the dust model in \cite{Zubko04}. Meaning, for every silicon atom, we add another six atoms, which together form the atomic composition of the non-carbonaceous bare silicate dust compound. With this approach, the relative abundance of atoms in $\mathrm{(Mg,Fe)_2SiO_4}$ roughly agrees with relative abundances in the ISM \citep{Zubko04}. \newline
In order to track the abundance of refractory carbon in dust conglomerates consisting of carbonaceous and non-carbonaceous components, we track the individual mass of each component, i.e. $M_c$ is the total mass of carbonaceous material and $M_s$ is the total silicate mass in a conglomerate. We define the \textit{carbon fraction} $f_c=M_c/(M_c+M_s)$ to trace the carbon mass fraction relative to the total refractory dust mass. In order to connect the carbon fraction $f_c$ to observational data, we convert the C/Si atomic ratios reported in sections \ref{sec:introduction} and \ref{sec:CoE}, to carbon fractions $f_c$, assuming every silicon atom is locked up in bare silicate. The obtained carbon fractions for Earth are listed in the last column of \autoref{table:table1}. In \autoref{fig:C_fraction_overview}, we visualize the carbon fractions of some Solar System bodies and the ISM. The data is based on the review of \cite{Bergin15} to which we have added the estimate for bulk Earth as discussed in section \ref{sec:CoE}. In \autoref{fig:C_fraction_overview}, the horizontal extent of the boxes corresponds to the  heliocentric distances of the expected formation region of individual objects. We place the formation location of carbonaceous chondrites at $1.5-2$ AU, the location of ordinary chondrites at $2.0-2.5$ AU and the location of enstatite chondrites at $2.5-3.0$ AU \citep[][]{Morbidelli12}. The range of carbon fractions in chondrites is based on the C/Si ratios from \cite{Bergin15}. For chondrites, the vertical extent of the boxes in \autoref{fig:C_fraction_overview} represents the ranges of measurements of different samples and not uncertainties. Unlike all the other boxes, which represent model and/or measurement uncertainties. We place the origin of comet 67P/Churyumov-Gerasimenko (67P) in the Kuiper belt ($30-50$ AU) where, according to a long-standing hypothesis, Jupiter-family comets originate \citep[][]{Duncan97}. We highlight that this hypothesis has somewhat weakened in recent years, and today it is also thought possible that Jupiter-family comets formed over a wider range of distances from the Sun \citep[][]{Altwegg2015}. It is also possible that comet 1P/Halley (Halley) originates from regions beyond the Kuiper belt \cite[][]{Jewitt2002}. Further, we arrive at a carbon fraction $f_c=0.29$ for the dust component of the ISM. This is close to the carbon fraction $f_c=0.25$ assumed in \cite{Klarmann2018}, who use an identical definition of the carbon fraction. The upper bound for Bulk Earth is $f_c=4.1\cdot 10^{-3}$, less than two orders of magnitude below ISM values. The estimated range based on geochemical modelling is in the range $f_c=3.7-7.7\cdot 10^{-3}$, roughly another order of magnitude lower.\newline
After detailing the total amount of refractory carbonaceous material expected to be found in presolar material, we now specify its composition. We divide the refractory carbonaceous component into five distinct carbonaceous compounds with different decomposition properties. For this, we follow the model of \cite{Gail2017} and divide the carbonaceous components into the moderately volatile aliphatic (3.0 per cent by mass) and aromatic compounds (3.0 per cent by mass), the refractory hydrocarbon compounds Kerogen I (10.0 per cent by mass) and kerogen II  (10.0 per cent by mass) and a fifth component, amorphous carbon (3.0 per cent by mass). The mass fractions in brackets are given relative to the total refractory dust mass and add up to the 29 per cent by mass of carbonaceous material. The relative abundances are thought to reflect our (limited) knowledge of the composition of cometary material and interplanetary dust particles (IDPs). In \autoref{fig:refarctory_dust_composition}, we illustrate composition and relative abundance of the refractory dust components considered in or model. 

\subsection{Disk Model}
\label{sec:disk_model}
With the goal of modelling the conditions in the early Solar System in mind, we assume a  a pre-main-sequence star located at the center of a circumstellar disk. We adopt stellar parameters for a Solar-mass pre-main-sequence star ($M_\star\:=\:1 M_\odot$) from \cite{Siess2000} at 1 Myr with a total luminosity of $L_\star\:=\:2.39 L_\odot$. We assume the FUV-luminosity to be $L_{UV} = 0.01L_*$, as proposed by, e.g., \cite{Siebenmorgen10}. We describe the disk around the central star using radial power-law dependencies for surface densities and temperature. The gas surface density follows a power law of the form 
\begin{equation}
    \Sigma_g = \Sigma_{g,0}\cdot\bigg(\frac{r}{r_0}\bigg)^{-p_g}
    \label{eq:gas_surface_density}
\end{equation}
where $r_0=1$ AU and $p_g=1$. We assume a total (gas) disk mass of $M_{g,\mathrm{tot}}=0.04M_\odot$ up to $r_\mathrm{out}=200$ AU. This results in a gas surface density of $\Sigma_g=283\:\mathrm{g\:cm^{-2}}$ at 1 AU. \newline
We assume the disk temperature to be set by external heating at all the heliocentric distances which are relevant for this study ($r \gtrsim 1$ au), and to be vertically isothermal. The disk temperature is then proportional to the fourth root of the stellar luminosity $L_*$ as \citep[e.g.][]{Dullemon18}
\begin{equation}\label{eq:rad_eq_temperature}
    T^4= \frac{\frac{1}{2}\Phi L_*}{4 \pi r^2 \sigma_{SB}}
\end{equation}
where $\Phi$ is the flaring angle of the disk and $\sigma_{SB}$ is the Stefan-Boltzmann constant. We take the flaring angle to be $\Phi=0.05$. Thus, the disk temperature profile takes a power law form 
 \begin{equation}
    T=T_0\cdot\bigg(\frac{r}{r_0}\bigg)^{-q}
    \label{eq:disk_temp}
\end{equation}
with $T_0=231$ K and $q=1/2$. Consequently, the vertical gas volume density follows a Gaussian distribution 
\begin{equation}
    \rho_g = \frac{\Sigma_g}{\sqrt{2\pi}h_g}\exp\bigg( -\frac{z^2}{2h_g^2}\bigg)
\label{eq:gasvolumedensity}
\end{equation}
where $h_g$ is the vertical gas pressure scale height given by $h_g = c_s/\Omega$ with $\Omega$ being the Keplerian frequency and $c_s = (k_B T/m_g)^{1/2}$ the isothermal sound speed, and $m_g = 3.9\cdot10^{-24}\:g$ denotes the mean mass of a gas molecule. For the initial dust surface density, we also assume a power law profile,
\begin{equation}
    \Sigma_d = \Sigma_{d,0}\cdot\bigg(\frac{r}{r_0}\bigg)^{-p_d}. 
    \label{eq:dust:surf_dens}
\end{equation}
where we set the power law index $p_d$ equal to $1.5$ and $\Sigma_{d,0}=19.7 \:\mathrm{g\:cm^{-2}}$. This slope corresponds to the equilibrium dust surface density in the fragmentation limit \citep[][]{Birnstiel12} and the prefactor to a dust mass accretion rate (in the fragmentation limit) of $10^{-5}\:\mathrm{M}_{\earth}/\mathrm{yr}$. With this profile, the dust-to-gas mass ratio is globally at 0.01, i.e. the canonical value expected in the ISM. With the term \textit{dust} we refer to any solid refractory disk component, which includes carbonaceous components and non-carbonaceous components (e.g. silicates). In table \ref{table:parametersummary}, we summarize the fiducial disk parameters.\newline
We divide the total dust population into six distinct refractory populations ($i=0...5$), according to the dust model described in section \ref{sec:dust_model}, with each population contributing with a mass-fraction $f_{c,i}$ to the total dust surface density $\Sigma_d$ such that
\begin{equation}\label{eq:sum_of_i_components}
    \Sigma_d = \sum_{i=0}^5 \Sigma_{d,i}
\end{equation}
and 
\begin{equation}
    \Sigma_{d,i} = f_{c,i} \Sigma_d
    \label{eq:f_i}
\end{equation}
are fulfilled. We consider the zeroth component ($i=0$) to be the silicate component and components $i=1...5$ to be the carbonaceous compounds (see \autoref{fig:refarctory_dust_composition}). From this, it follows directly that the total carbon fraction $f_c$, as introduced in section \ref{sec:dust_model}, is the sum of the individual mass fractions of components one to five
\begin{equation}
    f_c = \sum_{i=1}^5 f_{c,i}
    \label{eq:carbin_fraction_sum}
\end{equation}
and the mass fraction of silicates is 
\begin{equation}
    f_s = f_{c,0}
\end{equation}
Likewise, we denote the surface density of silicate grains with $\Sigma_s$. \newline

\begin{table}
\caption{\label{table:parametersummary} Fiducial model parameters. Here, \textit{p.l.i.} stands for power law index.}
\begin{center}
\begin{tabular}{l c c} 
 \hline
 \hline
  Quantity & symbol & value    \\ 
  \hline
 Stellar luminosity & $L_*$ & 2.39 $L_\odot$ \\
 Stellar UV luminosity & $L_\mathrm{UV}$ & 0.01 $L_\odot$ \\
 Total (gas) disk mass &  $M_\mathrm{g,tot}$  & 0.04 $M_\odot$ \\ 
 Gas surface density p.l.i. & $p_g$ & 1\\
 Initial dust accretion rate & $\dot{M}_d$ & $1\times 10^{-5}\:M_{\earth}/\mathrm{yr}$ \\
 Dust surface density p.l.i. & $p_d$ & 1.5\\

 Disk temperature at 1 AU & $T_0$ & 231 K \\
 Flaring angle & $\Phi$ & 0.05 \\
 
 Turbulence strength & $\alpha$ & $10^{-2}$ \\
 Fragmentation velocity & $v_\mathrm{frag}$ & 300 cm/s \\

 \hline
\end{tabular}
\end{center}
\end{table}

\subsection{Radial Dust Transport}
\label{sec:radial_dust_transport}
In this section, we describe the radial transport of dust, which happens as results of the loss of angular momentum caused by the drag interaction with the gas. When considering dust transport, we do not differentiate between the different dust compounds and assume all the compounds to have the same size distribution and the same average solid density $\rho_\bullet=3.0$ g cm$^{-3}$. This allows us to transport the dust distribution as a whole, and is also closer to reality, in which compounds are mixed within individual grains and do not exist as distinct populations. Similar to \cite{Birnstiel12}, we assign the total dust mass to two grain sizes, small and large grains, as a representation of a grain size distribution in coagulation-fragmentation equilibrium. The small dust grains are well coupled to the gas, and we assume their radius to be $a_s = 0.1 \micron$. The large grains of radius $a$ are generally only moderately coupled to the gas and thus subject to radial drift. In our model, we assume that collisions between dust grains are frequent enough that any radial transport happens at the drift speed of the large grains. We express the degree of coupling via the dimensionless Stokes number $St$ which we parametrize with a power law as: 
\begin{equation}
    St = St_0\cdot\bigg(\frac{r}{r_0}\bigg)^{s}.
    \label{eq:Stokes_number_para}
\end{equation}
The Stokes number at the disk midplane can be expressed as $St=\frac{\pi}{2} a \rho_\bullet/\Sigma_g$. 
We find for grains in the fragmentation limit
\begin{equation}\label{eq:fra_limit}
  St_0 = f_\mathrm{f}\frac{m_g }{3\alpha k_B} \frac{v_f^2}{T_0},\quad s = q  
\end{equation}
where $f_\mathrm{f}$ is a calibration factor of order unity \citep{Birnstiel12} and $\alpha$ is the dimensionless turbulence parameter \citep[][]{Shakura1973}. With the parametrization of the Stokes number as in equation (\ref{eq:Stokes_number_para}), we write the radial drift velocity of the dust grains with Stokes number St$\ll$1 as: 
\begin{equation}
    v_r = -\gamma St \frac{c_s^2}{v_k}
    \label{eq:drift_velr_para}
\end{equation}
With $\gamma=q/2+p_g+3/2$ being the modulus of the power-law exponent of the gas pressure $P_g=\rho_g c_s^2$.  We also express the radial drift velocity in power-law form. Thus, we rewrite equation (\ref{eq:drift_velr_para}) as  
\begin{equation}
    v_r = v_0\cdot\bigg(\frac{r}{r_0}\bigg)^{l}
    \label{eq:radial_velocity}
\end{equation}
with $v_0=-\gamma St_0 c_{s,0}^2/v_{k,0}$ and $l = s-q+1/2$

\subsubsection{Transport equations}
\label{sec:transport_equations}
Ultimately, we aim to find the radial distribution of the dust surface density (of carbonaceous and non-carbonaceous grains). Its evolution, we describe with a one-dimensional radial advection equation without diffusion and a source term that models the depletion of refractory carbon compounds
\begin{equation}
    \frac{\partial \Sigma_d}{\partial t} + \frac{1}{r}\frac{\partial}{\partial r}( r \Sigma_d v_r) = 
    -\dot{\Sigma}_d 
    \label{eq:dust_evolution}
\end{equation}
Using equation (\ref{eq:sum_of_i_components}), we rewrite equation (\ref{eq:dust_evolution}) as the sum of six individual equations
\begin{equation}
    \frac{\partial \Sigma_d}{\partial t} = \sum_{i=0}^5 \frac{\partial \Sigma_{d,i}}{\partial t}
\end{equation}
of which each summand has the general form 
\begin{equation}
    \frac{\partial \Sigma_{d,i}}{\partial t} + \frac{1}{r}\frac{\partial}{\partial r}( r \Sigma_{d,i} v_r) = 
    -\dot{\Sigma}_{d,i} 
    \label{eq:carbon_evolution}
\end{equation}
In our models, we assume silicate grains to not decompose, and their surface density to be conserved. Thus, we write $\dot{\Sigma}_{d,i=0} =0$. For all the carbonaceous components ($i=1..5$) we will discuss the explicit form of the source terms $\dot{\Sigma}_{d,i}$ in the following sections. Throughout this work, we use the subscript $c$ to refer to the sum of all the carbonaceous components ($i>0$) and the subscript $s$ to refer to the silicate component ($i=0$). Note, because we model the radial transport of the dust as one population, the radial drift velocity $v_r$ is identical for each component. 

\subsection{Carbon Depletion Timescale}
We assume that a carbonaceous compound, as introduced in section \ref{sec:dust_model}, can be gradually decomposed by an arbitrary (yet unspecified) carbon-depletion mechanism. We define the time $t_d$ to be the time it takes for a carbonaceous grain of radius $a$ to be completely decomposed by this mechanism. The efficiency of the mechanism can have a radial power-law dependence, thus, we define 
\begin{equation}
    t_d = t_{d,0}\cdot\bigg(\frac{r}{r_0}\bigg)^b
    \label{eq:power_law_depletion_time}
\end{equation}
where $t_{d,0}$ is the time it takes to decompose a carbon grain of size $a$ at radius $r_0$ in the disk. Further, we define the \textit{carbon depletion timescale},
\begin{equation}
    \tau_{c}=\frac{\Sigma_{c}}{\dot{\Sigma}_{c}}
    \label{eq:carbon_depl_timescale}
\end{equation}
which describes the characteristic carbon depletion time of the disk and is the ratio between surface density of the all the carbonaceous compounds $\Sigma_c$ and its depletion rate $\dot{\Sigma}_c$. Assuming the entire dust disk consists only of grains of size $a$, and the arbitrary carbon depletion mechanism is active throughout the entire disk, all the carbonaceous material will be destroyed within time $t_d$ and we find $\dot{\Sigma}_c=\Sigma_c/t_d$. Thus, in this simple case, the carbon depletion timescale is equal to the destruction time of a single grain $\tau_c = t_d$. However, it is possible that a given depletion mechanism is only active in certain layers of the disk, or the mechanism is only efficient in a certain fraction of the grain size distribution (or both). Therefore, we assume a given depletion mechanism is active only in a fraction $\Sigma_c^*$ of the total surface density $\Sigma_c$. This increases the carbon depletion timescale $\tau_c$, which is now generally larger than the time it takes to decompose a single grain $t_d$ because $\Sigma_c^*\leq\Sigma_c$. Thus, we write the carbon depletion timescale as
\begin{equation}
    \tau_c=\frac{\Sigma_c}{\Sigma_c^*}t_d
    \label{eq:intro_depl_timescale}
\end{equation}
where $\Sigma_c^*$ is the carbon surface density in which depletion is active. From the definition of the carbon fraction, we find $\Sigma_c = f_c\Sigma_d$, and if we assume for now that the vertical mixing timescale ($t_\mathrm{mix}=1/\alpha\Omega$) is small compared to the destruction time $t_\mathrm{mix}\ll t_d$, the carbon fraction $f_c$ is vertically uniform and $\Sigma_c^* = f_c\Sigma_d^*$ holds (compare to section \ref{sec:vertical_transport} where this assumption is lifted). Thus, we write the carbon depletion timescale as 
\begin{equation}
    \tau_c=\frac{\Sigma_d}{\Sigma_d^*}t_d
    \label{eq:destructio_rate}
\end{equation}
\newline
Depending on the detailed physics of the depletion mechanism, it is possible that the destruction time is better described by an exponential law, rather than a power-law
\begin{equation}
    t_d'= t_{d,0}'\exp(-t_d)
\end{equation}
where $t_d$ is the generic power law as defined in equation (\ref{eq:power_law_depletion_time}). Irrespective of the detailed functional dependence of $t_d$, the considerations in this section still hold.

\begin{figure}
\includegraphics[width=0.95\columnwidth]{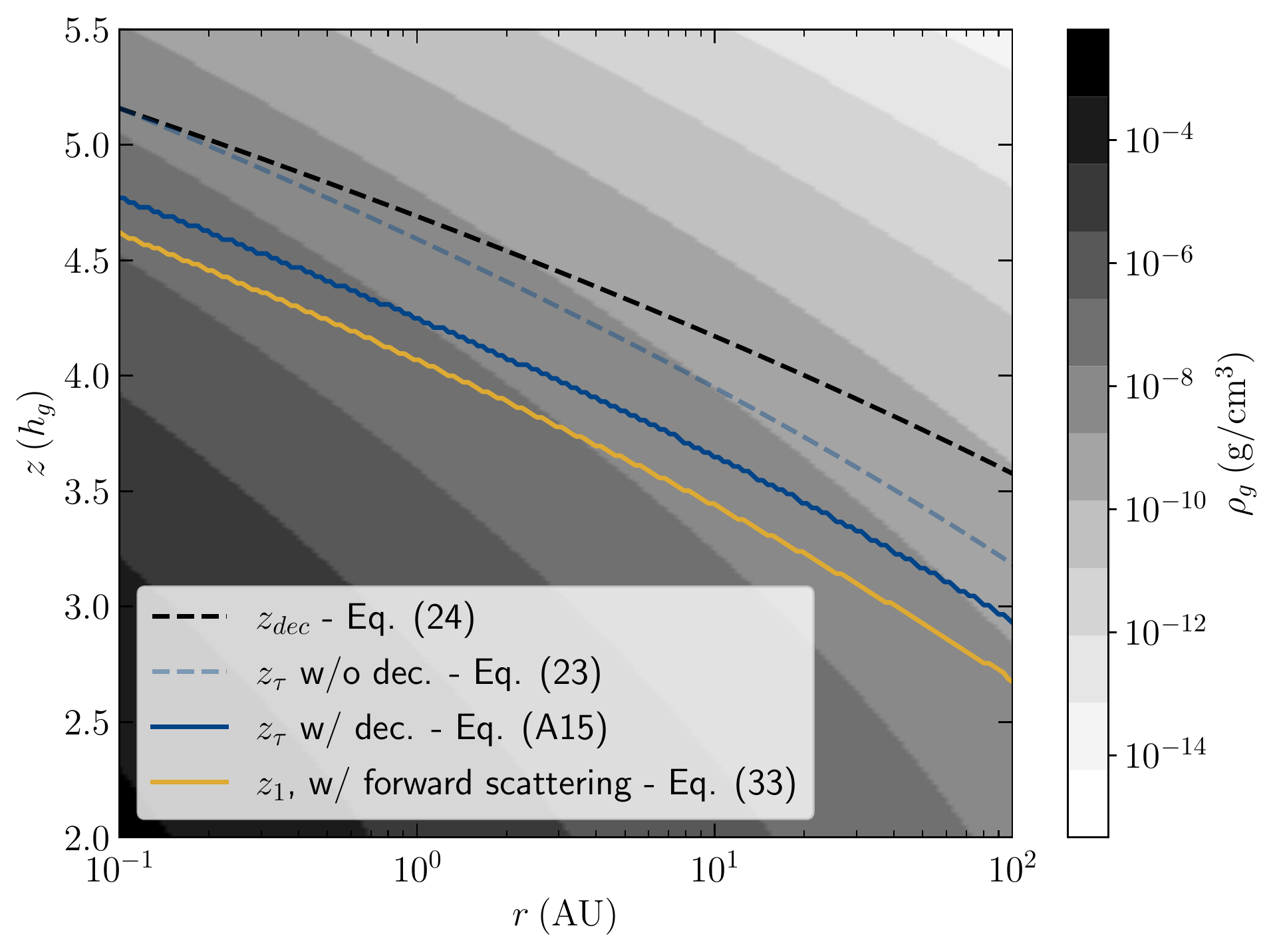}
\caption{Plotted is, with the black solid line, for the fiducial model, the height at which 0.1 $\mathrm{\mu m}$ sized grains decouple ($z_\mathrm{dec}$) as calculated with equation (\ref{eq:decoupling_height_main}). The blue lines represent the $\tau = 1$-surface at $z_\tau$. The dashed blue line is calculated using the analytical approximation, i.e., equation (\ref{eq:tau1_surface}). that does not consider the decoupling of grains at large z. The solid blue line does include the effects of grains decoupling and is the numerical solution to equation (\ref{eq:1=flz}). The yellow shows the solution to equation (\ref{eq:fwrdscattering}) which includes the effects of photon forward scattering. Thus, FUV photons penetrate deeper than the $\tau = 1$-surface. The background colors illustrate the gas volume density $\rho_g$, as in equation (\ref{eq:gasvolumedensity}).}
\label{fig:comparison}
\end{figure}

\begin{figure}
\includegraphics[width=\columnwidth]{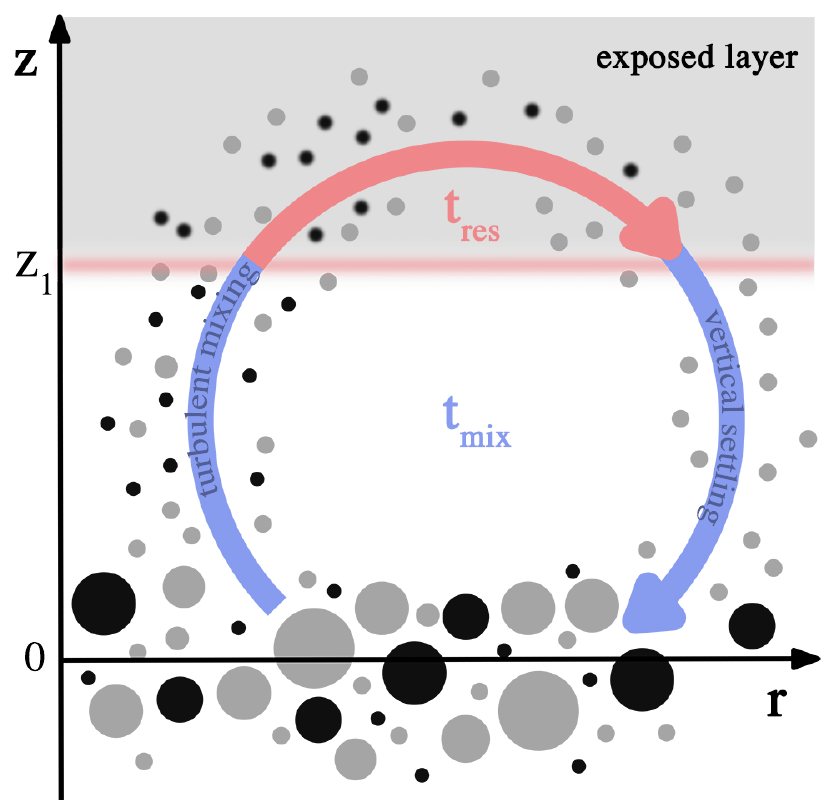}
\caption{Vertical profile of the disk above the midplane. Silicate grains are plotted in gray color, carbon grains in black. Large grains are confined to the midplane region. Small grains which are produced in collision in the high density region around the midplane are lifted to higher disk layers by turbulence. Above $z_1$, the grains reach the UV-irradiated exposed layer where carbon grains are destroyed by photolysis while silicate grains are unaffected. The longer the carbon grains spend in the exposed layer, the more carbon grains are destroyed and the carbon fraction drops. The residence time $t_\mathrm{res}$ is the total time the small grains spend in the exposed layer before settling toward the midplane where they collide and coagulate with other grains. During one of these mixing cycles, carbon depletion can become inefficient if the local carbon fraction in the exposed layer drops due to efficient carbon grain destruction. This is indicated by a gradient in the number of carbon grains which, in the sketch, move in the exposed layer from left to right. If $z_1$ moves closer to the midplane $t_\mathrm{res}$ takes up a larger fraction of the total mixing time $t_\mathrm{mix}$. At the same time, the surface density contained in the exposed layer $\Sigma^*$ takes up a larger fraction of the total dust surface density $\Sigma_d$.}
\label{fig:sketch}
\end{figure}

\section{Photodecomposition}
In this section, we consider the photodecomposition of refractory carbon. Specifically, we study the effects of photolysis via stellar UV radiation.  

\subsection{The Exposed Layer}
\label{sec:exposed_layer}
The far ultraviolet-flux (FUV) coming from the central disk region is largely unaffected by gas in the disk, but is mainly attenuated by small dust grains. Thus, the disk layers containing dust are generally very optically thick at FUV-wavelengths. Thus, potential photo-induced carbon depletion by FUV-photons can only be active in the layers of the disk in which stellar FUV-photons penetrate. Following \cite{Klarmann2018}, we call this the \textit{exposed layer} (\cite{Siebenmorgen10} call this layer the \textit{extinction layer}). The exposed layer extends vertically from height $z=z_1$ to $z=\infty$ and contains the dust surface density $\Sigma_d^*$. Our goal in this section is to quantify $\Sigma_d^*$ and $z_1$. In a first approach, we assume all the stellar FUV photons to be absorbed in the region where the optical depth is smaller than unity, i.e., we ignore effects of scattering. Further, we assume that the FUV flux has the form of a step function, where it is not attenuated above $z_1$ and zero below. This is a reasonable approximation because the bulk of the photons are absorbed close to $z_1$ due to the steep exponential increase of the dust density $\rho_d(z)$. Then, the height $z_1$ is equal to the height at which the (radial) optical path has optical depth $\tau=1$. Due to the flaring of the disk, a radial optical depth $\tau=1$ corresponds to a vertical optical depth of $\tau_z=\Phi$, where $\Phi$ is the flaring angle of the disk. This is nicely illustrated in Fig.~5 of \cite{Siebenmorgen10}.\newline
For the sake of simplicity, we assume a single dust grain size $a_s$ to be the dominant contributor to the opacity at FUV-wavelengths. This assumption is reasonable because for a given wavelength $\lambda$, grains smaller than $a=\lambda/2\pi$ are in the Rayleigh regime of scattering where the absorption of photons is a by-mass effect, meaning the grain size dominating the mass budget also dominates the opacity. In a coagulation-fragmentation equilibrium, larger grains generally contribute more to the total dust mass. Hence, in the Rayleigh regime, larger grains contribute most to the opacity. On the other hand, the opacity for grains larger than $a=\lambda/2\pi$, in the geometrical optics limit, is calculated as the ratio between the geometrical cross-section $\sigma$ and the mass of a grain $m$ as $\kappa =\sigma/m=3/(4\rho_\bullet a)$. Here, $\rho_\bullet$ is the solid density of a dust grains. Hence, in the geometrical optics limit, small grains contribute more to the opacity. To conclude, in this simple argument, the grain size that dominates the opacity is the size in between the two scattering regimes with an optical size of unity, i.e., the grains with size $a_s=\lambda/2\pi$ and opacity $\kappa_0 = 3/(4\rho_\bullet a_s)$. For $\lambda = 0.6 \micron$ and $\rho_\bullet=3 gcm^{-3}$, we find $a_s\sim0.1\micron$ and $\kappa_0\sim2.5\cdot10^{4} cm^2g^{-1}$ for the opacity. The surface density contained in the exposed layer in the disk is
\begin{equation}\label{eq:sigma_d_star_definition}
    \Sigma_\mathrm{d}^*=2\Phi/\kappa_0
\end{equation}
or equivalently $\Sigma_\mathrm{d}^*=8\Phi/3\rho_\bullet a_s$. The factor 2 in equation (\ref{eq:sigma_d_star_definition}) comes from the two sides of the disk. 
\newline
As for $z_1$, i.e., the lower edge of the exposed layer, an exact explicit expression can not be found. But because we define the lower boundary of the exposed layer to be the location at which the radial optical depth equals unity $z_1=z_{\tau}$. In appendix \ref{sec:tau=1_surface_derivation}, we derive the implicit equation (\ref{eq:1=flz}) that we solve numerically to find $z_1$ in all our quantitative analysis. In addition to that, in appendix \ref{sec:tau=1_surface_derivation}, we derive an explicit, but approximate, expression for $z_{\tau}$ under the assumption that the opacity dominating dust grains are perfectly coupled to the gas:
\begin{equation}
    \frac{z_{\tau}}{h_g} \simeq \sqrt{2\ln\frac{f_{\leq a_s}\Sigma_d\kappa_0}{2\sqrt{\pi}\Phi}}-\frac{1}{5}
\label{eq:tau1_surface}
\end{equation}
It is straightforward to see that $z_{\tau}$, in units of the gas scale height $h_g$, is farther away from the midplane if the flaring angle $\Phi$ is small because a larger fraction of the photon path lies inside the disk atmosphere. In addition to that, $z_{\tau}$ lies farther away from the midplane if the surface density of small dust grains ($f_{\leq a_s}\Sigma_d$) is large and if the opacity $\kappa_0$ is large. For the fiducial parameters, we obtain $z_{\tau,0}=4.6\:h_{g,0}$. Interestingly, $z_{\tau}$ is quite insensitive to changes of parameters because the density $\rho_d(z)$ changes rapidly with $z$. Decreasing the argument in the natural logarithm in equation (\ref{eq:tau1_surface}) by a factor of 10, e.g. by decreasing the total dust surface density $\Sigma_d$, results in $z_{\tau,0}=4.1\:h_{g,0}$, which is a decrease of only eleven per cent. In the top panel of \autoref{fig:comparison}, we show the radial dependence of the solution to equation (\ref{eq:tau1_surface}) using our fiducial model parameters (dashed blue line). We want to highlight that equation (\ref{eq:tau1_surface}) is only accurate to first-order, if $z_\tau>h_g$, and assumes small grains are perfectly coupled to the gas. For more accurate results, one should use higher-order terms, as in equation (\ref{eq:higher_orderX0}), or use a numerical approach. For our fiducial set of parameters, the solution lies well above the gas scale height $h_g$. To evaluate whether the well-coupling condition is fulfilled, we also plot the location at which the opacity dominating dust grains decouple from the gas 
\begin{equation}
    \frac{z_\mathrm{dec}}{h_g}=\sqrt{2\ln\frac{2\alpha\Sigma_g}{\pi \rho_\bullet a}}
\label{eq:decoupling_height_main}
\end{equation}
We derive the above equation in appendix \ref{sec:tau=1_surface_derivation} by finding the height at which the local Stokes number is equal to the turbulent alpha-parameter $\alpha$. In \autoref{fig:comparison}, we plot $z_\mathrm{dec}$ calculated with equation (\ref{eq:decoupling_height_main}) in black color with a dashed line. The solution to equation (\ref{eq:tau1_surface}) crosses $z_\mathrm{dec}$, i.e. the height at which we find the opacity dominating dust grains to decouple. Therefore, we expect the solution of equation (\ref{eq:tau1_surface}) to deviate slightly from the exact result for our fiducial choice of parameters. This difference becomes apparent when comparing the dashed and solid blue lines in \autoref{fig:comparison}. If the decoupling of small grains would be negligible, the two solution were identical. Nonetheless, equation (\ref{eq:tau1_surface}) serves as a valuable in the qualitative analysis of our results. For all the quantitative results, we use the exact numerical solution for $z_{\tau}$ as found by solving equation (\ref{eq:z_tau_result}). We show this solution in \autoref{fig:comparison} with a solid blue line. Indeed, we find the approximate solution to be about ten per cent above the exact solution because it does not account for grain decoupling. Further, we stress here that setting $z_1=z_{\tau}$ is not always a good approximation, as photons can reach deeper layers when scattering such that $z_1<z_{\tau}$. We briefly investigate the effects of forward scattering in section \ref{sec:forward_scattering}.  \newline

\begin{figure}
\includegraphics[width=\columnwidth]{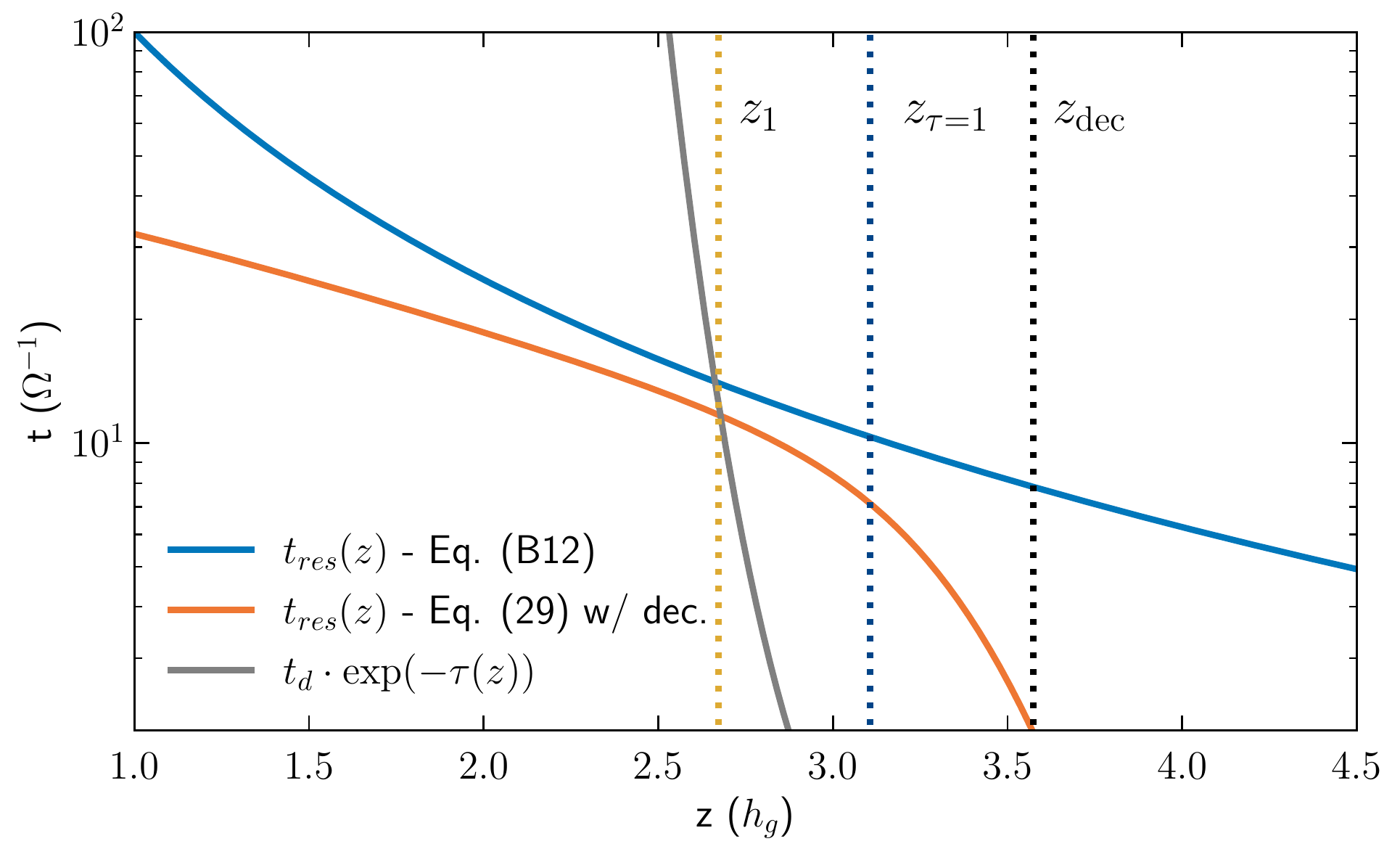}
\caption{Residence times $t_\mathrm{res}$ of 0.1 $\mathrm{\mu m}$ sized dust grains at different heights $z$ above the midplane calculated at 100 AU in units of the inverse of the Keplerian frequency $\Omega^{-1}$. The orange line shows the residence time calculated with equation (\ref{eq:residence_time_full_equation}). Due to grains decoupling, the residence time sharply decreases above about three gas scale height. We indicate the height at which grains decouple $z_\mathrm{dec}$, as calculated with equation (\ref{eq:decoupling_height_main}), with the black vertical dotted line. The blue line shows the approximate $(h_g/z)^{-2}$ dependency of the residence time, as introduced in equation (\ref{eq:t_res_k}). The solid gray line represents the grain destruction time $t_d$ multiplied with an exponential factor that accounts for the attenuation of the FUV field in the optically thick region of the disk. The blue vertical dotted line indicates the height of the $\tau=1$ surface as determined by solving equation (\ref{eq:1=flz}). The yellow dotted lines indicates the location of the solution to equation (\ref{eq:fwrdscattering}), i.e. the location where the gray and orange lines intersect. }
\label{fig:residence_time}
\end{figure}

\begin{figure}
\includegraphics[width=0.9\columnwidth]{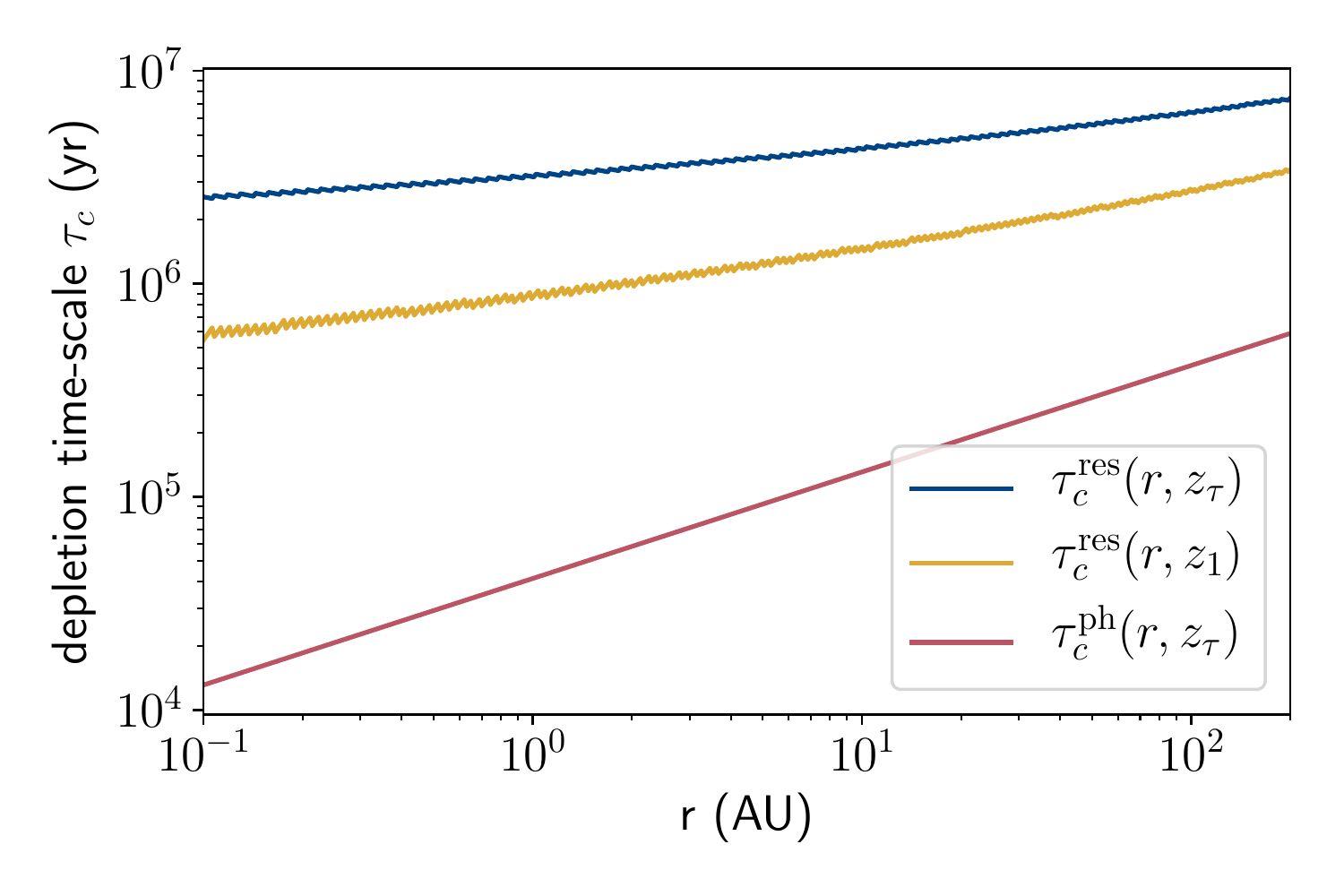}
\caption{Photolysis depletion timescale as a function of radius. The red line shows the unrestricted depletion timescale $\tau^\mathrm{ph}_c$ as defined in equation (\ref{eq:unrestricted_depl_timescale}). At 1 AU, it has a value of 40 kyr. The blue line shows the depletion timescale limited by vertical transport at the optical surface of FUV-photons as calculated with equation (\ref{eq:residencetimeliiteddesrate}). It is significantly larger, with a value of 2.6 Myr at 1 AU. The yellow line shows the resulting depletion timescale when FUV photons are allowed to penetrate the disk beyond the $\tau=1$-surface due to forward scattering. It has a value of 695 kyr at 1 AU. }
\label{fig:d_timescales}
\end{figure}

\subsection{Photolysis}
\label{sec:photolysis}
In this section, we discuss \textit{photolysis} as a specific example of a carbon depletion mechanism, which was also discussed in previous studies \citep[e.g.][]{Klarmann2018,Anderson2017}. Similarly, we use the term photolysis to refer to the photon-induced release of small hydrocarbons from the surface of a carbonaceous grain (as opposed to the photodissociation of a single molecule). When considering photolysis, \cite{Alata14,Alata15} found methane to be the C-bearing product of the highest yield. Therefore, we will only focus on this product here. The photolysis rate of C-bearing grains in a FUV field is $R_\mathrm{UV} = \sigma Y_\mathrm{ph} F_\mathrm{UV}$. Here, $R_\mathrm{UV}$ describes the number of C-atoms with mass $m_c$ released per unit time from a single grain where $\sigma = a^2\pi$ is the geometric cross-section of the carbon grain, $Y_\mathrm{ph}$ is the yield of the photolysis reaction per incoming photon ($\sim8\times10^{-4}$) and $F_\mathrm{UV}$ is the \textit{loca}l FUV flux ($[F_\mathrm{UV}]=cm^{-2}s^{-1}$). The time it takes to destroy a single carbon grain of radius $a$ and mass $m$ via photolysis is
\begin{equation}
    t_\mathrm{ph} = \frac{m}{m_c R_\mathrm{UV}}=\frac{4}{3}\frac{a \rho_c}{m_c Y_\mathrm{ph}F_\mathrm{UV}}
    \label{eq:c_destrcutiontime}
\end{equation}
and the depletion timescale, as introduced in equation (\ref{eq:intro_depl_timescale}), for unrestricted photolysis is
\begin{equation}
  \tau_{c}^\mathrm{ph} = \frac{\Sigma_{d}}{\Sigma_d^*}t_\mathrm{ph}  
  \label{eq:unrestricted_depl_timescale}
\end{equation}
By \textit{unrestricted}, we refer to the simplified assumption that the carbon fraction is vertically constant. This assumption only holds if carbon in the exposed layer is depleted on a much shorter timescale than the vertical mixing timescale of the disk. In reality, the carbon depletion can be restricted by inefficient vertical mixing when the carbon fraction is significantly lower in the exposed layer compared to the rest of the disk. Therefore, the unrestricted carbon depletion timescale in equation (\ref{eq:unrestricted_depl_timescale}) does only represent a lower limit. \newline
Combining equation (\ref{eq:sigma_d_star_definition}), equation (\ref{eq:c_destrcutiontime}) with equation (\ref{eq:unrestricted_depl_timescale}), we find
\begin{equation}
   \tau_{c}^\mathrm{ph}= \frac{\Sigma_{d}}{2 m_c Y_\mathrm{ph} \Phi F_\mathrm{UV}}
    \label{eq:C_unrestricted}
\end{equation}
where we have used $\kappa_0=3/(4\rho_\bullet a)$. Thus, photolysis, when not limited by other mechanisms such as transport, is more efficient in disks with a large flaring angle $\Phi$ and regions with a large UV flux ${F}_\mathrm{UV}$. It is also more efficient where the dust surface density $\Sigma_d$ is small. Interestingly, in equation (\ref{eq:C_unrestricted}) the opacity cancels out and equation (\ref{eq:C_unrestricted}) is independent of the amount of surface density contained in the exposed layer. In our fiducial model, the unrestricted depletion timescale at 1 AU is $\tau_{c,0}=40\:\mathrm{kyr}$ which is short compared to thy typical disk lifetimes ($\sim$Myr). \newline

\subsection{Vertical Transport}
\label{sec:vertical_transport}
The carbon-depletion efficiency can be limited by vertical transport when the stellar FUV photons do not penetrate the entire disk because the carbon grains must be vertically transported from the disk midplane to the exposed layer. To be specific, the carbon depletion process becomes inefficient if the local carbon fraction $f_c$ in the irradiated exposed layer becomes significantly smaller than the average carbon fraction below the exposed layer. This is the case whenever carbon grains in the exposed layer are destroyed faster than they are recycled through collisions and vertical transport with the rest of the grains in the lower dense disk layers. In this situation, the carbon grain destruction time $t_d$ is not a limiting factor to the depletion timescale anymore, as suggested by equation (\ref{eq:destructio_rate}). In such a situation, decreasing the carbon grain destruction time $t_d$ does not decrease the depletion timescale $\tau_c$ because any carbon that reaches the exposed layer is destroyed anyway. In other words, there is a lower limit to the carbon depletion timescale $\tau_c$, below which equation (\ref{eq:destructio_rate}) is not valid anymore. This limit is reached if the grain destruction time $t_d$ becomes equal to the typical time a grain spends in the exposed layer before it gets recycled through vertical transport and collisions in the lower disk layers. We call this time the \textit{residence time} $t_\mathrm{res}$. Hence, if $t_\mathrm{res}>t_d$, carbon depletion is \textit{residence-time limited} and the smallest possible depletion timescale is
\begin{equation}
   \tau_c^\mathrm{res} =  \frac{\Sigma_d}{\Sigma_d^*}t_\mathrm{res}.
    \label{eq:residencetimeliiteddesrate}
\end{equation}
In \autoref{fig:sketch}, we present a sketch of a mixing cycle of small dust grains between the disk midplane and the exposed layer where we illustrate the residence time as a fraction of the full mixing cycle. In appendix \ref{sec:residence_time_derivation}, we derive an expression for the residence time $t_\mathrm{res}$ taking into account the random turbulent motions of a grain in the disk, which we model as a stochastic Ornstein-Uhlenbeck process \citep[][]{Uhlenbeck1930}. For readability, we only report the results in this section and refer the reader to appendix \ref{sec:residence_time_derivation} for the full derivation. We find the residence time of a particle at height z to be calculated with an integral over the complementary error function and a dimensionless variable $\chi$ as
\begin{equation}
    t_\mathrm{res}(z)=\frac{1}{2}\int_0^{t_\mathrm{mix}} \mathrm{erfc}\big(\chi(z,t)\big)\mathrm{d}t
    \label{eq:residence_time_full_equation}
\end{equation}
The dimensionless variable $\chi$ is time-dependent and defined as 
\begin{equation}
    \chi(z,t)=\frac{z-\bar{z}(t)}{\sqrt{2}\upsilon(t)}
    \label{eq:chi}
\end{equation}
where we integrate up to the mixing time $t_\mathrm{mix}=1/\alpha\Omega$ and define a time-dependent mean $\bar{z}(t) = ze^{-\zeta t}$. Further, we define the time-dependent variance of the stochastic Ornstein-Uhlenbeck process $\upsilon^2(t)=\frac{D_d}{\zeta}\big(1-e^{-2\zeta t} \big)$ and the rate $\zeta$ as
\begin{equation}
    \zeta = St(z)\Omega +\frac{D_d}{h_g^2}
    \label{eq:zetaeq}
\end{equation}
In \autoref{fig:residence_time}, we plot the z-dependence of equation (\ref{eq:residence_time_full_equation}) in orange color. The residence time $t_\mathrm{res}$ has a shallow negative slope above the midplane up to the point at which dust grains decouple from the gas ($z_\mathrm{dec}$). Beyond this point, the residence time drops off sharply. When approaching the midplane, the residence time approaches half the mixing time and at for $z=0$ we find $t_\mathrm{res}=t_\mathrm{mix}/2$. This makes sense if one considers the mixing time to be the time a grain performs a full mixing cycle. Since the residence time only considers one side of the disk ($z>0$), at $z=0$ it must be equal to half a mixing time. \newline

\begin{figure*}
\includegraphics[width=2.1\columnwidth]{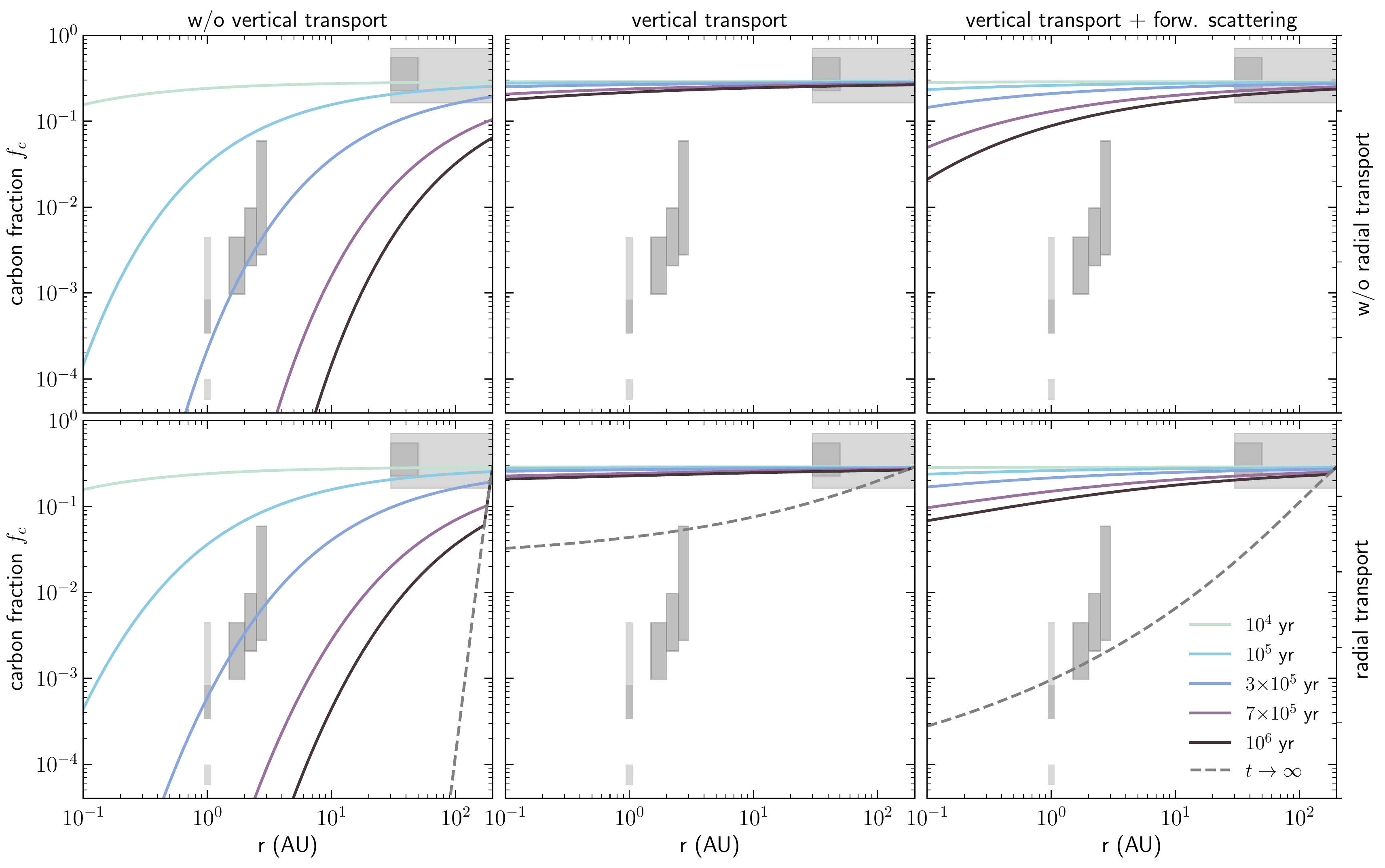}
\caption{Carbon fraction $f_c$ as a function of radius at different times over $10^6$ years using the fiducial model parameters. The gray boxes represent the estimated carbon fraction on Earth, meteorites and comets and are the same as in \autoref{fig:C_fraction_overview}. \textit{Upper row:} Carbon fraction without radial transport using equation (\ref{eq:nort_solution}) and (\ref{eq:final_solution}). \textit{Lower row:} Full model including radial transport using equation (\ref{eq:final_solution}) and  (\ref{eq:surface_density_ratio}). The gray dashed line represents the carbon fraction at $t\rightarrow \infty$. \textit{Left:} Photolysis without constraints by vertical transport, for which the depletion timescale is calculated using equation (\ref{eq:C_unrestricted}). At 1 AU we find $\tau_c^\mathrm{ph}(z_\tau)=40$ kyr. \textit{Middle:} Smallest possible carbon fraction if vertical transport is considered and the depletion is only active above the $\tau=1$-surface. The depletion timescale is calculated using equation (\ref{eq:residencetimeliiteddesrate}) which results in a value of $\tau_c^\mathrm{res}(z_\tau)=2.6$ Myr at 1 AU.  \textit{Right:} Same as the middle panels, but including forward scattering. The depletion timescale is calculated using equation (\ref{eq:fwrdscattering}) which results in a value of $\tau_c^\mathrm{eff}(z_1)=695$ kyr at 1 AU.}   
\label{fig:unrestricted_model}
\end{figure*}

\subsection{Photon Scattering}
\label{sec:forward_scattering}
In this section, we further investigate residence-time limited carbon depletion, i.e. $t_\mathrm{res}>t_\mathrm{ph}$ at the $\tau=1$-surface, and study what happens when we include the effects of photon scattering \citep[see also][]{Klarmann2018}. FUV-photons do not only penetrate the disk down to the $\tau=1$-surface, but penetrate optically thicker regions due to forward scattering \citep[][]{Zadelhoff2003}. Consequently, the lower edge of the exposed layer $z_1$ is lower than the $\tau=1$-surface and the FUV field extends deeper into the disk. Below the $\tau=1$-surface, with increasing optical depth, the UV-field is exponentially attenuated $F_\mathrm{UV}(z)\propto\exp\big(-\tau(z)\big)$. Hence, at a given radius, the lifetime of a carbon grain $t_\mathrm{ph}$ that is exposed to UV radiation, increases by a factor $\exp\big(-\tau(z)\big)$. In \autoref{fig:residence_time}, we plot the vertical dependence, i.e. the exponential increase towards the midplane, of the lifetime of a grain $t_d$ in gray color. At the same time, the surface density that is exposed to UV photons is larger by a factor $\tau(z)$, and we adapt the definition in equation (\ref{eq:sigma_d_star_definition}) when considering forward scattering to read
\begin{equation}
    \Sigma_d^*=\tau(z)2\Phi/\kappa_0.
    \label{eq:exposed_volume}
\end{equation}
Thus, below the $\tau=1$-surface, the unrestricted depletion timescale $\tau_c^\mathrm{ph}(z)$ increases with decreasing distance to the midplane. In fact, it increases faster with decreasing distance than the residence-time limited depletion timescale $\tau_c^\mathrm{res}(z)$. Thus, we find the effective depletion timescale by solving 
\begin{equation}
    \tau_c^\mathrm{res}(z_1)=\tau_c^\mathrm{ph}(z_1)
    \label{eq:fwrdscattering}
\end{equation}
for $z_1$ where now, $z_1<z_\tau$ holds. Then, all the grains above $z_1$ are destroyed faster than their residence time at $z_1$. Hence, when considering forward scattering in the residence-time limited case, we consider $z_1$, to be the lower boundary of the exposed layer. In \autoref{fig:comparison}, we visualize location of $z_1$ as a function of radius in our fiducial model (in yellow color). Throughout the entire disk, it lies less than half a gas scale height (0.5$h_g$) lower than the $\tau=1$-surface without forward scattering. \newline
In \autoref{fig:d_timescales} we visualize the depletion timescales (as a function of radius) for three different models with increasing level of sophistication. First, the red line shows the unrestricted carbon depletion timescale at the $\tau=1$-surface calculated with equation (\ref{eq:unrestricted_depl_timescale}). It is 41 kyr at 1 AU. The blue line shows the depletion timescale limited by vertical transport, i.e. the residence time limit, as calculated with equation (\ref{eq:residencetimeliiteddesrate}). It is 2.6 Myr at 1 AU. The yellow line shows the depletion timescale under the consideration of forward scattering, as calculated with equation (\ref{eq:fwrdscattering}). It has a value of 695 kyr at 1 AU.

\subsection{Analytic Solution of the Carbon Fraction}
\label{sec:analytic_solution}
In appendix \ref{sec:derivation}, we derive an analytic solution describing the carbon fraction as a function of radius and time $f_c(r,t)$, as described by the transport equations in section \ref{sec:transport_equations}. In this section, we only summarize the most important results. In the derivation, we study the movement of individual grains subject to radial drift at radial velocity $v_r$. Assuming we find grains at radius $r$ in the disk at a time $t>0$, and knowing the radial dependence of $v_r$ from equation (\ref{eq:radial_velocity}), we infer the initial radial position $r'$ of the grains at time $t=0$ 
\begin{equation}
r'(r,t)=
 \begin{cases}
    r\cdot\exp\big(-\frac{v_0}{r_0}t\big) &\text{if}\;l= 1 \\
    \Big(r^{1-l}+(l-1)r_0^{-l}v_0t \Big)^{1/(1-l)} & \text{else}
    \label{eq:initial_radius01}
 \end{cases}
\end{equation}
where time $t$ is the time when the grains are at radius $r$. Knowing the initial radius $r'(r,t)$ of grains at radius $r$ at time $t$, we calculate the surface density ratio $f_\Sigma=\Sigma_c/\Sigma_s$ at radius $r$ at time $t$ as
\begin{equation}
\label{eq:surface_density_ratio}
    f_\Sigma(r,t) = 
    \begin{cases}
       f_{\Sigma}(r',0)\cdot\bigg( \frac{r'}{r}\bigg)^{\frac{r_0}{\tau_{c,0}v_0}} & \text{if}\;\beta=0 \\
       f_{\Sigma}(r',0)\cdot\exp\bigg(\frac{r_0}{ \tau_{c,0}v_0\beta}\bigg[\Big(\frac{r'}{r_0}\Big)^\beta-\Big(\frac{r}{r_0}\Big)^\beta\bigg]\bigg) & \text{else}
    \end{cases}
\end{equation}
where
\begin{equation}
    \tau_{c,0}=\frac{\Sigma_{c,0}}{\dot{\Sigma}_{c,0}}
    \label{eq:first_def_C}
\end{equation}
is the carbon depletion timescale at $r=r_0$ and the power-law index $\beta$ is defined as $\beta= p_d-b-l+1$.
Here, we require $p_d=l+1$ because only then $\Sigma_{d}(r)$ is time independent (assuming $\Sigma_d\gg\Sigma_c$) and retains a power-law form, which is a requirement for our analytic analysis. Consequently, the carbon depletion timescale $\tau_{c,0}$ is also time-independent. This becomes clear if we rewrite equation (\ref{eq:first_def_C}) using equation (\ref{eq:destructio_rate}). The carbon depletion timescale then becomes 
\begin{equation}
  \tau_{c,0} = \frac{\Sigma_{d,0}}{\Sigma_d^*}t_{d,0}  
  \label{eq:unrestricted_destructionrate}
\end{equation}
and is a function of time-independent quantities only. From the equation above, one finds the carbon depletion timescale to be inversely proportional to the fraction of surface density in which the depletion mechanism is active ${\Sigma_{d,0}}/{\Sigma_d^*}$ and directly proportional to the carbon grain destruction time $t_d$. Ultimately, the carbon fraction $f_c$ as a function of radius and time is
\begin{equation}
    f_c(r,t)=\bigg(1+\Big(f_{\Sigma}(r,t)\Big)^{-1}\bigg)^{-1}
    \label{eq:final_solution} 
\end{equation}
or whenever $\Sigma_s \gg \Sigma_c$ holds, the carbon fraction is
\begin{equation}
    f_c(r,t)\approx f_{\Sigma}(r,t)
\end{equation}
From our set of governing equations (\ref{eq:initial_radius01}), (\ref{eq:surface_density_ratio}) and (\ref{eq:final_solution}), it becomes clear that the evolution of the carbon fraction $f_c$ at a given radius $r$ is mainly driven by the dimensionless quantity $\tau_c v_r/r$ which is the product of the carbon depletion timescale and the radial drift velocity normalized by the radius. Carbon depletion at a given radius will be more efficient the smaller this quantity is. The quantity $\tau_c v_r/r$ is small if the destruction mechanism is active in a large fraction of the disk surface density, i.e. $\Sigma_d^*/\Sigma_d$ is large, if the carbon grain  destruction time $t_d$ is small or if the radial drift velocity $v_r$ is small. In order to apply the analytic model, we must define $\Sigma_d^*$ and $t_d$, which we will do in the following sections.

\subsection{Connection to the Analytic Solution}
\label{sec:cttas}
Up to this point, we have introduced all the necessary ingredients to evaluate the time-evolution of the carbon fraction $f_c$. We plug in the relevant depletion timescale $\tau_{c,0}$ and the power law index $b$. For photolysis, unrestricted by vertical transport and without considering forward scattering, we use equation (\ref{eq:C_unrestricted}) and evaluate it at $r_0$=1 AU to find $\tau_{c,0}^\mathrm{ph}$. Because $F_{UV}\propto r^{-2}$ we immediately find the power-law index $b=2$. On the other hand, if depletion is limited by vertical transport, we generally do not find an explicit expression for $\tau_{c,0}^\mathrm{res}$ and $b$. Moreover, the carbon depletion timescale $\tau_c^\mathrm{res}$ does not necessarily have power-law dependence in $r$. Therefore, we approximate it with a power law as
\begin{equation}
     \tau^\mathrm{res}_{c}(r,z)\approx \tau_{c,0}^\mathrm{res}(z) \cdot\bigg(\frac{r}{r_0}\bigg)^{b-3/2}
     \label{eq:power-law_fit}
\end{equation}
and find the corresponding values at a given height $z$ for $\tau_{c,0}^\mathrm{res}$ and $b$ in a least-square fit. When combining photolysis and vertical transport in the disk, we calculate an \textit{effective} depletion timescale at every radius $r$. In cases when we do not consider forward scattering, we calculate the effective depletion timescale at the $\tau=1$-surface
\begin{equation}
    \tau_c^\mathrm{eff}(r,z_\tau) = \max\big[\tau^\mathrm{res}_c(r,z_\tau),\tau_c^\mathrm{ph}(r,z_\tau)\big].
    \label{eq:eff_depl_ts}
\end{equation}
When we consider forward scattering, and we find $z_1$ with equation (\ref{eq:fwrdscattering}), the effective depletion timescale is
\begin{equation}
    \tau_c^\mathrm{eff}(r,z_1) = \tau^\mathrm{res}_c(r,z_1) = \tau_c^\mathrm{ph}(r,z_1).
    \label{eq:eff_tau_c_w_scattring}
\end{equation}

\section{Thermal decomposition (pyrolysis / irreversible sublimation)}
\label{sec:thermal_decomposition}
The second carbon depletion mechanism we study, in addition to photolysis, is the process of \textit{pyrolysis} (this process is sometimes also referred to as \textit{irreversible sublimation} of the carbonaceous material) which is one of the prime suspects for depleting the inner protosolar disk of its refractory carbon \citep[][]{Li21,vanthoff2020,Gail2017,Nakano03}. While sublimation is a physical change of state that does involve chemical alterations, pyrolysis is a thermal decomposition process that transforms a solid into a gas via the decomposition of larger organic molecules into smaller molecules (e.g. CO, CO$_2$, CH$_4$ or hexane, toluene, phenol, heptane) and as such is generally irreversible. The pyrolysis temperatures of the large organic molecules are typically much larger than the sublimation temperatures of the volatile, newly produced molecules. Thus, the products directly enter the gas phase. In this section, we consider the \textit{soot line} \citep[][]{Kress10} to be the outer edge of a region within which thermally driven irreversible sublimation, i.e. the combined effect of breaking down molecular bonds within large organic molecules and the consequent release of the small molecules into the gas phase, irreversibly destroys carbonaceous material.  

\subsection{Depletion Timescales}
Similarly to section \ref{sec:photolysis}, in this section we define a depletion time $t_d'$ for irreversible sublimation by describing the sublimation process as a first order reaction process, i.e. the reaction rate is proportional to the local volume density of carbonaceous material in the disk, using kinetic theory. Thermogravimetric laboratory experiments on kerogen, a terrestrial analog to interstellar carbonaceous material, show that the rate coefficient $k_i$ of irreversible sublimation is best described by an Arrhenius-law equation \citep[][]{Chyba90} 
\begin{equation}
    k_i = A_ie^{-E_{i,a}/RT}
    \label{eq:rate_law}
\end{equation}
where $k_i$ is the rate, $A_i$ is the exponential prefactor, $E_{i,a}$ is the activation energy and T is the temperature of the carbonaceous material. Due to the expected chemical analogy of the carbonaceous material and the terrestrial kerogen, we also adopt this theory to describe the sublimation of carbonaceous material in our dust compounds in the presolar disk. During heating of kerogen, various volatile organic compounds are released, for which \cite{Oba2002} measured activation energies $E_a$ and prefactors $A$. For photolysis, limited laboratory data does not allow us to differentiate between different carbonaceous compounds. For pyrolysis, more laboratory data exists. However, for simplicity, we describe the sublimation process of each of our five carbonaceous compounds with only a single rate law, as in equation (\ref{eq:rate_law}). A more detailed analysis would not be justified given the uncertainty in the exact composition of presolar material. In \autoref{fig:refarctory_dust_composition}, we list the prefactors $A_i$ and activation energies $E_{a,i}$ for four carbonaceous components (i=1...4, see section \ref{sec:ther_dec_am_C} for $i=5$). Further, we assume that the internal temperature of a dust grains is uniform, and the decomposition process follows equation (\ref{eq:rate_law}) uniformly inside the grain volume (as opposed to e.g. evaporation which is a surface process). \cite{Patisson2000} confirm this uniformity, and as such the validity of our first order approach, for carbonaceous grains which are comparable in size and on temperature variation timescales which are much shorter than what we consider in this work. Moreover, we assume the dust grains to be in thermal equilibrium with their gaseous environment at all times. Assuming a vertically isothermal disk, dust grains only encounter different thermal environments via radial movements and \cite{Stammler17} argue that the radial motion of dust grains is slow enough that the instantaneous adaption of the grain temperature to the ambient temperature is well justified. Therefore, we write destruction time of an individual grain of compound $i$ at disk temperature $T$ as 
\begin{equation}
    t^\mathrm{sub}_{d,i} = A_i^{-1}e^{E_{i,a}/RT}
    \label{eq:subl_time}
\end{equation}
Because in our model, the disk is vertically isothermal, at every radius, sublimation is active throughout the entire vertical column of the disk. Thus, the depletion timescale is equal to the time it takes to deplete an individual grain, i.e.
\begin{equation}
    \tau^\mathrm{sub}_{c,i} = t^\mathrm{sub}_{d,i}
    \label{eq:subl_depl_time}
\end{equation}

\subsubsection{The Case of Amorphous Carbon}
\label{sec:ther_dec_am_C}
Compared to the other carbonaceous compounds considered in our model, amorphous carbon is more refractory. Also, it mainly consists of atomic carbon, without additional O, N, or S, which prevents it from being thermally decomposed into small hydrocarbons like the more volatile kerogen compounds at low temperatures. Instead, only at temperatures above $\sim 1500$ K it vaporizes via the release of chain molecules (mainly $C_1...C_5$) where the molecules readily react with oxygen \citep[][]{Duschl96}. However, in protoplanetary disk environments the amorphous carbon is eroded by chemical reaction with OH molecules, at temperatures below its sublimation temperature ($\sim 1200$ K). Thus, amorphous carbon can be destroyed by a combustion process before it sublimates \citep[][]{Duschl96,Gail01,Gail2017}. To model this thermochemical decomposition, a complex chemical model of the gas phase is required and is beyond the scope of this work. In a steady state, the temperature required for the oxidation of amorphous carbon via OH is only reached well at the midplane within Earth's orbit, a region which is not relevant for this paper. Alternatively, \cite{Anderson2017} and \cite{Klarmann2018} studied the effects of carbon depletion via oxidation in the hot disk atmosphere, but it was found to not significantly contribute to the carbon depletion in the inner Solar System.

\subsection{Analytic solution of the carbon fraction}
Based on the consideration in the previous section, and considering only the irreversible sublimation to affect the abundance of refractory carbonaceous material in the disk, we find the source term for the $i^{th}$ carbonaceous compound ($i=1..4$) in equation (\ref{eq:carbon_evolution}) to read
\begin{equation}
    \dot{\Sigma}_{d,i} = \Sigma_{d,i}A_i\exp\bigg(-\frac{E_{i,a}}{RT(r)}\bigg) 
    \label{eq:source_terM_sublimation}
\end{equation}
In appendix \ref{sec:sublimation_solution}, we derive the solution to equation (\ref{eq:carbon_evolution}) with the source term of the form (\ref{eq:source_terM_sublimation}) to model effects of irreversible sublimation. The solution to equation (\ref{eq:carbon_evolution}) for $i=1...4$ reads
\begin{equation}
 \begin{split}
    &\Sigma_{d,i}(r,t) = \Sigma_{d,i}(r',t')\Big( \frac{r}{r'}\Big)^{-(l+1)}\cdot\\ &\exp\bigg\{\frac{r_0}{v_0}\frac{A_iRT_0}{qE_{i,a}} \bigg[\exp\bigg(-\frac{E_{i,a}}{RT_0}\Big( \frac{r}{r_0}\Big)^q \bigg)-\exp\bigg(-\frac{E_{i,a}}{RT_0}\Big( \frac{r'}{r_0}\Big)^q \bigg)\bigg]\bigg\}
 \end{split}
\label{eq:sublimation_comp_i_solution}
\end{equation}
By plugging the solution of equation (\ref{eq:sublimation_comp_i_solution}) into equation (\ref{eq:sum_of_i_components}), we find the solution to the total dust surface density $\Sigma_d(r,t)$. By using equation (\ref{eq:f_i}) and equation (\ref{eq:carbin_fraction_sum}), we find the carbon fraction as a function of radius and time $f_c(r,t)$. We discuss these results in section \ref{sec:results_sublimation}. In section \ref{sec:Photolysis and Sublimation Combined}, we combine the results of irreversible sublimation with photolysis to study the combined effects of the carbon depletion mechanisms. 

\section{Results}
Here we present the results of our analysis. In section \ref{sec:results_photolysis}, we first present the effect of photolysis on the abundance of carbonaceous material in our disk model and discuss the influence of model parameters on the results. In section \ref{sec:results_sublimation}, we report the results when considering the effects of sublimation. In section \ref{sec:Photolysis and Sublimation Combined}, we combine photolysis and sublimation.  

\subsection{Photolysis}
\label{sec:results_photolysis}
\subsubsection{Depletion Timescales}
\label{sec:res1}
In our fiducial model, the depletion timescale of unrestricted photolysis in the exposed layer above $z=z_\tau$, i.e. without including restrictions by vertical transport, as calculated with equation (\ref{eq:C_unrestricted}), has a value of $\tau_{c}^\mathrm{ph}=40\;\mathrm{kyr}$ at 1 AU and is proportional to $r^{0.5}$. The radial profile of the unrestricted depletion timescale is plotted in \autoref{fig:d_timescales} in yellow color. When vertical transport is included, i.e. the depletion timescale is calculated with equation (\ref{eq:residencetimeliiteddesrate}) at $z=z_\tau$, the depletion timescale follows the blue line in \autoref{fig:d_timescales}. This profile is not strictly a power law, thus, as described in section \ref{sec:cttas}, we approximate it with a power law of the form of equation (\ref{eq:power-law_fit}). We find the proportionality factor of the depletion timescale to be equal to $\tau_{c,0}^\mathrm{res}=2.57\;\mathrm{Myr}$ and the power-law exponent to have a value of 0.22. When considering the limiting effects of vertical transport, the depletion timescale is almost two orders of magnitude larger compared to the unrestricted case. This is because, even though the exposed layer gets depleted in carbon quite quickly, it can not be replenished efficiently with undepleted material from the midplane by vertical mixing. This supports the findings of \cite{Klarmann2018} who find that vertical dust transport reduces the efficacy of carbon depletion. We also find the slope of $\tau_c^\mathrm{res}(z_\tau)$ to be shallower than the slope of $\tau_c^\mathrm{ph}(z_\tau)$, meaning the carbon depletion efficacy does not drop as much at larger distances from the star as compared to the unrestricted model. In a third model, we also include effects of photon forward scattering. When considering forward scattering, UV photons penetrate deeper into the disk than the $\tau=1$-surface, and we calculate the depletion timescale at height $z_1$ with equation (\ref{eq:eff_tau_c_w_scattring}). Because $z_1$ lies below the $\tau=1$-surface, the relevant residence time is larger than without scattering $t_\mathrm{res}(z_1)>t_\mathrm{res}(z_\tau)$ and grains spend more time in the exposed layer before being mixed back into the deeper layers of the disk. An increased residence time alone would result in less efficient carbon depletion. But as a result of forward scattering, a larger fraction of the surface density is exposed to stellar radiation. With decreasing height $z$, the inverse of the exposed surface density fraction ($\Sigma_d/\Sigma_d^*$) in equation (\ref{eq:residencetimeliiteddesrate}) decreases faster than the residence time ($t_\mathrm{res}$) increases. Therefore, the carbon depletion timescale with scattering (yellow line in \autoref{fig:d_timescales}) is overall smaller than the depletion timescale without scattering (blue line in \autoref{fig:d_timescales}). We find the power-law approximation to have a value of $\tau_{c,0}^\mathrm{res}(z_1)=695 \: \mathrm{kyr}$, almost four times smaller than without scattering, but still more than an order of magnitude larger than for unrestricted photolysis. Its dependence on $r$ is steeper than without scattering, $\propto r^{0.31}$.

\begin{figure*}
\includegraphics[width=1.5\columnwidth]{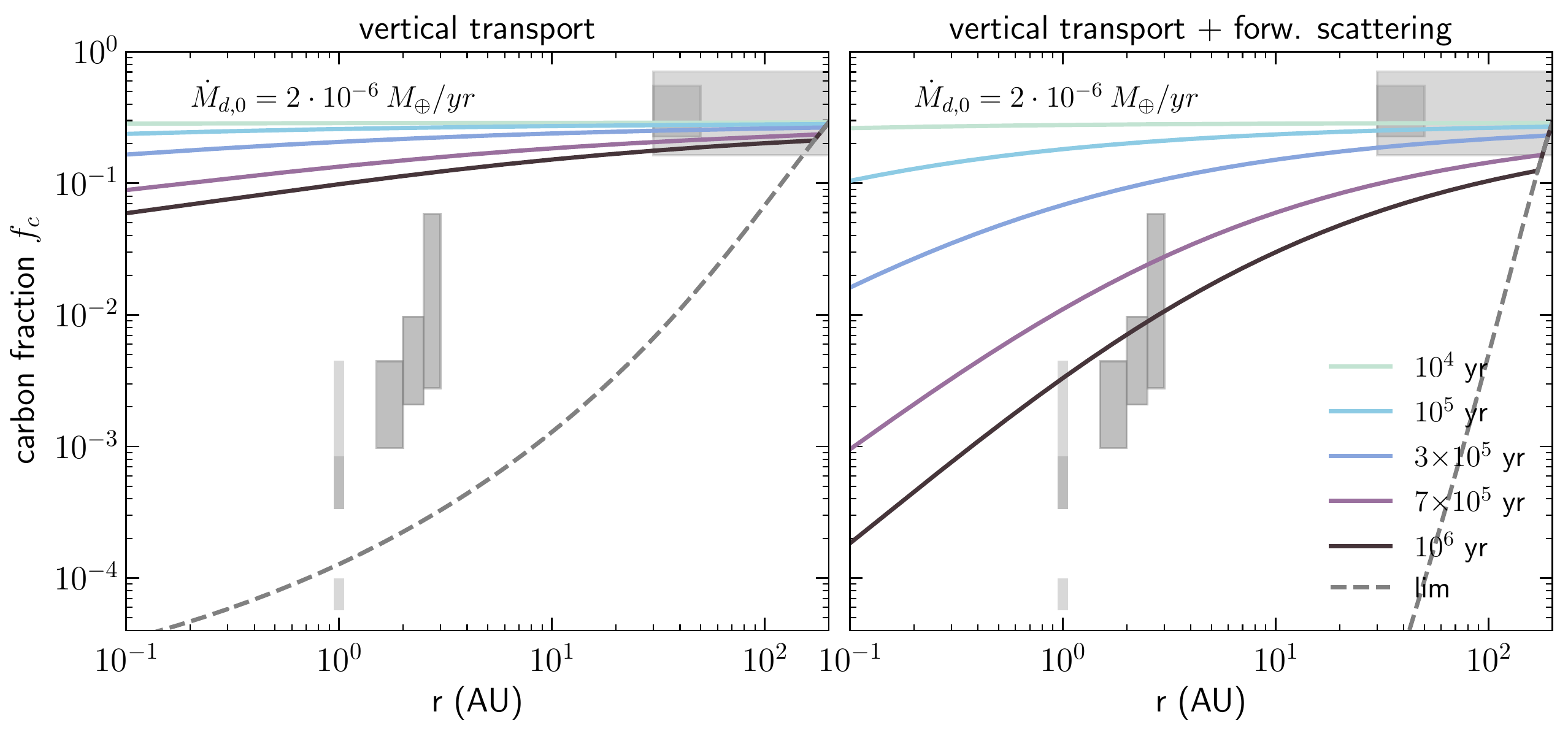}
\caption{Carbon fraction $f_c$ as a function of radius at different times over $10^6$ years, with dust surface density $\Sigma_d$ reduced by a factor of five compared to the fiducial model (including radial and vertical transport). \textit{left:} Carbon fraction as a function of radius and time for residence time limited depletion without forward scattering, i.e. the depletion timescale is calculated using equation (\ref{eq:residencetimeliiteddesrate}) which results in a value of $\tau_c^\mathrm{res}(z_\tau)=617$ kyr at 1 AU.  \textit{right:} Same as the left panel but including forward scattering. The depletion timescale is calculated using equation (\ref{eq:fwrdscattering}) which results in a value of $\tau_c^\mathrm{eff}(z_1)=158$ kyr at 1 AU.}
\label{fig:optimized_model}
\end{figure*}

\subsubsection{Photolysis without Radial Transport}
\label{sec:noradtrans}
We plug the three different depletion timescales, as reported in the previous section \ref{sec:res1}, into the analytical evolution equations of the carbon fraction $f_c$ in equation (\ref{eq:final_solution}). We first consider cases without radial transport, i.e., we use equation (\ref{eq:nort_solution}) to calculate the surface density ratio $f_\Sigma$ in equation (\ref{eq:final_solution}). In the upper row of \autoref{fig:unrestricted_model}, we plot, in the first panel, the carbon fraction $f_c$ as a function of disk radius at different times for photolysis without restrictions by vertical transport using $\tau_{c,0}^\mathrm{ph}=40\;\mathrm{kyr}$. In the second panel, we show the case in which the depletion is limited by vertical transport and $\tau_{c,0}^\mathrm{res}=2.6\;\mathrm{Myr}$, i.e. without scattering. In the third panel we show the case with scattering and $\tau_{c,0}^\mathrm{res}=695 \; \mathrm{kyr}$. The gray boxes in the background of each sub-panel in \autoref{fig:unrestricted_model} represent the estimated carbon fractions of Solar System objects, as discussed in section \ref{sec:dust_model}. We confirm the results of \cite{Klarmann2018} by showing that unrestricted photolysis (without restrictions by radial or vertical transport) depletes the inner disk of carbon to values of almost $10^{-4}$ at 1~AU within $<300$ kyr. The outer disk shows carbon depletion by more than a factor of ten within the same time, such that Solar System values can be reproduced within a few hundred thousand years. We also find that vertical transport significantly reduces the carbon depletion efficacy because the carbon fraction in the exposed drops and vertical transport can not efficiently replenish these layers with undepleted material from lower disk layers. As a result, the carbon fraction, in the second panel of \autoref{fig:unrestricted_model}, is barely reduced. Within 1 Myr, values below $10^{-1}$ are not reached anywhere in the disk. This result represents a lower limit not only for photolysis with vertical transport included, but more generally for any photo-induced carbon depletion mechanism that is active in the FUV irradiated layer of the disk. \newline Further, we find that including forward scattering improves the carbon depletion efficacy, but it does not sufficiently decrease the depletion timescale such that levels required to reproduce Solar System abundances are not reached. After 1 Myr, the carbon fraction at 1 AU is just slightly below $f_c=10^{-1}$.\newline

\subsubsection{Photolysis with Vertical and Radial Transport}
Next, we also include the effects of radial transport. Including radial transport does not change the depletion timescale $\tau_c$ in our models, but exposes grains to different environments with different depletion timescales as they radially drift inward. Because, the depletion timescales in all the three cases presented in \autoref{fig:d_timescales} have a positive radial gradient, including radial transport decreases the overall carbon depletion efficacy. This is because, at early times, grains experience environments with a lower depletion timescale at larger radial distances. This is not the case if radial transport is ignored and grains are exposed to the same depletion environment at all times. In the lower row of \autoref{fig:unrestricted_model}, we show the results of the full model, i.e. calculating the carbon fraction using equations (\ref{eq:initial_radius01}), (\ref{eq:surface_density_ratio}) and (\ref{eq:final_solution}). When radial transport is included, the carbon fraction $f_c$ does not decrease indefinitely, but eventually reaches a steady state. We indicate the steady-state solution with the gray dashed line in every sub-panel in the lower row of \autoref{fig:unrestricted_model}. The steady-state solution is the minimal possible carbon fraction when the dust disk is continuously replenished by undepleted material via the outer disk boundary. In all three solutions presented in the lower row of \autoref{fig:unrestricted_model}, the carbon fraction at 1 Myr is still far away from the steady-state solution throughout most of the disk because radial drift is relatively slow compared to the depletion timescale. At 200 AU the radial drift timescale ($r/v_r$) is $\sim6.5$ Myr. Overall, in our models, the influence of radial drift is significantly smaller than the restrictions by vertical transport. However, this is only true as long as the steady-state solution is not reached. At later times, the solutions with and without radial transport will diverge significantly. But this will only become relevant once the system approaches the drift timescale ($r/v_r$) of the grains at the outer disk boundary. In our models, this timescale is much larger than the timescale on which we expect planetesimal formation to occur. However, when radial transport is included in our fiducial model with scattering, the carbon fraction must come very close to the steady-state solution to explain to reproduce Solar System abundances. This happens on a timescale of a few Myr. 

\subsubsection{Overcoming Dust Transport}
\label{sec:Overc_d_t}
In the previous section, we have shown that inefficient vertical transport is the main limiting factor in the photo-induced carbon depletion via photolysis because the typical time a grain spends in the exposed layer, per mixing cycle (its residence time), is long compared to the time it takes to destroy and individual grain. In this section, we will study the properties of our analytical model to understand under which circumstances the carbon depletion timescale can be decreased sufficiently to reproduce carbon fractions $f_c$ as measured in the Solar System. It is not straightforward to understand how model parameters influence the carbon depletion timescale $\tau_c^\mathrm{res}$ as presented in equation (\ref{eq:fwrdscattering}), due to its implicit form. Therefore, we derive an explicit (but only approximate) expression of the depletion timescale. This allows us to understand the influence of individual parameters on our results. Later we confirm our findings by comparing with the exact result. We plug the explicit approximation of the residence time as derived in equation (\ref{eq:explicit_residence_time}) into the definition of the depletion timescale in equation (\ref{eq:residencetimeliiteddesrate}), and obtain an explicit expression of the carbon depletion timescale in the residence time limit, i.e. with vertical transport:
\begin{equation}
   \tau_{c}^\mathrm{res}\simeq\frac{\Sigma_{d}\kappa_0}{2\Phi\alpha \Omega}\cdot\Bigg(\sqrt{2\ln\frac{f_{\leq a_s}\Sigma_d \kappa_0}{2\sqrt{\pi}\Phi}}-\frac{1}{5}\Bigg)^{-2}
  \label{eq:explicit_C}
\end{equation}
In the above expression $f_{\leq a_s}\approx(a_\mathrm{max}/a_s)^{-0.5}$ is the mass fraction of grains below or equal to size $a_s$ which we derive in appendix \ref{sec:tau=1_surface_derivation}. The quantity $a_\textrm{max}$ is the upper limit of the grain size distribution and, in the fragmentation limit, is estimated using equations (\ref{eq:Stokes_number_para}) and (\ref{eq:fra_limit}). In the fragmentation limit, $a_\textrm{max}$ is inversely proportional to the turbulent alpha parameter  $a_\textrm{max} \propto \alpha^{-1}$. Equation (\ref{eq:explicit_C}) approximates the depletion timescale when limited by vertical transport, but without taking into account the effects of forward scattering. But, the results in equation (\ref{eq:explicit_C}) does serve as a good estimate of an upper limit for the solution which includes scattering. In our fiducial model, using equation (\ref{eq:explicit_C}), we find $\tau_{c}^\mathrm{res}(z_\tau)=3.6\;\mathrm{Myr}$ at 1 AU without decoupling of dust grains. For comparison, we calculated a value of $2.6\;\mathrm{Myr}$ with grain decoupling in \ref{sec:forward_scattering}. The lower value arises from the fact that more weakly coupled have a smaller residence $t_\mathrm{res}$ (see \autoref{fig:residence_time})

Due to the explicit nature of equation (\ref{eq:explicit_C}), we understand its dependence on model parameters. From the first factor in equation (\ref{eq:explicit_C}), we find that four parameters influence the depletion timescale ($\Sigma_d, \kappa_0, \Phi, \alpha$), plus the Keplerian frequency $\Omega$, but the latter is not influenced by the choice of model parameters. In equation (\ref{eq:explicit_C}), the dust surface density ($\Sigma_d$) also appears in the argument of the natural logarithm ($\ln$) in addition to the linear dependence of the first factor. Overall, both contributions counterbalance each other due to the negative exponent in the second factor. However, the argument of the natural logarithm is generally much larger than unity and, thus, its argument contributes less than linear to the overall depletion timescale. Therefore, a decrease in the dust surface density also decreases the depletion timescale because a lower dust surface density decreases the optical depth of the disk, which increases the surface density fraction contained in the exposed layer. The same scaling argument holds for changes in the opacity $\kappa_0$ or the flaring angle $\Phi$. A smaller opacity similarly decreases the optical depth of the disk and thus increases the carbon depletion efficacy. A larger flaring angle allows stellar photons to arrive at a steeper angle and thus penetrate deeper into the disk. On the other hand, the turbulence parameter ($\alpha$) does only appear in the first factor, but not explicitly in the argument of the natural logarithm. However, it implicitly acts on $f_{\leq a_1}$ via the maximum grain size in the fragmentation limit $a_\mathrm{max}$. The maximum grain size is proportional to $\alpha^{-1}$. Thus, $f_{\leq a_1} \propto \alpha^{0.5}$ and consequently, the turbulence parameter contributes slightly more than linear to the depletion timescale because more turbulence increases the efficiency of vertical transport and, at the same time, increases the number of small grains in the disk because large grains fragment more frequently. To conclude, we find carbon depletion to be favored by a low dust surface density, low opacity, high flaring angles and high turbulence. Moreover, it is favored by a large mass fraction in small grains, but to a lesser degree than the other factors because $f_{\leq a_1}$ only appears in the argument of the natural logarithm.\newline  
In our fiducial model, the turbulence parameter is already large, and the constraints on the flaring angle and the opacity are relatively tight in our model. The dust surface density $\Sigma_d$, on the other hand, is not well constrained. Decreasing the dust surface density by a factor of five compared to the fiducial model reduces the approximate depletion timescale at 1 AU, as calculated with equation (\ref{eq:explicit_C}), from $3.6\:\mathrm{Myr}$ to $860\:\mathrm{kyr}$. The exact result that also considers grain decoupling decreases from $2.6\:\mathrm{Myr}$ to $617\:\mathrm{kyr}$. In both cases, the decrease is a factor of 4.2, while the latter values are lower because the residence times of more weakly coupled grains are generally smaller (see \autoref{fig:residence_time}). We plot the evolution of the carbon fraction in the disk with the decreased depletion timescale (without forward scattering) in the left panel of \autoref{fig:optimized_model}. With a decreased dust surface density, the carbon fraction reaches levels of $10^{-2}$ at 1 AU within 1 Myr which is a depletion by almost a factor of 30. To compare, in the fiducial model, the carbon fraction was only decreased by a factor of 1.5.   \newline
When including forward scattering, and solving equation ($\ref{eq:fwrdscattering}$), we do not provide an explicit expression. But, the functional dependence of the depletion timescale will be expanded to include the stellar UV-flux. Forward scattering further decreases the depletion timescale at 1 AU to $87\;\mathrm{kyr}$. We show the evolution of the carbon fraction with scattering in the right panel of \autoref{fig:optimized_model}. For comparison, the depletion timescale for photolysis unrestricted by vertical transport is also smaller with lower dust surface density, with a value of $13\;\mathrm{kyr}$ at 1 AU. The lower dust surface density allows the carbon fractions in our model to reach levels comparable to Solar System abundances within a time of $\sim 7\times 10^5$ years, even when radial and vertical transport is included. The same is true for a decrease in the opacity or an increase in the turbulence or flaring angle. Or a change in all parameters which results in a combined decrease in the depletion timescale by a factor of eight compared to the fiducial model.

\subsection{Irreversible Sublimation}
\label{sec:results_sublimation}
\begin{figure}
\includegraphics[width=\columnwidth]{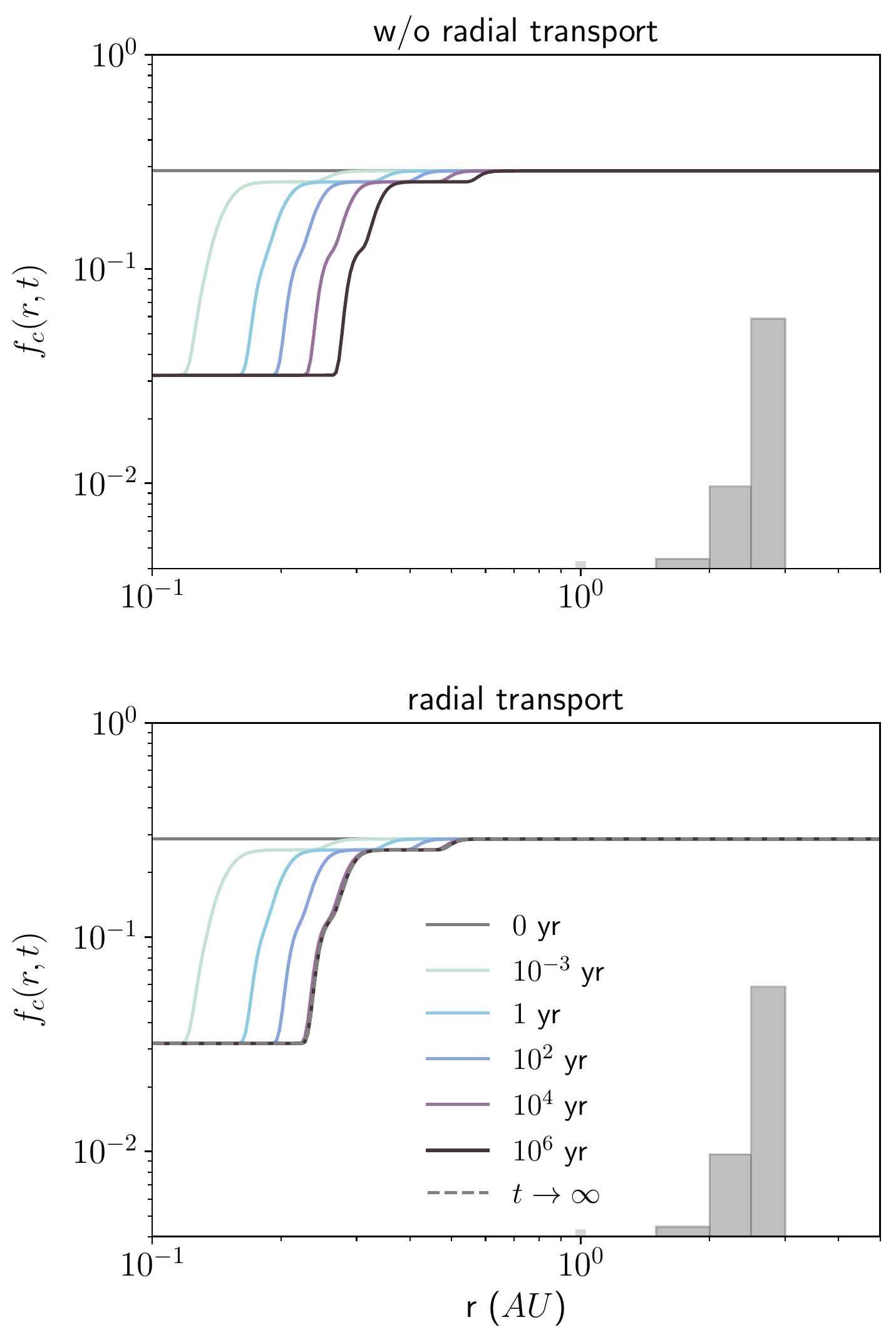}
\caption{Carbon fraction $f_c$ as a function of radius at different times under the assumption that irreversible sublimation is the sole contributor to the decomposition of refractory carbon. The upper panel shows the carbon fraction profile in the absence of radial drift. The individual soot lines continuously move outward, such that the region within about 0.3 AU only consists of amorphous carbon after about 1 Myr. The bottom panel shows the same situation, but including radial drift. At a time between 10 kyr and 100 kyr the outward motion of the soot lines is counterbalanced by the inward drift of the dust and a steady state forms. The gray boxes indicate the carbon fraction as found in chondrites.}
\label{fig:sublimation_only_results}
\end{figure}

In this section, we study the effects of irreversible sublimation on the carbon fraction $f_c$, as introduced in section \ref{sec:thermal_decomposition}. We first ignore contributions from photodecomposition processes. In \autoref{fig:sublimation_only_results}, we display the carbon fraction $f_c$ as a function of radius at various times, assuming that the refractory carbonaceous material is continuously decomposed by irreversible sublimation according to equation (\ref{eq:source_terM_sublimation}). Note, \autoref{fig:sublimation_only_results} only shows the radial range between 0.1 and 5 AU, and plot the carbon fraction $f_c$ at times at which are generally smaller compared to the previous figures in which we showed the results of photolysis. This is because the inner disk is depleted significantly faster due to the exponential dependence on temperature. The upper row of \autoref{fig:sublimation_only_results} shows the evolution of the carbon fraction in the absence of radial drift. At each point in time, one can identify multiple distinct steps in the radial profile of the carbon fraction $f_c$. These steps correspond to the soot lines of the individual carbonaceous compounds in our model (see \autoref{fig:refarctory_dust_composition}). The carbon fraction reaches a floor value at $f_c = 0.03$ when only the amorphous carbon compound is left. In the absence of radial drift, these soot lines continuously move outward because the depletion timescale is finite everywhere in the disk. At 1 Myr, all the carbonaceous compounds, except the amorphous carbon, have completely sublimated in the disk region within 0.3 AU from the star. \newline
In the bottom row of \autoref{fig:sublimation_only_results}, we include the effects of radial drift. The main difference compared to the no-drift situation is the fact that a steady-state distribution exists, which in our model is reached between 10 kyr and 100 kyr. After that, the carbon fraction profile does not change anymore and the individual soot-lines are stationary because the inward drift of the dust exactly cancels the outward motion of the soot-line. Due to the exponential dependence of the depletion timescale on temperature, the transition region in which an individual compound is only partially sublimated is very narrow (fractions of an AU). Thus, the radial gradient in the disk is more a result of the different sublimation temperatures of the individual compounds, rather than a product of the radial variation in the depletion timescale of an individual compound. This property is distinctly different from photolysis, where the shallow radial slope of the depletion timescale is responsible for the global radial carbon fraction gradient in the disk. Overall, we find irreversible sublimation in our passively heated steady state disk model to only deplete the innermost disk region. Furthermore, it decreases the carbon fraction by at most a factor ten due to the presence of a highly refractory amorphous carbon compound, which only decomposes at temperatures $> 1000$ K. The carbon fraction in the colder disk regions, where Earth or the chondrites are found, is not depleted. The lowest level that the inner disk can be depleted to by irreversible sublimation is strictly set by the abundance of amorphous carbon ($f_5=0.03$). The detailed composition of the carbonaceous material in the ISM is highly uncertain. However, the adopted value of 3 \% is close to the value of $4.2\:\%$ reported in \cite{Fomenkova94} for in situ measurements in the coma of comet Halley (however, the latter value is reported as a number fraction of measured dust grains rather than a mass fraction). In order to reach values relevant for bulk Earth with irreversible sublimation alone, the abundance of amorphous carbon must be a factor ten lower, i.e. as low as $f_4\lesssim 0.4\:\%$. \newline
In our model, the steady state soot line is located at a heliocentric distance of about $r\sim 0.25$ AU, which corresponds to a temperature  of about $\sim 450$ K in our passively heated, thermal equilibrium disk (see lower panel of \autoref{fig:sublimation_only_results}). The inclusion of viscous disk heating in our model could move the soot line radially outward as a result of the increased disk temperature. However, at 1 AU, we find the disk surface temperature to be dominated by viscous heating only if the mass accretion rate is larger than $\sim 2.7 \cdot 10^{-7} \mathrm{M_\odot/yr}$. In our disk model with $\alpha = 10^{-2}$, the accretion rate at 1 AU is with a value of $\sim 3.5 \cdot 10^{-8} \mathrm{M_\odot/yr}$ lower. Hence, the soot line does stay within Earth orbit even with the contribution of viscous heating. For comparison, \cite{Li21} show that the soot line in their model moves outward to about 1 AU with $\alpha = 10^{-3}$ when viscous dissipation does contribute to the local heating of the disk. Thus, in a steady state disk model with viscous heating, the disk region out to Earth's orbit might get depleted. But only by a factor of ten because the remaining amorphous carbon compounds survive up to temperatures of over 1200 K before the OH abundance level in the disk rises to levels sufficient for the amorphous carbon to oxidize \citep[][]{Gail2017,Gail01,Finocchi97}. Such temperatures are only reached well within Earth's orbit. Thus, with an amorphous carbon abundance above 0.4 \%, we fail to explain the two to three orders of magnitude depletion required to reproduce the carbon fraction of bulk Earth. Analogous arguments hold for the depletion of parent bodies of chondrites. 

\subsection{FU Ori-type Outbursts}
\label{eq:FU Ori-type Outbursts}
\begin{figure}
\includegraphics[width=\columnwidth]{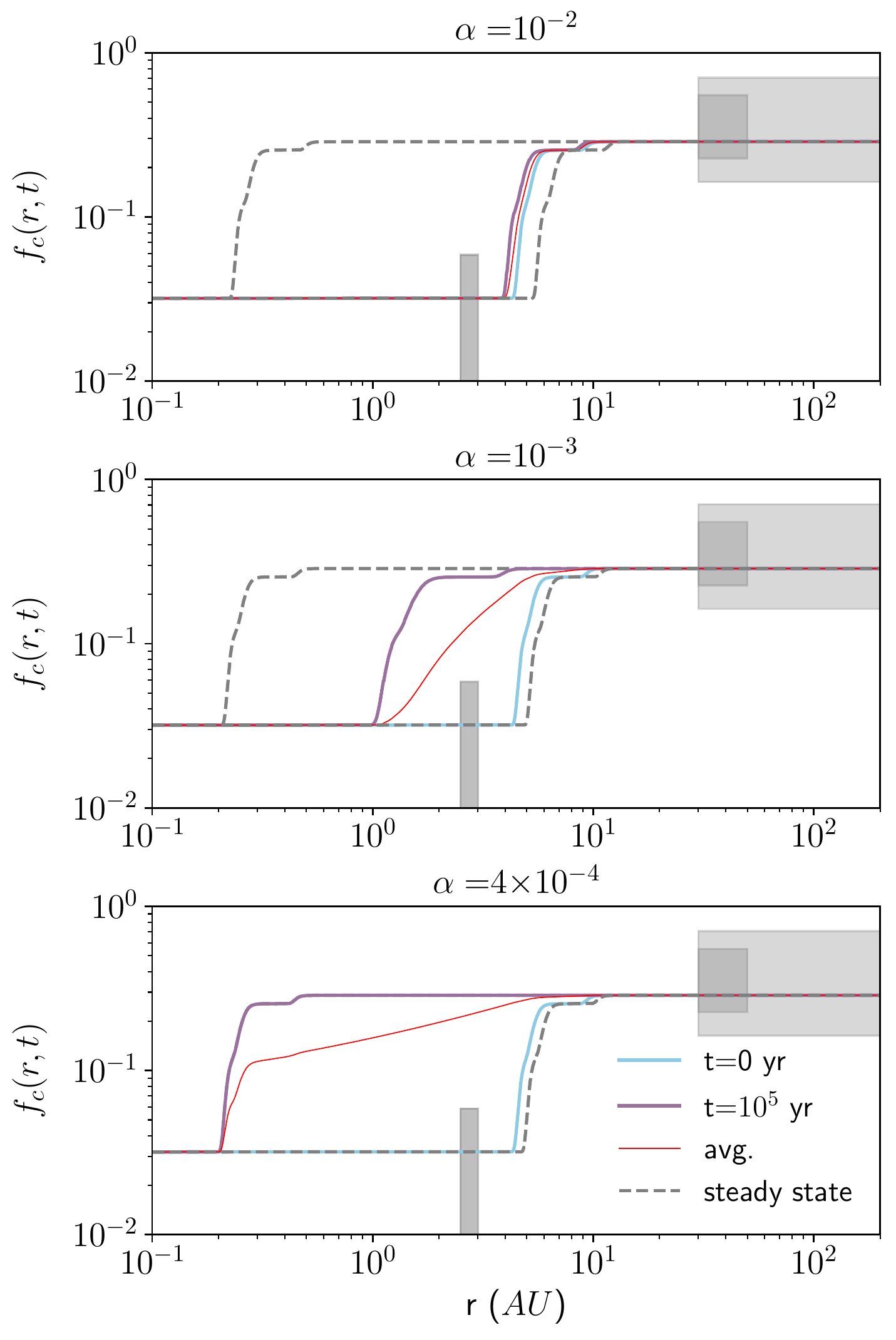}
\caption{Radial shift of the soot line as a result of an FU Ori-type outburst. Plotted is the carbon fraction $f_c$ as a function of radius for three different turbulent alphas. The dashed lines show the steady states in cases without outbursts and a state in which the luminosity in continuously increased by a factor $100$. The light blue line shows the carbon fraction after a 100-year-long outburst. The purple line shows the carbon fraction 100 kyr after an outburst when the dust had time to radially drift inward. The red line shows the time-averaged carbon fraction at every radius, assuming the outburst happen every 100 kyr. }
\label{fig:outbursts_1}
\end{figure}

In the previous section, we have shown that it is difficult to explain the carbon depletion of Earth and chondrites via irreversible sublimation in a steady state disk. In this section, we briefly study the effects of transient luminosity outbursts and irreversible sublimation on the carbon fraction. In the Solar System, there is considerable evidence that short-lasting and frequent temperature increases happened during the formation of chondrules \citep[see e.g.][ for a review]{Ciesla05}. Additional evidence against a steady state scenario during the early phases of planet formation come from FU Orionis-type objects, which are a class of objects containing a pre-main sequence star and show sudden increases in luminosity over a short period of time \citep[][]{Herbig66,Hartmann96}. We do not aim to imply a connection between the formation of chondrules and FU Ori-type outbursts here, but both phenomenon suggest a highly variable disk environment in the early formation phase of solids. Even though, it is not clear if the early Solar System has undergone any FU Ori-type outbursts, there are reasons to believe that frequent outbursts are common in the early phase of stellar evolution \citep[][]{Hartmann96}. Thus, in this section, we study the particular effects FU Ori-type outbursts on the destruction of carbonaceous material. While the underlying triggering mechanism of FU Ori-type outbursts are not well understood, the typical luminosity increase by a factor of $10^2-10^3$ within a timescale of a year to a decade that lasts about a century \citep[][]{Hartmann96} is thought to be the result of a burst of the accretion rate in the inner disk region \cite[$r<1$ AU,][]{Zhu07}. Even though no object has been observed which has undergone multiple such outburst, statistical arguments lead to the conclusion that they are repetitive, and an object undergoes at least ten outbursts, assuming that all (low-mass) young stars have FU Ori-type outbursts \citep[][]{Hartmann96}, i.e. one every $\sim 10^5$ years \citep[][]{Pena19}. \newline
In our model, we assume the disk to undergo episodic accretion events which last for 100 years and reoccur with a period of 100 kyr. We also assume that the region of increased accretion during an outburst is confined to the innermost disk region ($r\ll 1$ AU) which we do not explicitly model. We only model the disk outside the region of increased accretion, where we take into account passive heating via the increased accretion luminosity of the innermost disk. In FU Ori-type objects, the transition between active and passive region is derived to lie anywhere between 5 $R_\odot$ and $\sim0.5-1$ AU \citep[][]{Audard14} which is consistent with our model assumption. Further, we assume the passive portion of the disk to be heated instantly during outbursts and follows the temperature profile as in equation (\ref{eq:rad_eq_temperature}) where the stellar luminosity $L_*$ is replaced by the accretion luminosity $L_\mathrm{acc}$ of the inner disk. We set $L_\mathrm{acc} = 500 L_*$. Consequently, the resulting disk temperature during an outburst increases by a factor $T_\mathrm{burst}\approx 4.7\cdot T$. Alternatively, we find the disk temperature during the outbursts to shift radially by a factor $\sim 22.4$ like $T_\mathrm{burst}(r)\approx T(r / 22.4)$, i.e. the steady state soot line moves from $\sim 0.25$ AU to $\sim 5.6$ AU during an outburst. Naturally, any other sublimation line does move by the same factor. This is also roughly consistent when comparing the results of \cite{Cieza16}, who observed the water-snow line in the outbursting system V883 Ori at a distance of 42 AU, to the location of the steady state snow line in the early Solar System, which is expected around $\sim 3$ AU \citep[][]{Martin12}. \newline
In \autoref{fig:outbursts_1}, we show the radial shift of the soot line during one of our model outbursts in disks with a different turbulent alpha. The two dashed lines in each subplot indicate the steady-state carbon fraction $f_c$. The inner profile corresponds to the steady-state in the absence of any outburst, while the outer line corresponds to a steady state in a disk with 500 times increased luminosity. As mentioned above, the profile shifts to the outer disk by a factor about 22.4. The light-blue solid line shows the carbon fraction at the end of an outburst episode of 100 years, i.e. the expected duration of an FU Ori-type outburst. We label this with $t=0$. The purple line shows the carbon fraction at 100 kyr after the end of an outburst, when the dust had time to radially drift inward. The drift speed follows equation (\ref{eq:radial_velocity}) and is inversely proportional to the alpha parameter. During the relatively short outburst duration of 100 years, the soot-line moves outward by a considerable distance due to the short sublimation timescales (see equation (\ref{eq:rate_law})). From the steady-state location at $\sim 0.25$ AU, it moves out to $\sim 5.6$ AU. For $\alpha=10^{-2}$, the dust grains are so small that they barely drift inward in the 100 kyr time period in between outbursts, i.e. one outburst every 100 kyr is enough to permanently shift the soot line from $\sim 0.25$ AU to $\sim 5$ AU and the region within 5 AU remains permanently depleted. For $\alpha=10^{-3}$, the grains drift fast enough to just reach 1 AU, while for $\alpha=4\cdot 10^{-4}$ they reach the inner steady state radius slightly before the next outburst. \newline
Assuming a mechanism would continuously produce planetesimals at a given radius $r$, the produced planetesimals would show a different carbon fraction $f_c$ depending on whether the soot line was inside or outside their formation radius. We further assume the planetesimals are produced over an extended period of time and calculate the average carbon fraction of the entire population of planetesimals that has formed at a given radius and plot their average carbon fraction it in red color in \autoref{fig:outbursts_1}. Interestingly, for $\alpha\sim 10^{-3}$, the time averaged carbon fraction (red line) distributes the carbon more evenly in the system and does not show its step-like character. \newline
Note that, as a result of the different drift velocities, the location of the steady-state soot lines also change for different values of $\alpha$. However, as a result of the exponential dependence on temperature in equation (\ref{eq:rate_law}), the radial change is small. \newline

\subsection{Photolysis, Sublimation and Outbursts Combined}
\label{sec:Photolysis and Sublimation Combined}
\begin{figure*}
\includegraphics[width=1.5\columnwidth]{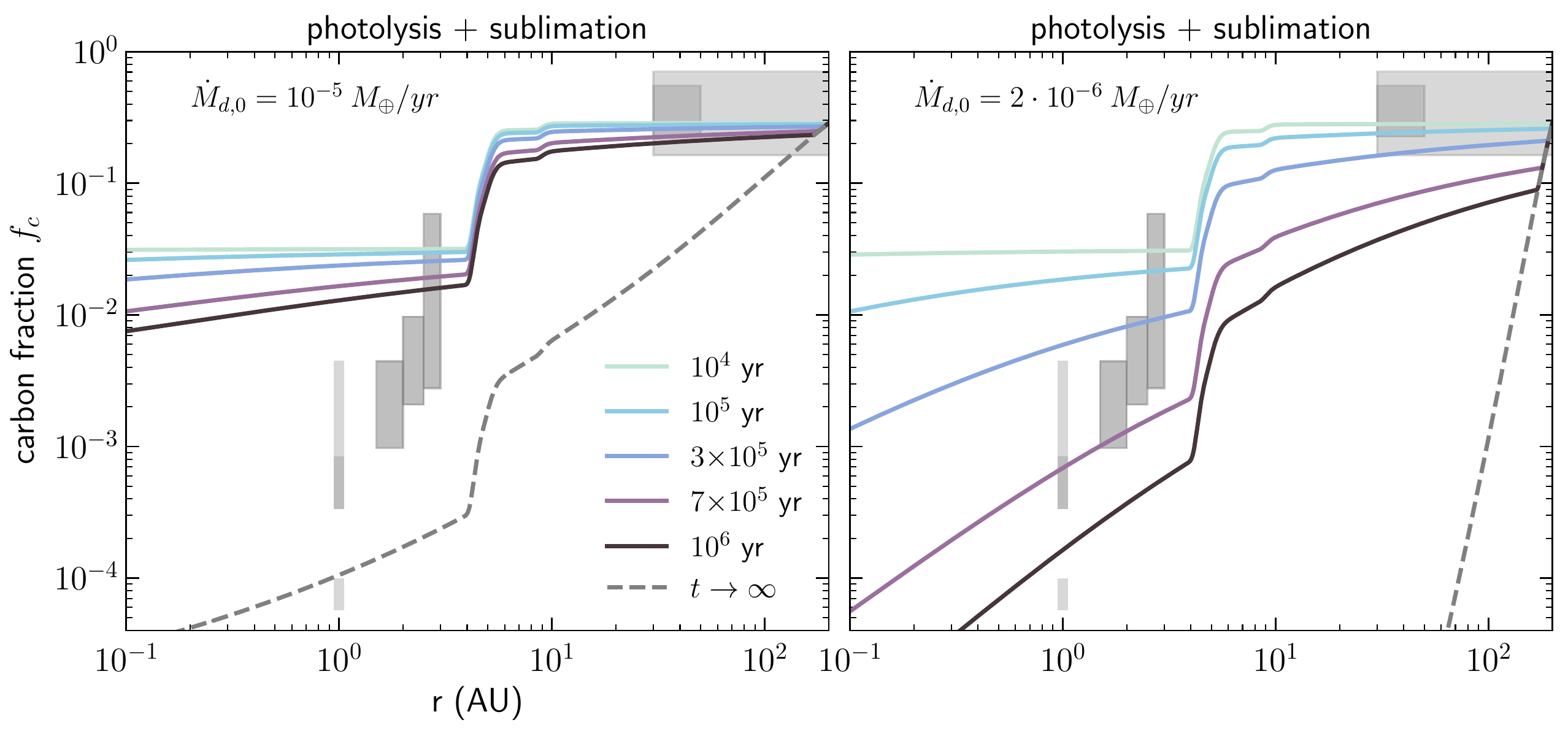}
\caption{Evolution of the carbon fraction $f_c$ under the influence of the combined effects of photolysis and time-averaged irreversible sublimation including FU Ori-type outbursts, with the fiducial dust surface density (left) and reduced by a factor of five (right). In the fiducial case, the carbon fraction does not reproduce the observed values for the inner Solar System within 1 Myr. On the other hand, when the dust-to-gas ratio is reduced by a factor of five, the observed Solar System values are reached within about $7\cdot 10^{5}$ years. }
\label{fig:results_photolysis_AND_sublimation}
\end{figure*}

In section \ref{sec:res1}, we showed that photolysis is residence-time limited in our models. And in section \ref{sec:Overc_d_t}, we showed with equation (\ref{eq:explicit_C}) that, in the residence time limit, the carbon depletion timescale $\tau_c^\mathrm{res}$ is independent of the incident radiation flux and consequently also of the underlying stellar luminosity. Therefore, we conclude that, at least to first order, carbon depletion via photolysis is not altered by an increase in luminosity during FU Ori-type outburst. The same is true for any photo-induced process that is active in the UV irradiated layers of the disk (e.g. oxidation). Nonetheless, we combine the results on stellar outbursts with the results of photolysis in section \ref{sec:results_photolysis} and study the combined effects of two depletion mechanisms of irreversible sublimation and photolysis.
In \autoref{fig:results_photolysis_AND_sublimation}, we show the combined effects of photolysis and time-averaged irreversible sublimation with the fiducial dust-to gas ratio (\textit{left subplot}) and reduced by a factor of five (\textit{right subplot}). We find the characteristic step-like shape at the soot lines beyond 4 AU as a result of FU Ori-type outbursts. Thus, at every time, there is a decrease by a factor ten in the inner disk compared to the outer disk. In the inner disk, photolysis further decomposes the refractory amorphous compounds. In the fiducial model, the carbon fraction reaches levels of carbonaceous chondrites within 1 Myr, but not the levels of the more depleted chondrites or of Earth. On the r.h.s. of \autoref{fig:results_photolysis_AND_sublimation}, the initial dust surface density $\Sigma_d$ is reduced by a factor of five, which decreases the photolysis depletion timescale by a factor of 4.2 (as described in section \ref{sec:Overc_d_t}). As a result, photolysis depletes the inner disk more efficiently and the carbon fraction $f_c$ reproduces Solar System values of chondrites and bulk Earth within 700 kyr. 

\section{Discussion}
\label{sec:discussion}

\subsection{Bouncing Collisions and Photolysis}
\label{sec:bouncing}
In this section, we consider the effect of bouncing collisions, which can alter the size distribution of the solid disk material\citep[][]{Blum10}, on the depletion timescale in the case of photolysis. Throughout this work, we have assumed the dust size distribution to be in a coagulation-fragmentation equilibrium in which the surface area of solids is dominated by the smallest grains. Bouncing collisions, rather than fragmenting collisions, reduce the number of small grains in the size distribution. We first assume bouncing and fragmenting collisions to coexists in a way that the net effect of bouncing is a reduction in the number of small grains, but fragmentation still replenishes small grains efficiently enough so that they are still the dominant contributor to the UV-opacity at the $\tau=1$-surface. The removal of small grains moves the $\tau=1$-surface closer to the midplane. If photolysis is still residence time limited, bouncing increases the carbon depletion timescale, i.e. photolysis becomes less efficient, because the residence time $t_\mathrm{res}$ increases with a shift of the $\tau=1$-surface towards the midplane (see equation (\ref{eq:residencetimeliiteddesrate}) and \autoref{fig:residence_time}). Photolysis becoming less efficient when bouncing is considered is also apparent when considering equation (\ref{eq:explicit_C}). As long as the smallest grains dominate the opacity at the $\tau=1$-surface, the inclusion of bouncing decreases the mass fraction $f_{\leq a_1}$ in equation (\ref{eq:explicit_C}), the remaining parameters remain unchanged. Because the depletion time $\tau_c^\mathrm{res}$ scales with the logarithm of the mass fraction $f_{\leq a_1}$, a change in the mass fraction $f_{\leq a_1}$, typically only has a small effect on the depletion time of photolysis. \newline
We further consider a more extreme example in which fragmentation does not replenish small grains efficiently, such that the smallest grains are no longer the main contributor to the UV-opacity at the optical surface and instead, the surface area is dominated by the largest particles at the bouncing barrier. As a result, the resident time $t_\mathrm{res}$ increases, but as we show in section \ref{sec:vertical_transport}, it always stays smaller than half the mixing time $t_\mathrm{mix}$, i.e. it increases by at most a factor five. The destruction time $t_\mathrm{ph}$, on the other hand, scales with the radius of the opacity dominating grain population (\ref{eq:c_destrcutiontime}). Thus, if the bouncing barrier is located at a large enough grain size, photolysis will no longer be residence time limited because $t_\mathrm{ph}>t_\mathrm{res}$. For an illustrative example, we assume the grains to grow until the bouncing barrier at size $a_B$. Then, the grain destruction time of a grain in the exposed layer is 
\begin{equation}
    t_\mathrm{ph,B} = \frac{m}{m_c R_\mathrm{UV}}=\frac{4}{3}\frac{a_B \rho_\bullet}{m_c Y_\mathrm{ph}F_\mathrm{UV}}
    \label{eq:bouncing_dest_time}
\end{equation}
For simplicity, we assume, the size $a_B$ of grains at the bouncing barrier, to be equal to the grains at the (hypothetical) fragmentation barrier, i.e. the only grain size we consider is 
\begin{equation}
    a_B = \frac{2}{3\pi}\frac{f_f m_g v_f^2}{\alpha k_B}\frac{\Sigma_g}{T}
    \label{eq:bouncing_barrier_radius}
\end{equation}
Then, the depletion time for bouncing $t_\mathrm{ph,B}$ scales with radius like the inverse of the Keplerian frequency
\begin{equation}
    t_\mathrm{ph,B} = t_\mathrm{ph,B,0}\bigg(\frac{r}{r_0}\bigg)^{2-p_g+q}
\end{equation}
with proportionality factor
\begin{equation}
    t_\mathrm{ph,B,0} = \frac{8}{9\pi} \frac{f_f m_g\rho_c v_f^2 \Sigma_{g,0}}{\alpha k_B m_c Y_\mathrm{ph}F_\mathrm{UV,0}T_0}
\end{equation}
In our fiducial model, we find $t_\mathrm{ph,B}=76.6\:\Omega^{-1}$ which is equal to 12.6 yr at 1 AU. The residence time is always smaller than half the mixing time, which in our fiducial model at 1 AU is equal to $50\:\Omega^{-1}$. Thus, $t_\mathrm{res}>t_\mathrm{ph,B}$ and photolysis is not limited by vertical transport in this example and the effective depletion timescale is 
\begin{equation}
  \tau_{c,B} = \frac{\Sigma_{d,0}}{\Sigma_d^*}t_{d,B}  
\end{equation}
which does simplify to 
\begin{equation}
    \tau_{c,B}^\mathrm{ph}= \frac{\Sigma_{d}}{2 m_c Y_\mathrm{ph} \Phi F_\mathrm{UV}}
    \label{eq:depl_time_bouncing}
\end{equation}
by plugging in equation (\ref{eq:bouncing_dest_time}) and using $\Sigma_d^*=\kappa_0^{-1}=4\rho_\bullet a_B/3$. Equation (\ref{eq:depl_time_bouncing}) is identical to the unrestricted carbon depletion timescale (\ref{eq:C_unrestricted}) in a coagulation-fragmentation distribution, i.e. $\tau_{c}^\mathrm{ph,B}=40\:\mathrm{kyr}$. \newline
To conclude, if bouncing does not heavily deplete the number of small grains, it does decrease the efficiency of photolysis. But, if it decreases the number density of small grains enough for large grains to become the dominant contributor to the FUV-opacity at the optical surface, it significantly decreases the depletion timescale of photolysis. The validity of this simple argument should be evaluated in a detailed study.

\subsection{Choice of Model Parameters}
\label{sec:Choice of Model Parameters}
\subsubsection{Dust Surface Density $\Sigma_d$}
We showed that the carbon depletion timescale is sensitive to the dust surface density $\Sigma_d$ and thus, indirectly also to the total dust mass contained in the disk. Hence, the carbon depletion efficacy is larger in less massive disks. However, the total dust mass in the early Solar disk is bound from below because there must be enough solid material in the inner disk to form the rocky objects in the Solar System. In our fiducial model, the total dust mass contained in the disk up to 200 AU is 131 $M_{\earth}$. \cite{Lenz20} have constrained the parameter space for the Solar nebula and propose a lower bound for the total mass of planetesimals at 76 $M_{\earth}$. If all the dust is converted into planetesimals, the initial global dust disk can not be less massive than 76 $M_{\earth}$, i.e. the dust surface density can, at most, be reduced by a factor of $1.7$ compared to our fiducial model. A reduction by a factor of 1.7 in the dust surface density alone can not reproduce the carbon abundance in the Solar System within 1 Myr. \newline
Dust traps, on the other hand, can locally decrease the dust surface density also in more massive disks. A dust trap caused by a Jupiter-mass planet can decrease the dust surface density inside the orbit of the planet by a factor of 10 \citep{Drazkowska2019}. The local decrease does not contradict \cite{Lenz20} who require a planetesimal mass of $0.1-8.77\;M_{\earth}$ in the inner Solar nebula between 0.7 and 4 AU. In our fiducial model, the total dust mass between 0.7 and 4 AU is $18.5\;M_{\earth}$. Hence, a decrease only in the inner disk by a factor 10 is consistent with constraints on the inner Solar Nebula. As a result, a dust trap is likely required to reproduce the observed carbon abundance in the inner Solar System. In other lower mass protostellar systems, a dust trap is not necessary to reproduce similar carbon depletion levels. Of course, while a dust trap decreases the dust surface density in the inner disk, it increases the dust surface density at the location of the dust trap, effectively switching off carbon depletion there. Thus, planetesimals which form inside the dust traps are likely less depleted in carbon. Therefore, if planetesimals mainly form in dust traps, carbon must be removed from the dust before it accumulates in the trap. \newline

\subsubsection{Opacity $\kappa$}
The depletion timescale $\tau_c$ sensitively depends on the UV-opacity of the disk atmosphere. Carbon depletion itself can reduce the opacity in the expose layer by releasing solid material into the gas. An effect that we do not take into account in our model. As carbonaceous material initially contributes 29 per cent to the total dust mass, it contributes about the same amount to the total opacity. Hence, the destruction of all the carbon would decrease the opacity in the exposed layer by 29 per cent. Furthermore, there are large uncertainties in the exact opacity of the solid disk material because it sensitively depends on the composition and the local size distribution of the material \citep[][]{Birnstiel2018}. We expect a detailed opacity treatment could easily introduce corrections by a factor of a few compared to our crude model assumptions.\newline

\subsubsection{Turbulent Alpha Parameter $\alpha$}
We use a relatively large value in the turbulent alpha parameter ($\alpha=10^{-2}$), which we show is needed for efficient carbon depletion in the exposed layer.  Observationally, it is difficult to directly infer the levels of turbulence in protoplanetary disks. However, measurements and modelling of the widths of dust rings in mm-continuum observations consistently report relatively strong levels of turbulence $\alpha/St \gtrsim 10^{-2}$ \citep[][]{Dullemon18,Rosotti20}. \newline

\subsection{Other Carbon Depletion Mechanisms}
In this work, we have only focused on photolysis and irreversible sublimation as refractory carbon depletion mechanisms. There are other mechanisms which could potentially increase the carbon depletion efficiency (e.g. photo-/thermochemically induced processes). However, they will likely only further decrease the carbon fraction if they are active in the denser disk regions close to the midplane. Any photo-induced mechanism that is active in the UV irradiated layers of the disk, suffers from the same limitations as photolysis and will be residence-time limited. As for thermal decomposition mechanisms, irreversible sublimation is already very efficient in the inner disk, but suffers from the problem that it does not decompose the amorphous carbon compounds. Amorphous carbon only decomposes at temperatures beyond 1200 K when enough oxygen is available for its oxidation \citep[][]{Gail2017}. In our models, the 1200 K line only reaches out to $\sim 0.85$ AU during an FU Ori-type outburst, which is well within the expected formation region of the solid objects in the asteroid belt. Further, it is not clear whether oxidation is efficient enough to significantly deplete the disk during the short lasting outbursts. 

\subsection{Timing of Planetesimal Formation}
In section \ref{sec:Photolysis and Sublimation Combined} we showed that it is in principle possible to deplete the disk in a coagulation-fragmentation equilibrium to values as observed in the Solar System. However, the coagulation-fragmentation equilibrium will only persist as long as planetesimal formation has not yet commenced. Refractory carbon depletion by externally induced photo- and/or thermal decomposition will likely cease to operate efficiently once the material is locked in planetesimals with a small surface-to-mass ratio and very little internal mixing. Thus, planetesimals inherit the local carbon fraction of the dust at their formation location in space and time. Depending on the details of the planetesimal and/or planet formation process, Solar System objects are built from planetesimals formed over an extended period of time at different location in space. The resulting carbon fraction of the body will then be an average over all the planetesimals the body has formed from. The timing and location of planetesimal formation is therefore critical \citep[][]{Lichtenberg21b}. Further decomposition of refractory compounds as a result of internal heating processes after the incorporation of the material into planetesimals is also possible, but beyond the scope of this project \citep[see e.g.][]{Lichtenberg21}.

In section \ref{sec:vertical_transport}, we have discussed the effects of vertical transport as a result of turbulent mixing. However, turbulence also causes radial dust transport in addition to radial drift. In our simplified model, we have ignored the effects of radial turbulent mixing. Turbulent diffusion, which is most effective at large values of $\alpha$, is capable of mixing carbon rich material from outside the soot line to the inner disk and replenish the inner disk after a FU Ori-type outburst even when radial drift is slow. With a simple timescale argument, we expect replenishment via radial diffusion to become relevant in the region between 1 AU and 5 AU when the radial component of the turbulent alpha is larger than $\sim 6\cdot10^{-3}$. Hence, the results in the first subplot of \autoref{fig:outbursts_1} are only reasonable if the underlying turbulent mixing is anisotropic, something that could be expected e.g. from hydrodynamic turbulence caused by vertical shear instability \citep[][]{Stoll17}.

\section{Summary}
In this work, we build upon the model of \cite{Klarmann2018} and study the depletion of refractory carbon in the solid material of the early Solar System by means of a analytic model of the young Solar disk. In section \ref{sec:introduction}, we provide an overview of the problem and, in section \ref{sec:CoE}, we estimate Bulk Earth to be depleted in carbon by at least two orders of magnitude (by mass) when comparing to the ISM. Based on detailed models, bulk Earth could even be depleted by almost three orders of magnitude (see \autoref{table:table1}). We expect solid refractory carbonaceous material to be selectively decomposed and lost to the gas phase, i.e., it is \textit{destroyed}, in the time between the infall of interstellar material and the formation of the parent bodies of today's rocky Solar System objects (terrestrial planets, asteroids, etc.), i.e., within the first $10^6$ years of Solar System formation. \newline
In section \ref{sec:dust_model}, we introduce a compositional model of the solid material that was delivered to the early Solar System based on literature values (see illustration in \autoref{fig:refarctory_dust_composition}) and illustrate how it is radially distributed among today's Solar System bodies in \autoref{fig:C_fraction_overview}. We aim to reproduce current Solar System abundances via the selective removal of carbonaceous material in the early Solar disk by photo-induced and/or thermally-induced decomposition processes.\newline
In section \ref{sec:photolysis}, we study \textit{photolysis} as a specific example of a photo-induced carbon decomposition mechanism that is active in the UV-irradiated \textit{exposed layer} of the disk atmosphere. We derive an analytic solution that describes the decreasing carbon fraction $f_c(r,t)$ as a function of heliocentric radius $r$ and time $t$ (see equations (\ref{eq:initial_radius01}), (\ref{eq:surface_density_ratio}) and (\ref{eq:final_solution})) and identify a characteristic carbon depletion timescale $\tau_c$ (see equation (\ref{eq:carbon_depl_timescale})). In our most detailed model, which considers both radial and vertical transport of the solid material as well as photon forward scattering, we find the carbon depletion timescale by solving equation (\ref{eq:exposed_volume}) which results in a value of $\tau_c^\mathrm{res}=695\:\mathrm{kyr}$ at 1 AU. I.e., a value too large to reproduce Solar System abundances within the disk lifetime (see third column of \autoref{fig:unrestricted_model}). While radial transport mainly determines the steady-state solution on long timescales (gray dashed lines in \autoref{fig:optimized_model} and \autoref{fig:results_photolysis_AND_sublimation}), it only marginally influences our result on the timescales relevant for our problem \cite[in contrast to the model in][]{Klarmann2018}. It is only the inefficient vertical transport that limits the carbon depletion in our model because the UV irradiated exposed layers of the disk are depleted in carbon faster than they can be replenished by vertical mixing (see sketch in \autoref{fig:sketch}). This limitation does not only hold for the specific example of photolysis, but for any photo-induced carbon depletion mechanism which is active in the UV-irradiated \textit{exposed layers} of the disk. \newline
In appendix \ref{sec:residence_time_derivation} and equation (\ref{eq:explicit_C}), we explicitly show that the carbon depletion timescale $\tau_c^\mathrm{res}$ can either be increased by decreasing the vertical mixing timescale $t_\mathrm{mis}=1/\alpha\Omega$ or increasing the extent of the exposed layer. Thus, to first order, the depletion timescale is directly proportional to the total dust surface density $\Sigma_d$ and the dust opacity at UV wavelengths $\kappa_0$, and inversely proportional to the disk flaring angle $\Phi$. Furthermore, there is a minor (higher-order) dependency on the mass fraction of small grains relative to large grains $f_{\leq a_s}$. We find no other parameters to influence the carbon depletion timescale $\tau_c^\mathrm{res}$ in our model, this includes the stellar UV flux $F_\mathrm{UV}$. Furthermore, we find a combined change by a factor of order ten compared to our fiducial values (see \autoref{table:parametersummary}) in any of the main parameters ($\alpha \uparrow,\Sigma_d\downarrow,\kappa_0\downarrow,\Phi\uparrow$) in a direction that decrease the depletion timescale $\tau_c^\mathrm{res}$, does reproduce Solar System abundances via photolysis, even when considering radial and vertical dust transport. In section \ref{sec:Choice of Model Parameters}, we discuss our choice of initial parameters and find them to be only weakly constrained such that, even though a combined change by a factor of ten is not well justified, it is entirely plausible based on our current limited understanding. \newline
The second depletion mechanism that we study is the thermally induced irreversible sublimation (pyrolysis) of carbonaceous material (see section \ref{sec:thermal_decomposition}). We describe the evolution of the carbon fraction $f_c(r,t)$ under the influence of irreversible sublimation in a kinetic approach with equation (\ref{eq:sublimation_comp_i_solution}) and different thermal decomposition parameters for each carbonaceous compound (see equation (\ref{eq:subl_time}) and \autoref{fig:refarctory_dust_composition}). We find a steady state to from within $10^4$ years when the outward radial motion of the soot line is balanced by inward drift, such that static soot lines, corresponding to the different carbonaceous compounds, in a radial range between 0.25~AU and 0.5~AU, divide the disk into a depleted and an undepleted region (see \autoref{fig:sublimation_only_results}). As a result of the disk temperature profile, the region, where we expect Earth and chondrite parent bodies to have formed ($\sim1-3$ AU), remains entirely undepleted. Further, the carbon fraction in the depleted region only decreases to a floor value of $f_c=0.03$ i.e., an order of magnitude above Earth's abundance, which corresponds to the abundance of highly refractory amorphous carbon compound. This compound does not thermally decompose at temperatures below $\sim 1200$~K (see section \ref{sec:ther_dec_am_C}). The initial abundance of amorphous carbon in our model is only constrained by a single source, based on in situ measurements on comet Haley. Thus, the general abundance in the material that was delivered to the early Solar System could very well be much lower, decreasing the resulting floor value. Otherwise, the amorphous carbon must be decomposed by other means to reproduce Earth's carbon abundance. Moreover, increased disk temperatures are required to shift the soot lines of the less refractory carbonaceous compounds radially outward such that disk regions beyond 1 AU are depleted. We expect viscous heating to not sufficiently increase the disk temperature to also affect the formation region of chondrites. Therefore, we additionally study the effects of frequent and short-lived stellar luminosity outbursts (\textit{FU Ori-type outbursts}) on the location of the soot lines in section \ref{eq:FU Ori-type Outbursts}. We find that in our fiducial setup, one outburst every 100 kyr is enough to permanently move the soot lines from the range between $\sim0.25-0.5$ AU to $\sim 4.5-10$ AU even when radial drift acts to replenish the inner disk in between individual outbursts (see \autoref{fig:outbursts_1}). The soot lines only return to their original steady state location if the radial drift speed is large enough, which is the case for $\alpha \lesssim 4\cdot 10^{-4}$. Further complication arises for $\alpha \gtrsim 6\cdot 10^{-3}$ and isotropic turbulence, for which we expect radial diffusion to replenish the inner disk in between individual outbursts. In our analytic approach, we do not include effects of radial turbulent diffusion. Nonetheless, there is a range around $\alpha \approx 10^{-3}$ in which neither radial drift nor turbulent diffusion can efficiently replenish the disk to the inside of 4.5 AU in between stellar outbursts. In conclusion, a reproduction of Solar System abundances via irreversible sublimation requires high temperatures ($>500$~K) that move the soot lines beyond the formation region of the parent bodies of the rocky objects of the inner Solar System ($> 3$ AU), in combination with a low abundance of the most refractory carbonaceous compounds ($f_c<0.4 \%$). If these temperatures are reached only temporarily like in FU Ori-type outbursts, the turbulent alpha parameter must be on the order of $10^{-3}$ such that the inner disk regions are not replenished by radial drift or diffusion. \newline
Another class of depletion mechanisms which can be active in the disk midplane (and thus not suffering from vertical transport restrictions like photolysis) are thermally induced chemical decomposition processes. However, the most promising example, oxidation via OH, only becomes efficient at midplane temperatures $>1200$~K, and thus, is not relevant in the formation region of Earth or chondrites.\newline
In section \ref{sec:Photolysis and Sublimation Combined}, we find the combined effects of photolysis, irreversible sublimation and FU Ori-type outbursts to reproduce Solar System abundances within 700~kyr when the depletion timescale of photolysis is decreased by a factor of five compared to our fiducial model. Specifically, we achieve this by decreasing the dust surface density $\Sigma_d$ by a factor of five (see \autoref{fig:results_photolysis_AND_sublimation}). \newline
In section \ref{sec:bouncing}, we argue that, if large dust grains become the dominant contributor to the UV-opacity as a result of bouncing, the carbon depletion timescale of photolysis can become as low as $\tau_c^{\mathrm{ph,B}}=40\:\mathrm{kyr}$. However, the effects of bouncing collisions are beyond the scope of this paper. \newline
Even though our fiducial setup does not reproduce carbon abundances of the Solar System, we demonstrate that, under specific but plausible conditions, photo- and/or thermally induced decomposition processes of carbonaceous material in the disk phase of the early Solar System, do reproduce carbon abundances that are observed today. 

\section*{Acknowledgements}
We thank Vignesh Vaikundaraman and Joanna Dr{\k{a}}{\.z}kowska for fruitful discussions and the anonymous referee for their thorough review and the helpful comments that lead to the improvements of this paper. T.B. acknowledges funding from the European Research Council (ERC) under the European Union's Horizon 2020 research and innovation programme under grant agreement No 714769. F.B. and T.B. acknowledges funding from the Deutsche Forschungsgemeinschaft under Ref. no. FOR 2634/1 and under Germany's Excellence Strategy (EXC-2094-390783311).

\section*{Data availability}
The data underlying this article will be shared on reasonable request to the corresponding author.

\bibliographystyle{mnras}
\bibliography{references} 

\appendix

\section{Derivation of the location of the optical surface}
\label{sec:tau=1_surface_derivation}
In this section, we derive an analytic (but approximate) expression of the location of the $\tau=1$-surface of stellar FUV photons. I.e., we solve the following equation 
\begin{equation}
    1=\tau(r,z)
    \label{eq:1=tau}
\end{equation}
We first use the disk flaring angle $\Phi$ and a geometric argument to convert the optical depth in radial direction ($\tau$) to the optical depth in vertical direction ($\tau_z$): 
\begin{equation}
    \tau=\frac{1}{\Phi}\tau_z
\end{equation}
To compute the optical depth at height z along an optical path in vertical direction, we assume grains of size $a_1$ at the geometrical optics limit dominate the FUV opacity $\kappa_0$. Then we integrate along the vertical axis from $z'=z_\tau$ to $z'=\infty$, to calculate the value of the vertical optical depth at height $z$ 
\begin{equation}
    \tau_z(z) = \kappa_0\int^\infty_{z}\rho_{a_1}\mathrm{d}z'
    \label{eq:integralXX}
\end{equation}
where $\rho_{a_1}$ is the local volume density of dust grains of size $a_1$ or smaller. The integral on the right-hand side of equation (\ref{eq:integralXX}) is the fraction of surface density of grains smaller or equal to $a_1$ above height $z$, which we set equal to $f_{\leq a_1}f_{\geq z}\Sigma_d$. Here, $f_{\leq a_1}$ is the mass fraction of grains of size $a_1$ or smaller. Assuming a grain size distribution between $a_\mathrm{min}$ and $a_\mathrm{max}$ that follows a power-law as $n\propto a^{-p}$ we calculate $f_{\leq a_1}$ with 
\begin{equation}
    f_{\leq a_1} = \frac{a_\mathrm{max}^{-p+4}-a_\mathrm{min}^{-p+4}}{a_1^{-p+4}-a_\mathrm{min}^{-p+4}} \approx \bigg( \frac{a_\mathrm{max}}{a_1}\bigg)^{-p+4}
\label{eq:mass_fractin_small_grains}
\end{equation}
where we have assumed $a_\mathrm{min}\ll a_1<a_\mathrm{max}$. We set $p=3.5$. \newline
The expression $f_{\geq z}$ stands for the dust surface density fraction in the layers between $z$ and $z=\infty$. We first assume that the vertical dust distribution is Gaussian, i.e., grains do not decouple from the gas in the upper disk layers. Then we calculate the dust surface density fraction in the layers between $z$ and $z=\infty$ as
\begin{equation}
    f_{\geq z} = \frac{1}{2}\mathrm{erfc}\bigg(\frac{z}{\sqrt{2}h_d}\bigg)
    \label{eq:f_geeq_z}
\end{equation}
Assuming, the grains smaller or equal to $a_1$ are well coupled to the gas, their vertical distribution is Gaussian with a scale height equal to the gas scale height $h_d\sim h_g$. The factor $1/2$ in equation (\ref{eq:f_geeq_z}) comes from the fact that we only consider one side of the disk. We define a new dimensionless variable $X:=z/(\sqrt{2}h_d)$ and rewrite equation (\ref{eq:1=tau}), with the definitions made above: 
\begin{equation}
    1=\frac{\kappa_0}{\Phi}f_{\leq a_1}f_{\geq z_\tau}\Sigma_d=\frac{\kappa_0}{2\Phi}f_{\leq a_1}\Sigma_d\mathrm{erfc}\big(X\big)
    \label{eq:1=erfcX}
\end{equation}
This, equation can easily be solved numerically, however, we will now make some approximations to obtain an explicit expression. For $X\gg1$, the complementary error function can be approximated using its asymptotic expansion 
\begin{equation}
    \mathrm{erfc}\big(X\big)\approx\frac{e^{-X^2}}{\sqrt{\pi}X}
    \label{eq:asymptotic_expansion}
\end{equation}
We replace the complementary error function in equation (\ref{eq:1=erfcX}) with this approximation and define a function 
\begin{equation}
    g(X):=\frac{\kappa_0}{2\sqrt{\pi}\Phi}f_{\leq a_1}\Sigma_d e^{-X^2}
\end{equation}
Then equation (\ref{eq:1=erfcX}) becomes 
\begin{equation}
    X-g(X)=0. 
    \label{eq:x-g_X}
\end{equation}
This equation is still not solvable explicitly. Therefore, we Taylor expand $g(X)$ about $X=X_0$ to second order. We find the most accurate results across a large range of parameters if we chose $X_0$ such that $g(X_0)=1$. Then, $X_0^2$ can be written as
\begin{equation}
    X_0^2=\ln \frac{f_{\leq a_1}\Sigma_d\kappa_0}{2\sqrt{\pi}\Phi}.
    \label{eq:higher_orderX0}
\end{equation}
The Taylor expansion of equation (\ref{eq:x-g_X}) and solving the quadratic equation in X gives us the following results for $X$:
\begin{equation}
    X=\frac{2X_0+\frac{1}{2}X_0^{-2}-\sqrt{2X_0^{-1}-X_0^{-2}+\frac{3}{2}X_0^{-4}}}{2-X_0^{-2}}
\end{equation}
For $X_0\gg1$, we simplify the above equation to 
\begin{equation}
    X\simeq X_0-\frac{1}{5}
     \label{eq:solution_to_capital_X}
\end{equation}
where we chose the subtrahend such that it minimizes the error in the range between 3 and 5 times the gas pressure scale height $h_g$. In terms of $z_\tau$ and model parameters, equation (\ref{eq:solution_to_capital_X}) is equivalent to 
\begin{equation}
    \frac{z_\tau}{h_d} \simeq \sqrt{2\ln\frac{f_{\leq a_1}\Sigma_d\kappa_0}{2\sqrt{\pi}\Phi}}-\frac{1}{5}
    \label{eq:z_tau_result}
\end{equation}
In the derivation up to now we have not considered the effects of dust grains decoupling from the gas at large $z$. Next, we also consider the effects of decoupling in the upper disk layers when calculating the $\tau=1$-surface. In that case, the vertical dust density profile does not follow a Gaussian anymore. In the upper disk layers, the dust density drops off steeper than a Gaussian profile. We consider the effects of decoupling by calculating $f'_{\geq z}$ with the following integral \citep[][]{Fromang2009}:
\begin{equation}
     f'_{\geq z} = \frac{f_0}{2}\int^\infty_z\exp\bigg[ -\frac{St_\mathrm{mid}}{\alpha}\bigg(\exp \bigg(\frac{z^2}{2h_g^2} \bigg)-1 \bigg)-\frac{z^2}{2h_g^2}\bigg]\mathrm{d}z
\end{equation}
where $f_0$ is a normalization constant such that $f'_{\geq 0} =1/2$. Similar to the case without decoupling in equation (\ref{eq:1=erfcX}), the solution to equation 
\begin{equation}
    1=\frac{\kappa_0}{\Phi}f_{\leq a_1}f'_{\geq z}\Sigma_d
    \label{eq:1=flz}
\end{equation}
will provide height of the $\tau=1$-surface when also considering decoupling. In \autoref{fig:comparison}, we plot the radial dependence of the solution to equation (\ref{eq:1=flz}) in our fiducial model with the solid blue line. Due to the effects of decoupling, the $\tau=1$-surface lies lower than when decoupling is ignored. \newline
To evaluate when the effects of decoupling become important, we evaluate at which height small grains decouple from the gas. The gas density drops away exponentially at large $z$, as shown in equation (\ref{eq:gasvolumedensity}). Hence, there is a height $z_\mathrm{dec}$ above which the gas density is low enough for even the smallest dust grains to decouple from the gas. We consider a grain to be decoupled if the local Stokes number of a grain is equal or smaller than the turbulent $\alpha$-parameter: $St(z_\mathrm{dec})\simeq\alpha$. Solving for $z_\mathrm{dec}$ provides an upper limit, above which equation (\ref{eq:z_tau_result}) is only accurate if one does not consider the decoupling of dust grains. The explicit expression for $z_\mathrm{dec}$ is the following: 
\begin{equation}
    \frac{z_\mathrm{dec}}{h_g}=\sqrt{2\ln\frac{2\alpha\Sigma_g}{\pi \rho_\bullet a}}
\label{eq:decoupling_height}
\end{equation}
The height above which small grains decouple, calculated with equation (\ref{eq:decoupling_height}) is shown in the top panel of \autoref{fig:comparison} in black color. Studying equation (\ref{eq:decoupling_height}), it becomes clear that larger particles with a larger solid density are more weakly coupled to the gas and therefore decouple at smaller $z/h_g$. On the other hand, particles decouple at larger $z/h_g$ if the gas surface density $\Sigma_g$ or the alpha turbulence parameter $\alpha$ are large. We find that the solution to equation (\ref{eq:z_tau_result}) deviates from the solution to equation (\ref{eq:1=flz}) when $z_\tau\gtrsim z_\mathrm{dec}$ and decoupling becomes important. In \autoref{fig:comparison}, $z_\mathrm{dec}$ lies close to the solution of equation (\ref{eq:z_tau_result}) which is the reason why the solution of equation (\ref{eq:z_tau_result}) and equation (\ref{eq:1=flz}) deviate slightly in our fiducial model. Nonetheless, equation (\ref{eq:z_tau_result}) provides valuable scaling relations for our analysis. However, we use the exact numerical solution of equation (\ref{eq:1=flz}) in all our quantitative results. 
\newline

\section{Derivation of the residence time }
\label{sec:residence_time_derivation}
In this section, we derive an expression to calculate the residence time, i.e., the time a grain spends in the exposed layer before being mixed back into the denser disk regions closer to the midplane. We model vertical motions of dust particles in the strong coupling approximation with a stochastic equation of motion \citep[e.g.][]{Ciesla2010,Ormel2018}
\begin{equation}
    \mathrm{d}z=-\zeta z\mathrm{d}t+\sqrt{2D_d}\mathrm{d}W_t
    \label{eq:stochastic_eom}
\end{equation}
where $D_d$ is the dust diffusivity and $W_t$ denotes a Wiener process. We define the dust diffusivity as $D_d=\alpha c_s h_g$. The differential of the Wiener process is $\mathrm{d}W_t=\sqrt{\mathrm{d}t}\mathcal{N}(0,1)$ where $\mathcal{N}(0,1)$ is the normal distribution with zero mean and unit variance. The effective velocity term $\zeta z$ is the sum of the vertical settling velocity at height z and a correction accounting for the gas density gradient
\begin{equation}
    \zeta = St\Omega +\frac{D_d}{h_g^2}
    \label{eq:def_zeta}
\end{equation}
For $z\ll z_\mathrm{dec}$ the correction term is generally much larger than the settling term. Hence, $\zeta\simeq D_d/h_g^2$ in these regions. \newline
If we assume $\zeta$ is independent of $z$ (e.g., by approximating $St(z)\sim St(z_{t=0})$), equation (\ref{eq:stochastic_eom}) denotes an Ornstein-Uhlenbeck process \citep[][]{Uhlenbeck1930} for which the probability density function $f$ as a function of z is a time dependent Gaussian
\begin{equation}
    f(z,t)=\frac{1}{\sqrt{2\pi \upsilon^2(t)}}\exp \bigg(\frac{(z-\bar{z}(t))^2}{2\upsilon^2(t)}\bigg)
    \label{eq:Gaussian_prob}
\end{equation}
with mean $\bar{z}(t) = z_{t=0}e^{-\zeta t}$ and variance $\upsilon^2(t)=\frac{D_d}{\zeta}\big(1-e^{-2\zeta t} \big)$ where $z_{t=0}$ is the location of the particle at time $t=0$.\newline
We now assume that $z_1$ is the lower boundary of the exposed layer in the disk. Then for every $t>0$, the following integral 
\begin{equation}
    P_{\geq z_1}(t)=\int_{z_1}^\infty f(z,t)\mathrm{d}z
\end{equation}
is the probability of finding a particle which had started at $z=z_{t=0}$ at $t=0$ above $z_1$ at time $t>0$. For an ensemble of N particles, all starting at $z=z_{t=0}$ at $t=0$, the average number of particles in the active layer above $z_{t=0}$ at time $t>0$ then becomes $N P_{\geq z_1}(t)$. The number of particles outside the exposed layer is $N (1-P_{\geq z_1}(t))$. In the time interval $(t,t+\mathrm{d}t)$, the particles in the exposed layer spend a time of $\mathrm{d}t$ in the exposed layer. The particles outside the active layer spend a total time $0$ in the exposed layer during that interval. Thus, the average time spent in the active layer in the time interval $(t,t+\mathrm{d}t)$ across the entire ensemble is
\begin{equation}
    \langle \mathrm{d}t_\mathrm{res}\rangle=\frac{N P_{\geq z_1}(t)\cdot \mathrm{d}t+N(1- P_{\geq z_1}(t))\cdot0}{N}
\end{equation}
Here, $\langle\cdot \rangle$ denotes the ensemble average. The above equation simplifies to 
\begin{equation}
    \langle \mathrm{d}t_\mathrm{res}\rangle=P_{\geq z_1}(t)\mathrm{d}t
    \label{eq:ensembleaveragedtres}
\end{equation}
In order to calculate the total time spent in the active layer across all times $t$, we integrate equation (\ref{eq:ensembleaveragedtres}) from $t=0$ to $t=\infty$
\begin{equation}
   \int_0^\infty P_{\geq z_1}(t)\mathrm{d}t
\end{equation}
However, this integral does generally not converge. This is not a problem because we are not interested in the total time spent in the active layer across all times $t$ but only up to the point when  the ensemble equalizes its carbon fraction in collisions, typically in the dense midplane regions. This typically happens within one mixing time $t_\mathrm{mix}$. Thus, we define the residence time as the integral between 0 and $t_\mathrm{mix}$
\begin{equation}
   \langle t_\mathrm{res}\rangle=\int_0^{t_\mathrm{mix}} P_{\geq z_1}(t)\mathrm{d}t
\end{equation}
One might choose a different value for the upper limit of the integral. However, the above integral is very insensitive to the exact choice of its upper limit as long as the upper limit is larger than $\sim t_\mathrm{mix}/2$. Next, we assume the ergodic hypothesis to be true. Then, the ensemble averaged residence time $\langle t_\mathrm{res}\rangle$ is equal to the time averaged residence time $t_\mathrm{res}$ i.e., the average time a single particle spends, in the active layer above $z_1$, before it is recycled through collisions under the condition that it reaches height $z_1$ at least once before the recycling happens. The residence time $t_\mathrm{res}$ is then defined as
\begin{equation}
   t_\mathrm{res}=\int_0^{t_\mathrm{mix}} P_{\geq z_1}(t)\mathrm{d}t
   \label{residence_time_definition}
\end{equation}
By defining a dimensionless variable 
\begin{equation}
    \chi(t)=\frac{z_1-\bar{z}(t)}{\sqrt{2}\upsilon(t)}
\end{equation}
we rewrite equation (\ref{residence_time_definition}) as an integral over the complementary error function
\begin{equation}
    t_\mathrm{res}=\frac{1}{2}\int_0^{t_\mathrm{mix}} \mathrm{erfc}\big(\chi(t)\big)\mathrm{d}t
\label{eq:residence_time_integral}
\end{equation}
where $t_\mathrm{mix}=1/\alpha\Omega$, $\bar{z}(t) = z_1e^{-\zeta t}$ is the time-dependent mean and $\upsilon^2(t)=\frac{D_d}{\zeta}\big(1-e^{-2\zeta t} \big)$ is the time-dependent variance. \newline
Unless otherwise stated, we will use equation (\ref{eq:residence_time_integral}) in our analysis to calculate the residence time. However, due to its implicit form, we can not gain much insight into the dependence on model parameters from equation (\ref{eq:residence_time_integral}). Therefore, we now aim to find an explicit, but approximate, expression of the residence time which we will use to study how the residence time depends on the chosen model parameters. This will be useful because the residence time is a limiting factor in the efficiency of photolysis of carbonaceous material. \newline
First, we assume dust grains to be perfectly coupled to the gas. Then, their residence time follows the green colored line in \autoref{fig:residence_time} and $D_d/h_g^2\gg St\Omega$ holds. Thus, the first term in equation (\ref{eq:def_zeta}) does not contribute, and we approximate equation (\ref{eq:def_zeta}) with $\zeta \simeq D_d/h_g^2$. Furthermore, in this perfectly coupled case, we approximate the residence time in equation (\ref{eq:residence_time_integral}) for $z\gg h_g$ as
\begin{equation}
t_\mathrm{res}\simeq\Big(\frac{h_g}{z}\Big)^2t_\mathrm{mix}    
\label{eq:t_res_k}
\end{equation}
We plot this dependence in blue in \autoref{fig:residence_time}. This is the same expression as the definition of the residence time in \cite{Klarmann2018} (see their equation (6)). However, in comparison to equation (\ref{eq:residence_time_full_equation}), equation (\ref{eq:t_res_k}) is divergent for $z\to0$. Further, we assume the lower edged of the exposed layer to be identical to the $\tau$=1-surface ($z_1\sim z_\tau$), which holds in the absence of photon forward scattering. Then we plug equation (\ref{eq:tau1_surface}) into equation (\ref{eq:t_res_k}) and find a rough scaling relation between the residence time of perfectly coupled grains and the model parameters
\begin{equation}
    t_\mathrm{res}\simeq\Bigg(\sqrt{2\ln\frac{f_{\leq a_1}\Sigma_d \kappa_0}{2\sqrt{\pi}\Phi}}-\frac{1}{5}\Bigg)^{-2}\alpha^{-1}\Omega^{-1}.
\label{eq:explicit_residence_time}
\end{equation}
In a residence-time-limited case, i.e., when $\tau_c^\mathrm{res}$ is lager than the unrestricted depletion time, we know from equation (\ref{eq:residencetimeliiteddesrate}), carbon depletion is more efficient if $t_\mathrm{res}$ is small. The residence time, as expressed in equation (\ref{eq:explicit_residence_time}), becomes small if the ratio inside the logarithm is large. However, this ratio is generally larger than unity and, because it is inside the natural logarithm, its impact on the residence time is small. The residence time is more sensitive to the alpha turbulence parameter, to which it is inversely proportional. It becomes clear that the residence time can be most efficiently decreased, and consequently carbon depletion increased, if the alpha parameter is increased. \newline
In \autoref{fig:residence_time}, we plot the z-dependence of the residence time. There, equation (\ref{eq:t_res_k}) is plotted in blue. The exact solution to equation (\ref{eq:residence_time_full_equation}), for perfectly coupled grains ($\zeta=D_d/h_g^2$), is plotted in green. The exact and the approximate solutions agree well for $z\gg h_g$. Using our fiducial parameters, equation (\ref{eq:t_res_k}) overestimates the residence time for $z<3h_g$. In \autoref{fig:residence_time}, we also plot the exact solution of the residence time for grains which decouple at large z in orange color. It clearly deviates above $z_\mathrm{dec}$ which is indicated with the vertical dashed black line. When including effects of decoupling, the residence time becomes more than an order of magnitude smaller for $z>z_\mathrm{dec}$ than without considering the decoupling. The dashed orange vertical line shows the location of the exposed layer. It is clearly below $z_\mathrm{dec}$. By choosing appropriate parameters, we could push $z_1$, as calculated with equation (\ref{eq:tau1_surface}), to larger z, beyond $z_\mathrm{dec}$ where the residence time becomes very low. However, when considering decoupling, it becomes very difficult to push $z_1$ beyond $z_\mathrm{res}$ (at least by choosing physically plausible parameters). At heights close to $z_\mathrm{dec}$, grains settle vertically so efficiently that only few grains are diffused beyond $z_\mathrm{dec}$. Hence, the dust density in layers beyond $z_\mathrm{dec}$ is so low that these layers remain optically thin and consequently $z_1<z_\mathrm{dec}$ holds. Thus, the no-decoupling assumption in the derivation of equation (\ref{eq:explicit_residence_time}) is generally justified.

\section{Derivation of the analytic solution of the carbon fraction}
\label{sec:derivation}
In this section, we derive the solutions given by equations (\ref{eq:initial_radius01}), (\ref{eq:surface_density_ratio}) and (\ref{eq:final_solution}) as discussed in section \ref{sec:analytic_solution} based on the radial transport equations presented in section \ref{sec:radial_dust_transport}.
\subsubsection{Solution without Radial Transport ($v_r=0$)}
First, we derive the solution to equation (\ref{eq:carbon_evolution}) without radial transport, i.e., by setting $v_r=0$. For the silicate component ($i=0$), the solution is simply 
\begin{equation}
    \Sigma_{s}(r,t)=\Sigma_{s}(r,t')
\end{equation}
where $t'$ is an arbitrary time with $t'<t$, most commonly identified as the initial time $t'=0$. Due to the absence of a mechanism that depletes silicate in our model, the silicate surface density remains constant in the absence of radial transport. Assuming the carbon depletion timescale for all the carbonaceous components is identical ($\tau_{c,i}=\tau_c$), the solution to equation (\ref{eq:carbon_evolution}) summed up over all the carbonaceous components is 
\begin{equation}
    \Sigma_c(r,t)=\Sigma_c(r,t')\exp\Big(-\frac{t-t'}{\tau_c}\Big( \frac{r}{r_0} \Big)^{p_d-b}\Big)
\end{equation}
where we have assumed $\Sigma_c\ll\Sigma_s$. The surface density ratio $f_{\Sigma}$ between carbon and silicate grains then becomes 
\begin{equation}
    f_{\Sigma}(r,t)=\frac{\Sigma_c(r,t)}{\Sigma_s(r,t)}=f_{\Sigma}(r,t')\cdot\exp\Big(-\frac{t-t'}{\tau_c} \Big( \frac{r}{r_0} \Big)^{p_d-b}\Big)
    \label{eq:nort_solution}
\end{equation}
and the carbon fraction $f_c$ as a function of radius and time is
\begin{equation}
    f_c(r,t)=\bigg(1+\Big(f_{\Sigma}(r,t)\Big)^{-1}\bigg)^{-1}
    \label{eq:cfraction_from_sig_fraction}
\end{equation}

\subsubsection{Steady State}
The steady-state solution of the silicate components in equation  (\ref{eq:carbon_evolution}) can be found by setting $d\Sigma_s/dt =  0$. Thus, $r\Sigma_s v_r=const.$ and we identify the constant as $\dot{M}_s/2\pi$. Hence, the steady-state solution of the silicate component in equation (\ref{eq:carbon_evolution}) is: 
\begin{equation}
    \Sigma_s=\frac{\dot{M}_s}{2\pi r_0 \abs{v_0}}\bigg(\frac{r}{r_0}\bigg)^{-(l+1)}
\end{equation}
Hence, $p_d=l+1=s-q+3/2$. In the fragmentation limit (s=q), we find $p_d=3/2$.

\subsubsection{Grain Trajectory}
We now aim to solve equation (\ref{eq:carbon_evolution}) for the carbonaceous components, including radial transport. For this, we first follow a trajectory of an arbitrary grain in the disk, moving at radial velocity described by equation (\ref{eq:radial_velocity}). We rewrite equation (\ref{eq:radial_velocity}) as
\begin{equation}
    v_0dt = \bigg(\frac{r}{r_0}\bigg)^{-l}dr
    \label{eq:radial_velocity2}
\end{equation}
and assume a grain to start at radius $r'$ at time $t'\geq0$ and radially drift towards $r(t)$ where the grain arrives at time $t>t'$. We integrate equation (\ref{eq:radial_velocity2}) from $r'$ to $r$
\begin{equation}
    \int_{t'}^t v_0dt = \int_{r'}^{r} \bigg(\frac{r''}{r_0}\bigg)^{-l}dr''
    \label{eq:radial_velocity2.1}
\end{equation}
where $r''$ is the integration variable, and solve for $r'$ to find the initial radius as a function of the grain position $r$ at time $t$. For $l\neq1$ we find
\begin{equation}
    r'(r,t,t')=\Big(r^{1-l}-(1-l)r_0^{-l}v_0(t-t')\Big)^{1/(1-l)}.
    \label{eq:initial_radius}
\end{equation}
Note that this equation has a physical solution for all times $t$ only if $l<1$ (assuming $v_0<0$). For $l>1$, there is a maximum time $t_\mathrm{max}$ for any given $r$ and $t'$ at which the initial radius diverges $r'=\infty$: 
\begin{equation}
    t_\mathrm{max}(r,t') = t'+\frac{r^{1-l}r_0^l}{(1-l)v_0}
\end{equation}
meaning, grains can drift from infinite distance to radius $r$ in a finite amount of time. This solution can still be physical if the disk itself is finite. The solution to equation (\ref{eq:radial_velocity2.1}) for $l=1$ is:
\begin{equation}
    r'(r,t)=r\cdot\exp\big(-\frac{v_0}{r_0}(t-t')\big)
    \label{eq:initial_radius2}
\end{equation}
We want to highlight that for most cases, it is sufficient to chose $t'=0$ which simplifies the solution. However, carrying on with $t'$ will allow us to apply this solution to problems which are piece-wise defined by power-law dependencies instead of a single, global power law.

\subsubsection{Characteristic Equations}
For the next steps in the derivation, it is convenient to introduce the dimensionless variable $x=r/r_0$ such that equation (\ref{eq:carbon_evolution}), summed over all the carbonaceous components, becomes

\begin{equation}
    \frac{\partial \Sigma_c}{\partial t} + \frac{1}{xr_0}\frac{\partial}{\partial x}( x \Sigma_c v_r) = \dot{\Sigma}_c
    \label{eq:carbon_evolution2}
\end{equation}
Following the approach in \cite{Birnstiel14} appendix A, we solve the partial differential equation (PDE) in equation (\ref{eq:carbon_evolution2}) with the method of characteristics. Along characteristic trajectories, the PDE simplifies to a system of ordinary differential equations (ODE) which we solve analytically. The three characteristic equations of equation (\ref{eq:carbon_evolution2}) are:
\begin{equation}
    \frac{dt}{ds}=1
    \label{eq:first_characteristic}
\end{equation}
\begin{equation}
   \frac{d x}{d s}=\frac{v_r}{r_0}
   \label{eq:second_characteristic}
\end{equation}
\begin{equation}
    \frac{d \Sigma_c}{d s} =-\frac{\Sigma_c}{xr_0}\frac{\partial(v_r x)}{\partial x}+\dot{\Sigma}_c
    \label{eq:third_characteristic}
\end{equation}

\subsubsection{Homogeneous Solution}
The homogeneous solution ($\dot{\Sigma}_c=0$) of equation (\ref{eq:carbon_evolution2}) can be found by using $d(v_r x)=\partial (v_r x)/\partial x \cdot dx$ and rewriting equation (\ref{eq:third_characteristic}) as
\begin{equation}
    \frac{d\Sigma_c}{\Sigma_c}=-\frac{1}{xr_0}d(v_r x)\frac{d s}{d x} 
\end{equation}
and using equation (\ref{eq:second_characteristic}) to write the above expression as 
\begin{equation}
    \frac{d\Sigma_c}{\Sigma_c}=-\frac{d(v_r x)}{v_r x}
\end{equation}
which is equivalent to 
\begin{equation}
    d\ln\Sigma_c=-d\ln(v_r x).
    \label{eq:thisarbitraryequation}
\end{equation}
From equation (\ref{eq:thisarbitraryequation}) we see that the solution is $\Sigma_c\propto 1/(v_r x)$ and thus: 
\begin{equation}
    \Sigma_c(r,t)=  \Sigma_c(r',t')\frac{v_r(r')r'}{v_r(r)r}
\end{equation}
With using definition in equation (\ref{eq:radial_velocity}), this simplifies to:
\begin{equation}
    \Sigma_{c,\mathrm{hom.}}(r,t,t')=\Sigma_c(r',t')\cdot\bigg( \frac{r'}{r}\bigg)^{l+1}
\end{equation}
where $r'$ is the initial radius of the particle at time $t'$ as found in equation (\ref{eq:initial_radius}).
Coincidentally, this is also the general form of the solution to the silicate component in equation (\ref{eq:carbon_evolution})
\begin{equation}
    \Sigma_s(r,t,t')=\Sigma_s(r',t')\cdot\bigg( \frac{r'}{r}\bigg)^{l+1}
    \label{eq:dust_solution_hom}
\end{equation}

\subsubsection{Inhomogeneous Solution}
We continue with equation (\ref{eq:third_characteristic}) to find the inhomogeneous solution and rewrite the equation as 
\begin{equation}
\begin{split}
d ln\Sigma_c  &=-\frac{1}{xr_0}\Big( x\frac{\partial v_r}{\partial x}+v_r\Big)ds+\frac{\dot{\Sigma}_c}{\Sigma_c}ds \\
& =-\Big( l\frac{v_0}{r_0}x^{l-1}+\frac{v_0}{r_0}x^{l-1}+x^{p_d-b}/\tau_{c,0}\Big)ds
\end{split}
\label{eq:inhomo_2}
\end{equation}

\paragraph*{Case I} ($l-1=p_d-b$)\newline
First, we consider the case when $l-1=p_d-b$, this is equivalent to an (initial) steady-state surface density distribution. In this case, we can rewrite equation (\ref{eq:inhomo_2}) as
\begin{equation}
\begin{split}
d ln\Sigma_c  &=-x^{l-1}\frac{v_0}{r_0}\Big( l+1+\frac{r_0}{v_0\tau_{c,0}}\Big)ds \\
&=-x^{-1}\Big( l+1+\frac{r_0}{v_0\tau_{c,0}}\Big)dx
\end{split}
\label{eq:inhomo_3}
\end{equation}
where we have used the relation $ds=(r_0/v_0)x^{-l}dx$ in the second line. With defining 
\begin{equation}
    B = l+1+\frac{r_0}{v_0\tau_{c,0}}
    \label{eq:def_B}
\end{equation}
we can rewrite equation (\ref{eq:inhomo_3}) as
\begin{equation}
    d ln\Sigma_c = -B \:d ln x
    \label{eq:inhomo_4}
\end{equation}
The solution to equation (\ref{eq:inhomo_4}) is:
\begin{equation}
    \Sigma_c(r,t)=C'\cdot\bigg( \frac{r}{r_0}\bigg)^{-B}
\end{equation}
where $C'$ is a constant (in $r$) that can depend on time $t$. We require $\Sigma_c(r,t=t')=\Sigma_c(r',t')$ because $r=r'(t=t')$. Hence, $C'=\Sigma_c(r',t')(r_0/r')^{-B}$ and the solution to the inhomogeneous equation for $l-1=p_d-b$ is: 
\begin{equation}
    \Sigma_c(r,t)=\Sigma_c\big(r'(r,t,t'),t'\big)\cdot\bigg( \frac{r}{r'(r,t,t')}\bigg)^{-B}
    \label{eq:inhomo_5}
\end{equation}
Combining equation (\ref{eq:inhomo_5}) and equation (\ref{eq:dust_solution_hom}), we can also find the solution for the surface density ratio
\begin{equation}
    f_\Sigma(r,t) = \frac{\Sigma_c(r,t)}{\Sigma_d(r,t)} = f_{\Sigma}\big(r'(r,t,t'),t'\big)\cdot\bigg( \frac{r}{r'(r,t,t')}\bigg)^{-\frac{r_0}{v_0\tau_{c,0}}}
    \label{eq:fc_caseI}
\end{equation}
from which the carbon fraction $f_c$ follows directly from equation (\ref{eq:cfraction_from_sig_fraction}).

\paragraph*{Case II} ($l-1\neq p_d-b$)
We start again with equation (\ref{eq:inhomo_2}) and use the relation $ds=(r_0/v_0)x^{-l}dx$ to write the equation as following: 
\begin{equation}
    d ln \Sigma_c = -\Big( l+1+\frac{r_0}{v_0\tau_{c,0}}x^{p_d-b-l+1}\Big)d ln x
\end{equation}
We now substitute $x=\exp(\tilde{x})$ ($x>0$ holds) and define $\beta:=p_d-b-l+1$ and find 
\begin{equation}
    \frac{d ln \Sigma_c}{d\tilde{x}} = -(l+1)-\frac{r_0}{v_0\tau_{c,0}}\exp\big(\tilde{x}\beta\big)
    \label{eq:cas2_eq2}
\end{equation}
The solution to equation (\ref{eq:cas2_eq2}) can be identified as 
\begin{equation}
    ln \Sigma_c = -(l+1)\tilde{x}-\frac{r_0}{v_0\tau_{c,0}\beta}\exp\big(\tilde{x}\beta\big) +ln C''
\end{equation} where $C''$ is a (yet unidentified) constant which can depend on $t$. Then we substitute back by using $\tilde{x}=ln x$ and find 
\begin{equation}
    ln \Sigma_c = ln\Big( x^{-(l+1)}\Big)-\frac{r_0}{v_0\tau_{c,0}\beta}x^\beta +ln C''
\end{equation}
which we solve for $\Sigma_c$ by applying the exponential function
\begin{equation}
    \Sigma_c = x^{-(l+1)}\exp\Big(-\frac{r_0}{v_0\tau_{c,0}\beta}x^\beta\Big)\cdot C''
    \label{eq:inhom_eq25}
\end{equation}
Next, we evaluate this equation at $r=r'$ and $t=t'$ to find an expression for the constant $C''$:
\begin{equation}
    C''=\Sigma_c(r',t')\Big(\frac{r_0}{r'}\Big)^{-(l+1)}\exp\bigg(\frac{r_0}{v_0\tau_{c,0}\beta}\Big(\frac{r'}{r_0}\Big)^\beta\bigg)
    \label{eq:constant_c_dash}
\end{equation}
inserting equation (\ref{eq:constant_c_dash}) back into equation (\ref{eq:inhom_eq25}) leads to the inhomogeneous solution for the surface density $\Sigma_c$:
\begin{equation}
    \Sigma_c(r,t)=\Sigma_c(r',t')\Big( \frac{r}{r'}\Big)^{-(l+1)}\exp\bigg(\frac{r_0}{v_0\tau_{c,0}\beta}\bigg[\Big(\frac{r'}{r_0}\Big)^\beta-\Big(\frac{r}{r_0}\Big)^\beta\bigg]\bigg)
\end{equation}
And from this, we can also find the solution for the surface density ratio:
\begin{equation}
\begin{split}
    f_\Sigma(r,t) &= \frac{\Sigma_c(r,t)}{\Sigma_s(r,t)} \\
    & = f_{\Sigma}(r',t')\cdot\exp\bigg(\frac{r_0}{v_0\tau_{c,0}\beta}\bigg[\Big(\frac{r'}{r_0}\Big)^\beta-\Big(\frac{r}{r_0}\Big)^\beta\bigg]\bigg)
    \label{eq:fc_caseII}
\end{split}
\end{equation}
where \newline
$l=s-q+1/2$\newline
\begin{equation}
    \beta= p_d-b-l+1= p_d-b-s+q+1/2
    \label{eq:beta}
\end{equation}
Note; by taking the limit $v_0\to0$ in equation (\ref{eq:fc_caseII}) we obtain equation (\ref{eq:nort_solution}), i.e., the solution without radial transport. The same is true for taking the limit $v_0\to0$ of equation (\ref{eq:fc_caseI}).

\section{Derivation of the analytic solution for irreversible sublimation}
\label{sec:sublimation_solution}
In this section, we derive the analytic solution to the carbon fraction considering the carbon depletion via irreversible sublimation, analogous to section \ref{sec:derivation}, but with a source term that follows an exponential law rather than a power law. 
\subsubsection{Steady-State Solution}
First, we search for the steady state solution of equation (\ref{eq:carbon_evolution}) with a source term of the form as in equation (\ref{eq:source_terM_sublimation}). I.e. the equation we want to solve is
\begin{equation} \label{eq:solve_this_eq}
    \frac{\partial \Sigma_{d,i}}{\partial t} + \frac{1}{r}\frac{\partial}{\partial r}( r \Sigma_{d,i} v_r) = 
    -\Sigma_{d,i}A_i\exp\bigg(-\frac{E_{a,i}}{RT(r)}\bigg) 
\end{equation}
In the steady state, we require the time derivative to vanish. With using the dimensionless radius $x=r/r_0$, the steady state condition becomes
\begin{equation}
    \frac{\partial}{\partial x}( x \Sigma_{d,i} v_r) = 
    -\Sigma_{d,i} r_0 xA_i\exp\bigg(-\frac{E_{a,i}}{RT_0}x^q\bigg) 
\end{equation}
By defining $g(x)=x \Sigma_{d,i} v_r$, we rewrite the equation above to read
\begin{equation}
    \frac{\partial}{\partial x}( \ln g) = 
    \frac{r_0}{v_0}A_ix^{-l}\exp\bigg(-\frac{E_{a,i}}{RT_0}x^q\bigg) 
\end{equation}
\paragraph*{Case I} ($1-q-l=0$)\newline
If we assume $1-q-l=0$, the above equation can easily be integrated:
\begin{equation}
   \ln g(x)=-\frac{r_0}{v_0 q}\frac{A_iRT_0}{E_{a,i}}\exp\bigg(-\frac{E_{a,i}}{RT_0}x^q\bigg) +\mathrm{const.}
\end{equation}
Hence,
\begin{equation}
    r\Sigma_{d,i}v_r=\exp \bigg[-\frac{r_0}{v_0 q}\frac{A_iRT_0}{E_{a,i}}\exp\bigg(-\frac{E_{a,i}}{RT(r)}\bigg)\bigg]\cdot\mathrm{const.}
\end{equation}
We require $\Sigma_{d,i}=\mathrm{const.}$ as $T_0\to 0$, thus, 
\begin{equation}
    \Sigma_{d,i}(r,t)=\Sigma_{d,i}(r,t') \exp \bigg[-\frac{r_0}{v_0 q}\frac{A_iRT_0}{E_{a,i}}\exp\bigg(-\frac{E_{a,i}}{RT(r)}\bigg)\bigg]
\end{equation}
which results in the steady state surface density ratio 
\begin{equation}
    f_{\Sigma_{d,i}}(r,t) = f_{\Sigma_{d,i}}(r',t')\exp \bigg[-\frac{r_0}{v_0 q}\frac{A_iRT_0}{E_{a,i}}\exp\bigg(-\frac{E_{a,i}}{RT(r)}\bigg)\bigg]
\end{equation}

\subsubsection{Inhomogeneous Solution}
In the next step, we search for the inhomogeneous solution of equation (\ref{eq:solve_this_eq}). We use dimensionless radius $x=r/r_0$ and the characteristic equations:
\begin{equation}
    \frac{dt}{ds}=1
\end{equation}
\begin{equation}
   \frac{\partial x}{\partial s}=\frac{v_r}{r_0}
\end{equation}
\begin{equation}
    \frac{\partial \Sigma_{d,i}(s)}{\partial s} =-\frac{\Sigma_{d,i}}{xr_0}\frac{\partial(v_r x)}{\partial x}-\Sigma_cA_i\exp\bigg(-\frac{E_{a,i}}{RT(x)}\bigg) 
\end{equation}
With the equations above, we find 
\begin{equation}
    ds=dx\cdot r_0/v_r
\end{equation}
and thus
\begin{equation}
    d ln\Sigma_{d,i} = -\bigg[(l+1)+\frac{r_0}{v_0}x^{-l+1}A_i\exp\bigg(-\frac{E_{a,i}}{RT_0}x^q \bigg)\bigg]d\ln x
\end{equation}
which we rewrite to
\begin{equation}
    d ln\Sigma_c = -\bigg[(l+1)\frac{1}{x}+\frac{r_0}{v_0}x^{-l}A_i\exp\bigg(-\frac{E_{a,i}}{RT_0}x^q \bigg)\bigg]dx
\end{equation}
We define a variable $\tilde{x}=x^q$ and write the equations in terms of $\tilde{x}$:
\begin{equation}
    d ln\Sigma_c = -(l+1)\tilde{x}^{-1/q}d\tilde{x}-\frac{r_0}{v_0}\tilde{x}^{(1-q-l)/q}A_i\exp\bigg(-\frac{E_{a,i}}{RT_0}\tilde{x} \bigg)d\tilde{x}
\end{equation}
\paragraph*{Case I} ($1-q-l=0$)\newline
Then the equation simplifies to:
\begin{equation}
    d ln\Sigma_{d,i} = -(l+1)x^{-1}dx-\frac{r_0}{v_0}\frac{A_i}{q}\exp\bigg(-\frac{E_{a,i}}{RT_0}\tilde{x} \bigg)d\tilde{x}
\end{equation}
which can be integrated to give: 
\begin{equation}
    ln\Sigma_{d,i} = -(l+1)\ln x+\frac{r_0}{v_0}\frac{A_iRT_0}{qE_{a,i}}\exp\bigg(-\frac{E_{a,i}}{RT_0}x^q \bigg)+const.
\end{equation}
and then 
\begin{equation}
    \Sigma_{d,i}(x,t) = C'\cdot x^{-(l+1)}\exp\bigg[\frac{r_0}{v_0}\frac{A_iRT_0}{qE_{a,i}}\exp\bigg(-\frac{E_{a,i}}{RT_0}x^q \bigg)\bigg]
\end{equation}
Next, we evaluate the above equation at $r=r'$ and $t=t'$ to find an expression for the constant $C'$:
\begin{equation}
    C'=\Sigma_{d,i}(r',t')\cdot \Big( \frac{r'}{r_0}\Big)^{l+1}\exp\bigg[-\frac{r_0}{v_0}\frac{A_iRT_0}{qE_{a,i}}\exp\bigg(-\frac{E_{a,i}}{RT_0}\Big( \frac{r'}{r_0}\Big)^q \bigg)\bigg]
\end{equation}
which leads to the full solution
\begin{equation}
 \begin{split}
    &\Sigma_{d,i}(r,t) = \Sigma_{d,i}(r',t')\Big( \frac{r}{r'}\Big)^{-(l+1)}\cdot\\ &\exp\bigg\{\frac{r_0}{v_0}\frac{A_iRT_0}{qE_{a,i}} \bigg[\exp\bigg(-\frac{E_{a,i}}{RT_0}\Big( \frac{r}{r_0}\Big)^q \bigg)-\exp\bigg(-\frac{E_{a,i}}{RT_0}\Big( \frac{r'}{r_0}\Big)^q \bigg)\bigg]\bigg\}
 \end{split}
\end{equation}
and also the solution of the surface density ratio:
\begin{equation}
\begin{split}
    &f_{\Sigma_{d,i}}(r,t) = f_{\Sigma_{d,i}}(r',t')\cdot \\
    &\exp\bigg\{\frac{r_0}{v_0}\frac{A_iRT_0}{qE_{a,i}} \bigg[\exp\bigg(-\frac{E_{a,i}}{RT_0}\Big( \frac{r}{r_0}\Big)^q \bigg)-\exp\bigg(-\frac{E_{a,i}}{RT_0}\Big( \frac{r'}{r_0}\Big)^q \bigg)\bigg]\bigg\}
\end{split}
\end{equation}

\begin{table*}
\caption{List of notations}
\begin{center}
\begin{tabular}{l l}
 \hline
  symbol & description [cgs unit]    \\ 
   \hline
   \hline
 $\Sigma_c$             & surface density of carbonaceous material [$\mathrm{g\:cm^{-2}}$]\\
 $\Sigma^*$           & surface density in the exposed layer, Eq. (\ref{eq:sigma_d_star_definition}) [$\mathrm{g\:cm^{-2}}$]\\
 $\dot{\Sigma}_c$       & destruction rate of carbonaceous material [$\mathrm{g\:cm^{-2}s^{-1}}$]\\
 $\Sigma_d$             & dust surface density, Eq. (\ref{eq:dust:surf_dens}) [$\mathrm{g\:cm^{-2}}$]\\
 $\dot{\Sigma}_d$       & destruction rate of the $i$-th dust compound [$\mathrm{g\:cm^{-2}s^{-1}}$]\\
 $\dot{\Sigma}_d$       & dust destruction rate [$\mathrm{g\:cm^{-2}s^{-1}}$]\\
 $\Sigma_g$             & gas surface density, Eq. (\ref{eq:gas_surface_density}) [$\mathrm{g\:cm^{-2}}$]\\
 $\Sigma_s$             & surface density of silicate material [$\mathrm{g\:cm^{-2}}$] \\
 $\Phi$                 & disk flaring angle [$\mathrm{rad}$] \\
 $\Omega$               & Keplerian angular frequency [$\mathrm{s^{-1}}$]\\
 
 $\alpha$               & dimensionless turbulence parameter [1] \\
 $\beta$                & power-law exponent defined in Eq. (\ref{eq:beta}) [1]\\
 $\gamma$               & modulus of the power-law exponent of the gas pressure [1]\\
 $\zeta$                & settling rate as defined in Eq. (\ref{eq:def_zeta}) [$\mathrm{s^{-1}}$]\\
 $\kappa$               & opacity [$\mathrm{cm^2g^{-2}}$]\\
 $\kappa_0$             & UV-opacity [$\mathrm{cm^2g^{-2}}$]\\
 $\lambda$              & wavelength [$\mathrm{cm}$]\\
 $\rho_g$               & gas volume density, Eq. (\ref{eq:gasvolumedensity}) [$\mathrm{g\: cm^{-3}}$]\\
 $\sigma$               & geometrical cross-section [$\mathrm{cm^2}$]\\
 $\tau$                 & optical depth in radial direction [1]\\
 $\tau_c$               & carbon depletion timescale, Eq. (\ref{eq:carbon_depl_timescale}) [s] \\
 $\tau_{c}^\mathrm{ph}$ & unrestricted photolysis carbon depletion timescale, Eq. (\ref{eq:unrestricted_depl_timescale}) [s] \\
 $\tau_{c}^\mathrm{res}$ & residence time limited carbon depletion timescale, Eq. (\ref{eq:residencetimeliiteddesrate}) [s] \\ 
 $\tau_{c}^\mathrm{eff}$ & effective carbon depletion timescale, Eq. (\ref{eq:eff_depl_ts}) [s] \\ 
 $\tau^\mathrm{sub}_{c}$ & sublimation depletion timescale, Eq. (\ref{eq:subl_depl_time}) [s] \\ 
 $\tau_z$               & vertical optical depth [1]\\
 $\upsilon$             & standard deviation (variance) of Eq. (\ref{eq:Gaussian_prob}) [$\mathrm{cm}$] \\
 $\chi$                 & dimensionless vertical coordinate defined in Eq. (\ref{eq:chi}) [1] \\

 $A$                    & exponential prefactor in the Arrhenius law [$s^{-1}$]\\
 $B$                    & dimensionless constant defined in equation (\ref{eq:def_B}) [1]\\
 $C', C''$              & integration constants [$\mathrm{g\:cm^{-2}}$] \\
 $D_d$                  & dust diffusivity [$\mathrm{cm^2\:s^{-1}}$]\\
 $E_a$                  & activation energy of sublimation [erg]\\
 $F_\mathrm{UV}$        & FUV flux [$\mathrm{cm^{-2}s^{-1}}$]\\
 $L_*$                  & stellar luminosity [erg s$^{-1}$]\\
 $L_\mathrm{UV}$        & stellar UV luminosity [erg s$^{-1}$]\\
 $M_*$                  & stellar mass [g] \\
 $M_c$                  & total mass of carbonaceous material [g] \\
 $\dot{M}_d$            & dust mass accretion rate [$\mathrm{g\:s^{-1}}$]\\
 $M_s$                  & total silicate mass [g]\\
 $\dot{M}_s$            & non-carbon grains mass accretion rate [$\mathrm{g\:s^{-1}}$]\\
 $M_{g,\mathrm{tot}}$   & total gas mass [g]\\
 $\mathcal{N}$          & normal distribution\\
 $P_g$                  & gas pressure [$\mathrm{dyn\:cm^{-2}}$]\\
 $R$                    &gas constant [$\mathrm{erg\: K^{-1}mol^{-1}}$]\\
 $R_\mathrm{UV}$        & photolysis rate [$\mathrm{s^{-1}}$]\\
 $St$                   & Stokes number, Eq. (\ref{eq:Stokes_number_para}) [1]\\
 $T$                    & disk temperature, Eq. (\ref{eq:disk_temp}) [K]\\
 $T_\mathrm{burst}$     & elevated disk temperature during an FU Ori-type outburst [K]\\
 $W_t$                  & Wiener process \\
 $Y_\mathrm{ph}$                    & photolysis yield [1]\\
 
   \end{tabular}
  \hspace{1em}
  \begin{tabular}{l l}
 $a$                    & dust grain radius [cm]\\
 $a_B$                  & dust grain radius at the bouncing barrier, Eq. (\ref{eq:bouncing_barrier_radius}) [cm] \\
 $a_s$                  & radius of opacity dominating small grains [cm] \\
 $b$                    & p.l.i. of the carbon grain destruction time $t_d$ [1] \\
 $b_\mathrm{res}$       & fit parameter power law index of the residence time [1]\\
 $c_s$                  & isothermal sound speed [$\mathrm{cm\:s ^{-1}}$]\\
 $f$                    & probability density function \\
 $f_\Sigma$             & surface density ratio, Eq. (\ref{eq:surface_density_ratio}) [1]\\
 $f_c$                  & carbon fraction [1] \\
 $f_\mathrm{f}$         & fragmentation limit calibration factor [1]\\
 $f_s$                  & silicate mass fraction [1] \\
 $f_{\leq a_s}$         & mass fraction of grains smaller or equal to $a_s$ \\
 $f_{\geq z}$           & dust surface density fraction above height $z$ \\
 $h_g$                  & vertical gas scale height [cm]\\
 $k$                    & rate of sublimation, Eq. (\ref{eq:rate_law}) [$\mathrm{s^{-1}}$] \\
 $k_B$                  & Boltzmann constant [$\mathrm{cm^2\:g\:s^{-2}\:K^{-1}}$]\\
 $l$                    & p.l.i. of the radial drift velocity $v_r$ [1] \\
 $m$                    & mass of a dust grain [g]\\
 $m_c$                  & mass of a carbon atom [g]\\
 $m_g$                  & mean weight of a gas molecule [g]\\
 $\rho_\bullet$         & dust grain solid density [$\mathrm{g\:cm^{-3}}$]\\
 $p_d$                  & dust surface density p.l.i., Eq. (\ref{eq:dust:surf_dens}) [1]\\
 $p_g$                  & gas surface density p.l.i., Eq. (\ref{eq:gas_surface_density}) [1]\\
 $q$                    & gas temperature p.l.i., Eq. (\ref{eq:disk_temp}) [1]\\
 $r$                    & radial coordinate (distance from the star) [cm]\\
 $r'$                   & position of a drifting grain at $t=t'$, Eq. (\ref{eq:initial_radius01}) [cm]\\
 $r''$                  & integration variable [cm]\\
 $r_0$                  & reference radius (1 AU) [$\mathrm{cm}$] \\
 $r_\mathrm{out}$       & outer edge of the disk [$\mathrm{cm}$]\\
 $s$                    & Stokes number p.l.i., , Eq. (\ref{eq:Stokes_number_para}) [1] \\
 $t$                    & time [s]\\
 $t'$                   & initial time [s]\\
 $t_d$                  & carbon grain destruction time, Eq. (\ref{eq:power_law_depletion_time}) [s]\\
 $t_\mathrm{ph}$        & photolysis destruction time, Eq. (\ref{eq:c_destrcutiontime}) [s]\\
 $t_\mathrm{ph,B}$      & photolysis destruction time for bouncing, Eq. (\ref{eq:bouncing_dest_time}) [s]\\
 $t_\mathrm{res}$       & residence time, Eq. (\ref{eq:residence_time_full_equation}) [s]\\
 $t_\mathrm{mix}$       & mixing time [$\mathrm{s}$]\\
 $t^\mathrm{sub}_{d}$   & sublimation destruction time, Eq. (\ref{eq:subl_time}) [s]\\
 $v_f$                  & fragmentation velocity [$\mathrm{cm\:s ^{-1}}$]\\ 
 $v_r$                  & radial drift velocity, Eq. (\ref{eq:drift_velr_para}) [$\mathrm{cm\:s ^{-1}}$]\\
 $x$                    & dimensionless radial coordinate [1]\\
 $z$                    & vertical coordinate (distance above the midplane) [cm]\\
 $z_1$                  & lower boundary of the exposed layer [cm]\\
 $z_\tau$               & height of the $\tau=1$-surface, Eq. (\ref{eq:tau1_surface}) [cm]\\
 $z_\mathrm{dec}$       & height at which dust grains decouple, Eq. (\ref{eq:decoupling_height_main}) [cm]\\
 $[.]_0$                & arbitrary quantity $[.]$ evaluated at the reference radius $r_0$\\
$[.]_i$                & arbitrary quantity $[.]$ describing the $i$-the dust compound\\

 \hline
\end{tabular}
\label{table:notations}
\end{center}
\end{table*}

\bsp	
\label{lastpage}
\end{document}